\documentclass[aps,pra,superscriptaddress,twocolumn,10pt,floatfix,nofootinbib]{revtex4-2}

\usepackage[utf8]{inputenc}
\usepackage{amsmath}
\usepackage{amsfonts}
\usepackage{latexsym}
\usepackage{amssymb}
\usepackage{bm}
\usepackage{graphicx} 
\usepackage{xcolor}
\usepackage{hyperref}
\hypersetup{colorlinks=true,linkcolor=blue,citecolor=blue,urlcolor=cyan}
\usepackage{array}
\usepackage{tabularx}

\begin{document}

\title{Supersymmetry of dissipative Bose-Fermi systems with application to Jaynes-Cummings and Dicke models}

\author{Colin V. Coane}
\affiliation{Center for Theoretical Physics, Sloane Physics Laboratory,
Yale University, New Haven, CT 06520-8120, USA}

\author{Francesco Iachello}
\affiliation{Center for Theoretical Physics, Sloane Physics Laboratory,
Yale University, New Haven, CT 06520-8120, USA}

\date{\today}

\begin{abstract}
We demonstrate how supersymmetries of Hamiltonians for coupled
Bose-Fermi systems can be used to place the Hamiltonians of the
Jaynes-Cummings model and Dicke model under the rotating wave
approximation in matrix form and provide
explicit analytic solutions for their eigenvalues. We then use this
supersymmetry to place the Liouvillians of the associated
Markovian open systems in matrix form and provide explicit solutions
for their eigenvalues. These results are a consequence of the fact that
the Hamiltonian of the Jaynes-Cummings model commutes with the
linear Casimir invariant of the superalgebra $u(1|1)$ and that the
Hamiltonian of the Dicke model commutes both with the linear
invariant of $\sum_{i} u_{i}(1|1)$ and with the invariant of
an additional $su(2)$ algebra. Our methods apply to various coupled
Bose-Fermi systems with $u(1|1)$ and more generally with $u(n|m)$
dynamical superalgebras, and may provide efficient tools for 
studying more complicated examples.
\end{abstract}

\maketitle

\section{Introduction} \label{sec:intro}

Symmetries have provided a useful tool for solving Hamiltonian problems
in closed form. Dynamic symmetries, situations in which the Hamiltonian $H$
is written in terms of Casimir operators $C_{i}$ of a chain of algebras
$g_{1} \supset g_{2} \supset \ldots$, $H = f\left( C_{i} \right)$, provide solutions
to the eigenvalue problem in terms of the expectation values of the Casimir
operators, $\left\langle C_{i} \right\rangle$, in the representation of the algebras
$g_{1} \supset g_{2} \supset \ldots$, with energies $E = f\left( \left\langle C_{i} \right\rangle \right)$.
A related situation occurs when $H$ is not written in terms of Casimir operators
but it commutes with them, $[H,C_{i}] = 0$, in which case the eigenvalue problem
splits into diagonalization of matrices in the basis provided by irreducible
representations $\Lambda$ of the algebra, with dimensions $\dim \left[\Lambda\right]$.

In 1974, for applications to particle physics, the concept of symmetry was
enlarged to that of supersymmetry \cite{neveu,ramond,volkov,wess}.
Supersymmetries have found applications in a variety of fields, especially
in nuclear physics \cite{iac1,balantekin,iac5,iac3} where
the concept of ``dynamical supersymmetry'' was introduced \cite{iac1}.
Dynamic supersymmetries are situations in which the Hamiltonian of
a Bose-Fermi system is written in terms of Casimir operators of a
superalgebra $g^{\star}$ \cite{kac,ramond2,bars},
providing a closed form solution for the eigenvalues of $H$. These types
of supersymmetry have been particularly useful in nuclear physics
\cite{iac1,balantekin,iac5,iac3}. Also here, when $H$ is not written in
terms of the Casimir operators $C_{i}^{\star}$ but it commutes with them,
$[H,C_{i}^{\star}] = 0$, the eigenvalue problem splits into diagonalization
of matrices in the basis provided by the irreducible representations of
$g^{\star}$. This basis is characterized by conserved quantum
numbers, which are the expectation values $\left\langle C_{i}^{\star} \right\rangle$ of $C_{i}^{\star}$.

In this article, we further enlarge the use of symmetries and supersymmetries
to the study of Markovian open quantum systems described by Lindblad master
equations \cite{lindblad,sudarshan,breuer} and show with specific examples
how this concept produces explicit analytic solutions for the eigenvalues of
the Liouvillian superoperator, the generator of the dynamics. 

The class of supersymmetries discussed here applies to all coupled
Bose-Fermi systems composed of $n$ bosons and $m$ fermions. Introducing
boson creation, $a^{\dag}_{\alpha}$, and annihilation, $a_{\alpha}$, operators
$(\alpha = 1, ... , n)$, satisfying $[a_{\alpha}, a^{\dag}_{\beta}] = I \delta_{\alpha \beta}$,
and fermion creation, $f^{\dag}_{i}$, and annihilation, $f_{i}$, operators $(i = 1, ... , m)$,
satisfying $\{ f_{i}, f^{\dag}_{k} \} = I \delta_{i k}$, and taking all bilinear products
of creation and annihilation operators, one can construct a superalgebra $u(n|m)$
\cite{iac3} with elements written in matrix form
\begin{equation}
\label{eq:unm-alg}
u(n|m) \doteqdot
\left( 
\begin{array}{cc}
G_{\alpha \beta}=a^{\dag}_{\alpha} a_{\beta} & F^{\dag}_{i \alpha}=f^{\dag}_{i} a_{\alpha} \\ 
F_{\alpha i}=a^{\dag}_{\alpha} f_{i} & G_{i k}=f^{\dag}_{i} f_{k}
\end{array}
\right)
\end{equation}
with bosonic sector $u(n|m) \supset  u_{B}(n) \oplus u_{F}(m)$ and
appropriate commutation and anticommutation relations between elements.

A particularly simple coupled Bose-Fermi system is a
coupled qubit-photon system. The Hamiltonian of one such system, the
Jaynes-Cummings (JC) model \cite{jaynes,larson} can be written as
\begin{equation}
\label{eq:jc-ham}
H_{JC}=\omega_{c}a^{\dag}a+\omega_{q}\frac{\sigma_{z}}{2}+g\left(
\sigma_{+}a+\sigma_{-}a^{\dag}\right) .
\end{equation}
Here, $a^{\dag},a$ are boson creation and annihilation operators satisfying 
$\left[ a,a^{\dag}\right] =1$. The operators $a,a^{\dag},I$ form an algebra,
called the oscillator algebra $os_{B}(1)$. By adding the number operator
$a^{\dag}a$, one has the Heisenberg algebra $h(2)$ \cite{iac4} with
commutation relations $\left[ a,a^{\dag}\right] =I$,
$\left[ a,I\right] =[a^{\dag},I]=0$, $[a,a^{\dag} a]=a$,
$\left[ a^{\dag},a^{\dag}a\right] =-a^{\dag}$.
The spin operators $\sigma_{+},\sigma_{-},\sigma_{z}$
satisfy the usual commutation relations of the $su(2)$
algebra $\left[ \sigma_{+},\sigma_{-}\right] =\sigma_{z}$,
$\left[\sigma_{z},\sigma_{\pm}\right] =\pm 2\sigma_{\pm}$.
However, since Pauli matrices anticommute,
$\sigma_{+}$ and $\sigma_{-}$ can also be viewed
as fermion operators $\sigma_{+} \equiv f^{\dag}$,
$\sigma_{-} \equiv f$, with anticommutation relations
$\{ f, f^{\dag} \} = I$. The three operators $f^{\dag},f,I$
form an algebra, called the fermion oscillator algebra,
$os_{F}(1)$. By adding the operator $f^{\dag}f$, one has
the Holstein-Primakoff realization of $su(2)$ with elements
$\sigma_{+} = f^{\dag}$, $\sigma_{-} = f$, and
$\sigma_{z} = 2 f^{\dag}f - 1$ \cite{holstein}.
Furthermore, boson operators commute with fermion operators,
$\left[a,f\right] =\left[ a^{\dag},f\right] = \left[a,f^{\dag}\right] =\left[ a^{\dag},f^{\dag}\right] =0$.
The spectrum generating algebra of
$H_{JC}$ can thus be recast as a $u(1|1)$ superalgebra. 
The JC model, describing light-matter interactions between
a two-level quantum system and a single bosonic mode,
has been the subject of many theoretical investigations
\cite{buck,sukumar,tomka,shen,gutierrez}, and can be
experimentally studied in platforms such as with trapped ions
\cite{pedernales,leibfried}, cavity QED \cite{walther,blais1,fink},
and circuit QED \cite{wallraff,gambetta,mezzacapo,felicetti1,blais2}.

It has been known for some time that the Hamiltonian problem for
the JC model can be solved in closed form through the conservation
of the total excitation number \cite{jaynes}. In the first part of this article,
we recast this conservation law in terms
of a $u(1|1)$ supersymmetry, then use this supersymmetry to derive explicit
expressions for the eigenvalues of the Liouvillian of the JC model, and of all models
obtained from it by simple transformations and generalizations. While the analytic
results for the eigenvalues of the Hamiltonian of these models were previously known,
those for the eigenvalues of the Liouvillian are novel and previously unknown.

For applications to quantum optics, one may also study
many qubits coupled to photons. One such model is
the Dicke Hamiltonian under the rotating wave approximation,
also known as the Tavis-Cummings model,
\begin{equation}
\label{eq:dicke-ham}
H_{D} = \omega_{c}a^{\dag}a + \omega_{q} \sum_{i=1}^{N} \frac{\sigma_{z,i}}{2}
+ g  \sum_{i=1}^{N} \left( \sigma_{+,i}a+\sigma_{-,i}a^{\dag}\right) .
\end{equation}
It has also been known for some time that this Hamiltonian
is solvable in the resonant case, $\omega_{c} = \omega_{q}$
\cite{tavis}. In the second part of this
article, we recast the Hamiltonian of the Dicke model in terms of the
superalgebra $\sum_{i} u_{i}(1|1)$, show that $H_{D}$ has an additional
quasi-spin $su(2)$ symmetry, and then derive explicit
results for the eigenvalues of the Hamiltonian and Liouvillian.
In particular, we show explicit results for the entire spectrum of
eigenvalues of the Hamiltonian and of the Liouvillian for two
qubits coupled to photons in the resonant case, a system which
can be investigated experimentally in cavity QED \cite{gambetta},
and explicit results for the eigenvalues of the Hamiltonian and
Liouvillian for low-lying states
for any number of qubits $N$ coupled to photons.
The eigenvalues of the Liouvillian
here are also novel and previously unknown, and
the latter result is particularly important for quantum computation.

\section{Hamiltonian Eigenvalues} \label{sec:jc-ham}

Already in 1989, it was shown that the Hamiltonian of the JC model has a $u(1|1)$
dynamical superalgebra \cite{rasetti} with elements written in matrix form
\begin{equation}
\label{eq:u11-alg}
u(1|1) \doteqdot
\left( 
\begin{array}{cc}
G_{B}=a^{\dag}a & F^{\dag}=\sigma_{+}a \\ 
F=\sigma_{-}a^{\dag} & G_{F}=\sigma_{+}\sigma_{-}
\end{array}
\right)
\end{equation}
and linear invariant Casimir operator $C=a^{\dag}a+\sigma_{+}\sigma_{-}$
(or $C^{\prime}=a^{\dag}a+\frac{\sigma_{z}}{2}$).
$G_B, G_F$ form the bosonic sector of $u(1|1)$ and the subalgebra
$u_{B}(1) \oplus u_{F}(1) \subset u(1|1)$, while $F^{\dag}, F$ form the
fermionic sector of $u(1|1)$. 
In terms of the elements of $u(1|1)$, the Hamiltonian is
\begin{equation}
\label{eq:jc-susy}
H_{JC}=-\frac{\omega_{q}}{2}+\omega_{c}G_{B}+\omega_{q}G_{F}+g(F^{\dag
}+F)
\end{equation}
and $C$ can be viewed as the excitation number operator. $H_{JC}$
has a $u(1|1)$ supersymmetry as it commutes with $C$,
$\left[ H_{JC},C\right] =\left[ H_{JC},C^{\prime}\right] =0$,
which leads to conservation of the total excitation number.
In the $u_{B}(1) \otimes u_{F}(1)$ basis
$\left\vert n_{B},s\right\rangle =\left\vert n_{B}\right\rangle \otimes\left\vert s\right\rangle$,
where $a^{\dag} a \left\vert n_{B}\right\rangle = n_{B} \left\vert n_{B}\right\rangle$,
$n_{B}=0,1,2,...,\infty$ and
$\frac{\sigma_{z}}{2} \left\vert s\right\rangle = s \left\vert s\right\rangle$, $s = \pm \frac{1}{2}$,
$C$ is diagonal with eigenvalues $n = \left( n_{B}+s+\frac{1}{2}\right)$
counting the total number of bosonic plus fermionic excitations.
The Hamiltonian is then block diagonal in subspaces with fixed $n$.
The ground state subspace, $n=0$, is one-dimensional,
while $n>0$ subspaces are two-dimensional.
For $n=0$, the eigenvalues and eigenfunctions are trivially given, 
\begin{equation}
\label{eq:jc-ham-eig-gnd}
E_{0} = -\frac{\omega_{q}}{2} , \quad
\left\vert \psi_{0}\right\rangle = \left\vert 0,-\frac{1}{2} \right\rangle ,
\end{equation}
while for $n > 0$ they can be obtained by diagonalizing the
$2\times 2$ matrix representation of the Hamiltonian projected onto
the $n$ subspace, 
\begin{align}
\label{eq:jc-ham-eig-exc}
E_{n}^{\mp} &= \omega_{c}(n - \frac{1}{2})\mp \frac{1}{2}\sqrt{\left(
\omega_{c}-\omega_{q}\right)^{2}+4g^{2}n} , \\
\left( 
\begin{array}{c}
\left\vert \psi_{n}^{-}\right\rangle \\ 
\left\vert \psi_{n}^{+}\right\rangle
\end{array}
\right)
&= \left( 
\begin{array}{cc}
\cos \theta_{n} & \sin \theta_{n} \\ 
-\sin \theta_{n} & \cos \theta_{n}
\end{array}
\right) \left( 
\begin{array}{c}
\left\vert n-1,+\frac{1}{2}\right\rangle \\ 
\left\vert n,-\frac{1}{2}\right\rangle
\end{array}
\right) . \nonumber
\end{align}
\begin{figure}[t!]
    \centering
    \includegraphics[width=0.35\textwidth]{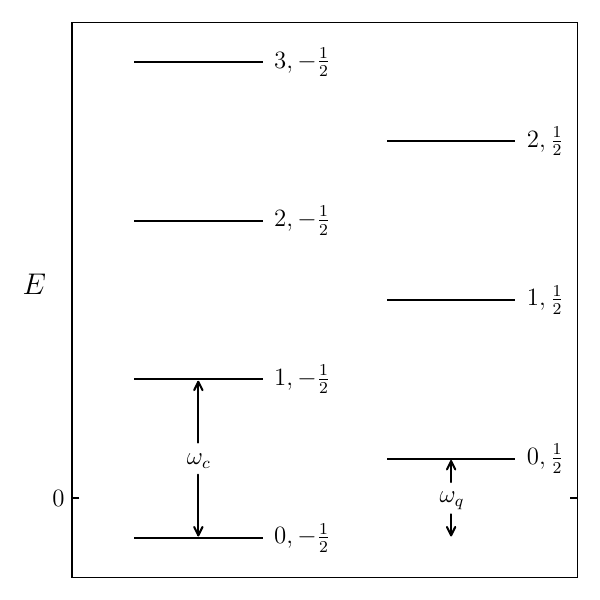}
    \caption{Spectrum of the JC Hamiltonian for $g=0$. States
    are labelled by $n_{B},s$.}
    \label{fig:jc-uncoupled-spectrum}
\end{figure}
Here, eigenfunctions are written in terms of a $so(2)$ rotation in the $n$ subspace, with
$\tan 2\theta_{n}= - \frac{2g\sqrt{n}}{\omega_{c}-\omega_{q}}$.
When $\omega_{c}=\omega_{q}$ ($\theta_{n}=\frac{\pi}{4}$), $H_{JC}$ has an additional
symmetry since its first two terms
are themselves the Casimir invariant, $C^{\prime}$, and $\cos \theta_{n} = \sin \theta_{n} = \frac{1}{\sqrt{2}}$.

\begin{figure*}[t!]
    \centering
    \includegraphics[width=0.75\textwidth]{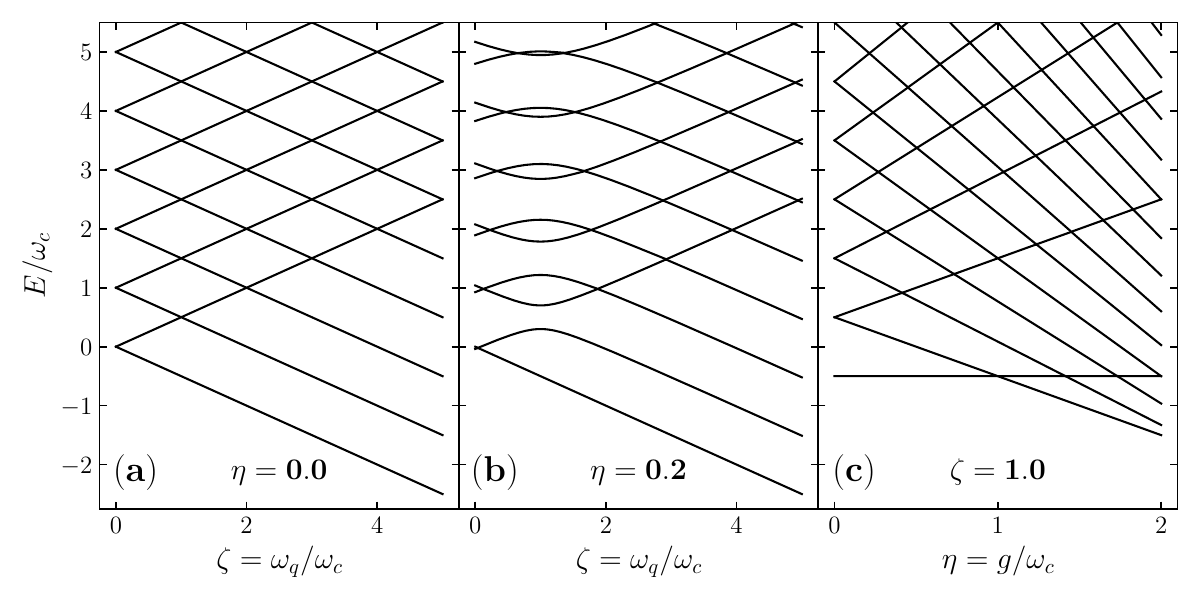}
    \caption{(a) Behavior of the eigenvalues of $H_{JC}$ as a
    function of $\zeta = \omega_{q} / \omega_{c}$ for $\eta = g / \omega_{c} = 0$.
    (b) Behavior of the eigenvalues of $H_{JC}$ as a function of $\zeta$ for $\eta = 0.2$.
    (c) Behavior of the eigenvalues of $H_{JC}$ as a function of $\eta = g / \omega_{c}$
    for $\zeta = 1$ (resonant case).}
    \label{fig-jc-scaled-spectrum}
\end{figure*}
The spectrum of the uncoupled Hamiltonian Eq.~(\ref{eq:jc-ham})
with $g=0$ is shown in Figure \ref{fig:jc-uncoupled-spectrum}.
The JC Hamiltonian scaled by $\omega_{c}$ has two parameters
$\zeta = \omega_{q} / \omega_{c}$ and $\eta = g / \omega_{c}$.
The behavior of the eigenvalues of $H / \omega_{c}$ is shown
in Figure \ref{fig-jc-scaled-spectrum} as a function of $\zeta$ and
$\eta$. Panel (a) shows that additional degeneracies occur in the
spectrum for integer values of $\zeta$. The degeneracy for $\zeta = 1$
is the well-known resonant case. These degeneracies are removed when
$g \neq 0$. Panel (b) shows the situation at $\eta = 0.2$. In this case,
the behavior of the first few levels, $g$, $1^{\pm}$, $2^{\pm}$, as a function of $\zeta$
has been measured and verified experimentally in a cavity QED system \cite{fink}.
Panel (c) shows the behavior as a function of $\eta$. At $\eta = 1$,
$g = \omega_{c}$, the ground state changes nature, flipping from
$\left\langle \frac{\sigma_{z}}{2}\right\rangle = -\frac{1}{2}$ to
$\left\langle \frac{\sigma_{z}}{2}\right\rangle = +\frac{1}{2}$. For
larger values of $\eta$, further changes occur, indicating that
the ultrastrong regime, $g \gg 1$, is somewhat complex. Also,
at these strong values, additional terms may appear in the JC
Hamiltonian.
$H_{JC}$ in Eq.~(\ref{eq:jc-ham}) is linear in the elements of
the superalgebra $u(1|1)$. To second order in the elements of
$u(1|1)$, one may have terms such as
$\left( a^{\dag}a \right) \frac{\sigma_{z}}{2}$, $\left( a^{\dag}a \right)^{2}$,
$\left( a^{\dag}a \right) \left( \sigma_{+}a + \sigma_{-}a^{\dag} \right) +
\left( \sigma_{+}a + \sigma_{-}a^{\dag} \right) \left( a^{\dag}a \right)$,
$\frac{\sigma_{z}}{2} \left( \sigma_{+}a + \sigma_{-}a^{\dag} \right) +
\left( \sigma_{+}a + \sigma_{-}a^{\dag} \right) \frac{\sigma_{z}}{2}$,
$\left( \sigma_{+}a + \sigma_{-}a^{\dag} \right)^{2}$. The Casimir
operator $C$ commutes with all these terms,
so the eigenvalues of the JC Hamiltonian with quadratic terms can also
be obtained in explicit analytic form.

\section{Liouvillian Eigenvalues} \label{sec:jc-liou}

Markovian dissipative dynamics
are described by the Lindblad equation \cite{lindblad,sudarshan,breuer},
\begin{align}
\label{eq:liouvillian-general}
\partial_{t}\rho \equiv \mathcal{L}\rho &= -\mathrm{i}\left[ H,\rho\right] +\sum_{\mu}
\kappa_{\mu}
\mathcal{D}[\Gamma_{\mu}]\rho , \nonumber \\
\mathcal{D}\left[ \Gamma_{\mu}\right] \rho &= \Gamma_{\mu}\rho\Gamma_{\mu}^{\dag}-\frac{1}{2}
\left[ \Gamma_{\mu}^{\dag}\Gamma_{\mu}\rho+\rho\Gamma_{\mu}^{\dag}\Gamma_{\mu}\right] ,
\end{align}
where $\rho$ is the system density matrix,
$\mathcal{L}$ is the Liouvillian superoperator, and $\mathcal{D}[\Gamma_{\mu}]$
are dissipators. Jump operators $\Gamma_{\mu}$ are each associated with a specific
dissipation channel occurring at a rate $\kappa_{\mu} \geq 0$ and characterize
interactions of the system with the environment \cite{breuer}.
The Liouvillian $\mathcal{L}$ has eigenvalues $\lambda_{i}$ and associated
eigenmatrices $\rho_{i}$, $\mathcal{L}\rho_{i}=\lambda_{i}\rho_{i}$.
$\mathcal{L}$ is not necessarily Hermitian and may have complex eigenvalues,
and its eigenmatrices need not be Hermitian nor orthogonal,
$\left\langle \rho_{i}, \rho_{j \neq i} \right\rangle = \mathrm{Tr} \left[\rho_{i}^{\dag} \rho_{j \neq i} \right] \neq 0$.
Eigenmatrices of the Liouvillian are not generally density matrices, but if the
Liouvillian is diagonalizable, one can construct density matrices
from Hermitian linear combinations of eigenmatrices $\rho_{i}$.

\begin{figure*}[ht!]
    \centering
    \includegraphics[width=0.75\textwidth]{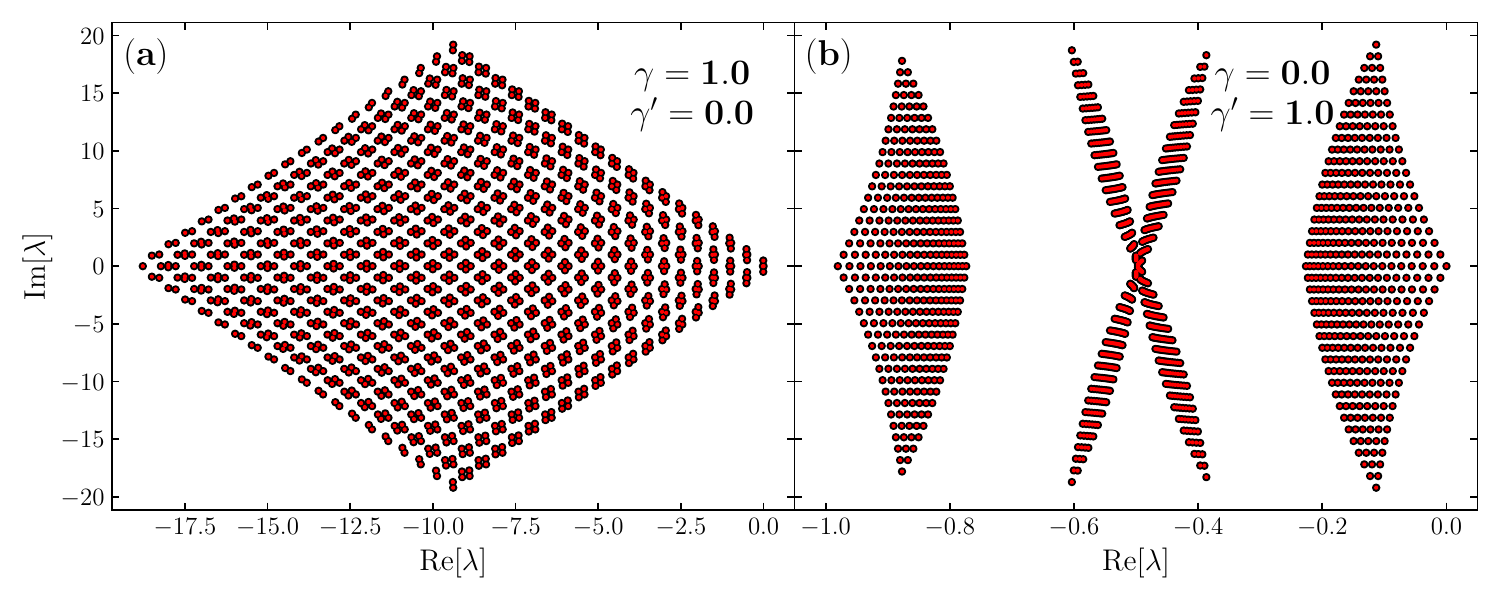}
    \caption{Scatterplot of the Jaynes-Cummings model Liouvillian eigenvalues $(\omega_{c} =1.0,\omega_{q}=0.5,g=0.1)$
    for (a) $\gamma=1.0,\gamma^{\prime}=0$, and (b) $\gamma=0,\gamma^{\prime}=1.0$. 
    Red points are evaluated from the analytic formula, Eq.~(\ref{eq:liou-eigs-freq})
    and black points are numerical results with $N_{Fock}=20$.}
    \label{fig:jc-liou-scatterplots}
\end{figure*}
For the JC model at zero temperature, we consider
single photon dissipation and qubit damping,
\begin{align}
\label{eq:liouvillian}
\mathcal{L\rho} &= -\mathrm{i}\left[ H,\rho \right] +\gamma \mathcal{D}[a]\rho+
\gamma^{\prime}\mathcal{D}[\sigma_{-}]\rho , \nonumber \\
\mathcal{D}[a]\rho &= a\rho a^{\dag}-\frac{1}{2}\left( a^{\dag}a\rho +\rho
a^{\dag}a\right) , \nonumber \\
\mathcal{D}[\sigma_{-}]\rho &= \sigma_{-}\rho \sigma_{+}-\frac{1}{2}\left(
\sigma_{+}\sigma_{-}\rho +\rho \sigma_{+}\sigma_{-}\right) .
\end{align}
A detailed derivation of the eigenvalues of the Liouvillian
is given in Appendix~\ref{appx:jc-liou}. Here we provide the main features of the
derivation. We expand $\rho $ into eigenstates of the Hamiltonian, 
$\left\{ \left\vert \psi_{n}^{s}\right\rangle \right\}$,
defining $\left\vert \psi_{0}^{-} \right\rangle \equiv \left\vert \psi_{0} \right\rangle$,
\begin{equation}
\label{eq:liou-wfn}
\rho =\sum_{n m s t} \rho_{nm}^{st}\left\vert \psi_{n}^{s}\right\rangle \left\langle
\psi_{m}^{t}\right\vert ; s,t=\pm ; n,m=0,1,2,...
\end{equation}
and consider the action of operators on $\left\vert \psi_{n}^{s}\right\rangle $.
Both $a\left\vert \psi_{n}^{s} \right\rangle$ and $\sigma_{-}\left\vert \psi_{n}^{s} \right\rangle$
decrease the number of excitations $n \rightarrow n-1$, with
$a \left\vert \psi_{0}\right\rangle = \sigma_{-} \left\vert \psi_{0}\right\rangle = 0$
annihilating the ground state. Thus, the first terms in $\mathcal{D}[a]$,
$a\left\vert \psi_{n}^{s}\right\rangle \left\langle \psi_{m}^{t}\right\vert a^{\dag}$,
and in $\mathcal{D}[\sigma_{-}]$, 
$\sigma_{-}\left\vert\psi_{n}^{s}\right\rangle \left\langle \psi_{m}^{t}\right\vert \sigma_{+}$,
either take states in the $(n,m)$ subspace into the $(n-1,m-1)$ subspace or annihilate them.
On the contrary, 
\begin{align*}
a^{\dag} a\left\vert \psi_{n}^{-} \right\rangle &=
\left( n - \cos^{2} \theta_{n} \right) \left\vert \psi_{n}^{-} \right\rangle +
\cos \theta_{n} \sin \theta_{n} \left\vert \psi_{n}^{+} \right\rangle \\
a^{\dag} a\left\vert \psi_{n}^{+} \right\rangle &=
\left( n - \sin^{2} \theta_{n} \right) \left\vert \psi_{n}^{+} \right\rangle +
\cos \theta_{n} \sin \theta_{n} \left\vert \psi_{n}^{-} \right\rangle \\
\sigma_{+}\sigma_{-}\left\vert \psi_{n}^{-}\right\rangle &=
\cos^{2} \theta_{n} \left\vert \psi_{n}^{-}\right\rangle -
\cos \theta_{n} \sin \theta_{n} \left\vert \psi_{n}^{+}\right\rangle \\
\sigma_{+}\sigma_{-}\left\vert \psi_{n}^{+}\right\rangle &=
\sin^{2} \theta_{n} \left\vert \psi_{n}^{+}\right\rangle -
\cos \theta_{n} \sin \theta_{n} \left\vert \psi_{n}^{-}\right\rangle .
\end{align*}
conserve $n$. In this case, the terms
$a^{\dag}a\left\vert \psi_{n}^{s}\right\rangle\left\langle \psi_{m}^{t}\right\vert 
+\left\vert \psi_{n}^{s}\right\rangle\left\langle \psi_{m}^{t}\right\vert a^{\dag}a$
and
$\sigma_{+}\sigma_{-}\left\vert \psi_{n}^{s}\right\rangle \left\langle \psi_{m}^{t}\right\vert 
+\left\vert \psi_{n}^{s}\right\rangle \left\langle \psi_{m}^{t}\right\vert \sigma_{+}\sigma_{-}$
leave the subspace invariant, $(n,m)\rightarrow (n,m)$. We also note that
$-\mathrm{i}\left[ H,\left\vert \psi_{n}^{s}\right\rangle \left\langle \psi_{m}^{t}\right\vert \right]=
-\mathrm{i}(E_{n}^{s}-E_{m}^{t})\left\vert \psi_{n}^{s}\right\rangle \left\langle\psi_{m}^{t}\right\vert$
always leaves $(n,m)$ subspaces invariant.

Overall, the Liouvillian acts on $(n,m)$ subspaces as $ \mathcal{L}: (n,m) \rightarrow (n,m) \oplus (n-1,m-1)$
(or $(n,m) \rightarrow (n,m)$ if $n$ or $m=0$).
$\mathcal{L}$ can be represented in a block triangular vectorized form,
where $\mathcal{L}^{(n,m)} : (n,m) \rightarrow (n,m)$
are blocks along the diagonal and $\rho_{(n,m)}$
are components of $\rho$ in the $(n,m)$ subspace,
\begin{equation}
\label{eq:liouvillian-matrix}
\mathcal{L} \rho = \left( 
\begin{array}{cccc}
\mathcal{L}^{(0,0)} & \times & & \\
 0 & \ddots & \times & \\
  & 0 & \mathcal{L}^{(n,m)} & \times \\
  & & 0 & \ddots
\end{array}
\right)
 \left( 
\begin{array}{c}
\rho_{(0,0)} \\
\vdots \\
\rho_{(n,m)} \\
\vdots
\end{array}
\right) .
\end{equation}
Eigenvalues of $\mathcal{L}$ can then be found by diagonalizing $\mathcal{L}^{(n,m)}$ blocks,
which are $1\times 1$ $(n=0,m=0)$, $2\times 2$ $(n>0,m=0),(n=0,m>0)$ and $4\times 4$ 
$(n>0,m>0)$ matrices. These submatrices, derived in Appendix~\ref{appx:jc-liou}, are
\begin{align}
\label{eq:liou-block-00}
\mathcal{L}^{(0,0)} &= (0) , \quad
\mathcal{L}^{(n,0)} = \left( 
\begin{array}{cc}
K_{n}^{+} & A_{n} \\ 
A_{n} & K_{n}^{-}
\end{array}
\right) , \\
\mathcal{L}^{(0,m)}&=\left( 
\begin{array}{cc}
K_{m}^{+} & A_{m} \\ 
A_{m} & K_{m}^{-}
\end{array}
\right) , \\
\label{eq:liou-block-nm}
\mathcal{L}^{(n,m)} &= \left( 
\begin{array}{cccc}
B^{++}_{nm} & A_{m} & A_{n} & 0 \\ 
A_{m} & B^{+-}_{nm} & 0 & A_{n} \\ 
A_{n} & 0 & B^{-+}_{nm} & A_{m} \\ 
0 & A_{n} & A_{m} & B^{--}_{nm}
\end{array}
\right) ,
\end{align}
with coefficients
\begin{align*}
K_{n}^{s} &= -\mathrm{i}(E_{n}^{s}-E_{0})-\frac{\gamma}{2}n-\frac{\gamma
^{\prime}-\gamma}{2}
\left\{ 
\begin{array}{c}
\delta_{s+} \sin^{2}\theta_{n} + \\ 
\delta_{s-} \cos^{2}\theta_{n}
\end{array}
\right\} , \\
K_{m}^{t} &= -\mathrm{i}(E_{0}-E_{m}^{t})-\frac{\gamma}{2}m-\frac{\gamma
^{\prime}-\gamma}{2}
\left\{ 
\begin{array}{c}
\delta_{t+} \sin^{2}\theta_{n} + \\ 
\delta_{t-} \cos^{2}\theta_{n}
\end{array}
\right\} , \\
B^{st}_{nm} &= -\mathrm{i}(E_{n}^{s}-E_{m}^{t})-\frac{\gamma}{2}(n+m) \nonumber \\
&- \frac{\gamma
^{\prime}-\gamma}{2}
\left( \left\{ 
\begin{array}{c}
\delta_{s+} \sin^{2}\theta_{n} + \\ 
 \delta_{s-} \cos^{2}\theta_{n}
\end{array}
\right\} +\left\{ 
\begin{array}{c}
\delta_{t+} \sin^{2}\theta_{m} + \\ 
\delta_{t-} \cos^{2}\theta_{m}
\end{array}
\right\} \right) , \\
A_{n} &= \frac{\gamma^{\prime}-\gamma}{2}\cos \theta_{n}\sin \theta_{n} , \;
A_{m}=\frac{\gamma^{\prime}-\gamma}{2}\cos \theta_{m}\sin
\theta_{m} .
\end{align*}
Each block, $\mathcal{L}^{(0,0)}$, $\mathcal{L}^{(n,0)}$, $\mathcal{L}^{(0,m)}$,
$\mathcal{L}^{(n,m)}$, can be diagonalized individually. Denoting eigenvalues
corresponding to $(n,m)$ blocks as $\lambda^{s,t}_{(n,m)}$ where $s,t = \pm, 0$,
we obtain,
\begin{align}
\label{eq:liou-eigs}
\lambda^{\pm,0}_{(n,0)} &= 
-\mathrm{i}\left[ \left(\omega_{c} -\mathrm{i}\frac{\gamma}{2} \right)\left(n - \frac{1}{2}\right) + \left(\omega_{q} -\mathrm{i}\frac{\gamma^{\prime}}{2} \right)\frac{1}{2} \right. \nonumber \\
&\pm \left. \frac{1}{2} \sqrt{\left[\left(\omega_{c} - \mathrm{i} \frac{\gamma}{2}\right) - \left(\omega_{q} - \mathrm{i} \frac{\gamma^{\prime}}{2} \right)\right]^{2} + 4g^{2}n} \right] , \nonumber \\
\lambda^{0,\pm}_{(0,m)} &= 
\mathrm{i}\left[ \left(\omega_{c} +\mathrm{i}\frac{\gamma}{2} \right)\left(m - \frac{1}{2}\right) + \left(\omega_{q} +\mathrm{i}\frac{\gamma^{\prime}}{2} \right)\frac{1}{2} \right. \nonumber \\
&\pm\left. \frac{1}{2} \sqrt{\left[\left(\omega_{c} + \mathrm{i} \frac{\gamma}{2}\right) - \left(\omega_{q} + \mathrm{i} \frac{\gamma^{\prime}}{2} \right)\right]^{2} + 4g^{2}m} \right] , \nonumber \\
\lambda^{\pm,\pm}_{(n,m)} &= \lambda^{\pm,0}_{(n,0)} + \lambda^{0,\pm}_{(0,m)} , \quad \lambda^{0,0}_{(0,0)} = 0 . 
\end{align}
Eigenvalues can be further simplified by introducing complex frequencies
$\widetilde{\omega}_{c} \equiv \omega_{c} -\mathrm{i}\frac{\gamma}{2}$ and
$\widetilde{\omega}_{q} \equiv \omega_{q} -\mathrm{i}\frac{\gamma^{\prime}}{2}$,
\begin{align}
\label{eq:liou-eigs-freq}
\lambda^{\pm,0}_{(n,0)} &= 
-\mathrm{i}\left[ \widetilde{\omega}_{c}\left(n - \frac{1}{2}\right) + \frac{\widetilde{\omega}_{q}}{2}
\pm \frac{1}{2} \sqrt{\left(\widetilde{\omega}_{c} - \widetilde{\omega}_{q}\right)^{2} + 4g^{2}n} \right] \nonumber \\
\lambda^{0,\pm}_{(0,m)} &= 
\mathrm{i}\left[ \widetilde{\omega}_{c}^{\ast}\left(m - \frac{1}{2}\right) + \frac{\widetilde{\omega}_{q}^{\ast}}{2}
\pm \frac{1}{2} \sqrt{\left(\widetilde{\omega}_{c}^{\ast} - \widetilde{\omega}_{q}^{\ast}\right)^{2} + 4g^{2}m} \right] \nonumber \\
\lambda^{\pm,\pm}_{(n,m)} &= \lambda^{\pm,0}_{(n,0)} + \lambda^{0,\pm}_{(0,m)} , \quad \lambda^{0,0}_{(0,0)} = 0
\end{align}
We observe that $\lambda^{\pm,0}_{(n,0)} = -\mathrm{i} \left( \widetilde{E}_{n}^{\pm} - \widetilde{E}_{0} \right)$
and $\lambda^{0,\pm}_{(0,m)} = \mathrm{i} \left( \widetilde{E}_{m}^{\pm \ast} - \widetilde{E}_{0}^{\ast} \right)$ where
$\widetilde{E}_{n}^{\pm}, \widetilde{E}_{0}$ are eigenvalues of the JC model Hamiltonian with 
complex frequencies $\widetilde{\omega}_{c}$ and $\widetilde{\omega}_{q}$.

\begin{figure}[ht!]
    \centering
    \includegraphics[width=0.45\textwidth]{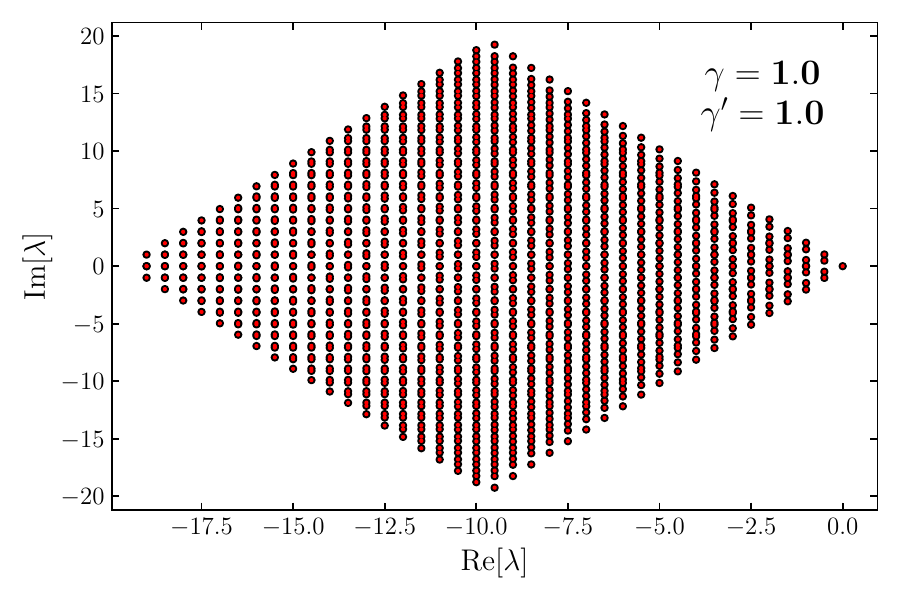}
    \caption{Scatterplot of the Jaynes-Cummings model Liouvillian eigenvalues $(\omega_{c} =1.0,\omega_{q}=0.5,g=0.1)$
    for $\gamma=\gamma^{\prime}=1.0$, and $N_{Fock}=20$. Red points are evaluated from
   Eq.~(\ref{eq:liou-susy-eig}) and black points are numerical results.}
    \label{fig:jc-liou-sym-scatterplot}
\end{figure}

The importance of this result, which is a consequence of the supersymmetry
of the JC model, is that the problem of finding eigenvalues of its Liouvillian
has been reduced to independent diagonalization of matrices which are at most $4\times 4$
and can be diagonalized in explicit analytic form.
The eigenvalues obtained from
Eqs.~(\ref{eq:liou-block-00}-\ref{eq:liou-block-nm}) have been verified
by numerical diagonalization of $\mathcal{L}$ in Eq.~(\ref{eq:liouvillian})
in a Hilbert space truncated to $N_{Fock}$ boson states.
To perform numerical calculations, we employ a vectorized
representation of operators and a matrix
representation of superoperators \cite{schmutz,minganti}.
Since $\mathcal{L}$ only
conserves or lowers excitations $(n,m)$, numerical eigenvalues are exact for
all states except when $n$ or $m=N_{Fock}$. These states can then
be identified and removed from results.
We consider first the cases
$\gamma \neq 0,\gamma^{\prime}=0$ and $\gamma =0,\gamma^{\prime}\neq 0$.
Eigenvalues for $N_{Fock}=20$ are plotted in the complex plane in Fig.~\ref{fig:jc-liou-scatterplots}.
We next consider equal dissipation strengths, $\gamma =\gamma^{\prime}$, for which
a further simplification of Eqs.~(\ref{eq:liou-block-00}-\ref{eq:liou-eigs-freq}) occurs. Here,
the excitation-conserving part of the dissipator,
\begin{align}
\label{eq:liou-susy}
&\gamma \left( a^{\dag}a\rho +\rho a^{\dag} a\right) +\gamma^{\prime}\left(
\sigma_{+}\sigma_{-}\rho +\rho \sigma_{+}\sigma_{-}\right)  \\
&= \gamma \left[ \left( a^{\dag}a+\sigma_{+}\sigma_{-}\right) \rho +\rho
\left( a^{\dag}a+\sigma_{+}\sigma_{-}\right)\right] =\gamma [ C\rho +\rho
C] , \nonumber
\end{align}
is written in terms of the $u(1|1)$ Casimir operator, $C$. This is the definition of a
dynamical supersymmetry \cite{iac1}. As a result, all blocks, Eqs.~(\ref{eq:liou-block-00}-\ref{eq:liou-block-nm}),
are diagonal and eigenvalues of $\mathcal{L}$ have the simple form
\begin{equation}
\label{eq:liou-susy-eig}
\lambda_{n,m}^{\pm ,\pm}=-\mathrm{i}\left( E_{n}^{\pm}-E_{m}^{\pm}\right) -\frac{
\gamma}{2}(n+m) .
\end{equation}
This result is verified numerically and shown in Fig.~\ref{fig:jc-liou-sym-scatterplot}.

\section{Transition Rates} \label{sec:transitions}

The Liouvillian $\mathcal{L}$ can also be used to calculate transition rates between
individual states induced by the operator $\Gamma_{\mu}$ in
Eq.~(\ref{eq:liouvillian-general}). For the JC model at zero temperature,
these are the transitions induced by single photon $a$ and qubit spin
flip $\sigma_{-}$, which are the excitation non-conserving terms
$\Gamma_{\mu} (\cdot) \Gamma_{\mu}^{\dag}$ in $\mathcal{D}\left[ a \right]$
and $\mathcal{D}\left[ \sigma_{-} \right]$ in Eq.~(\ref{eq:liouvillian}).
As discussed in the previous section and in Appendix~\ref{appx:jc-liou}, both
$a \left\vert \psi_{n}^{s}\right\rangle \left\langle \psi_{m}^{t}\right\vert a^{\dag}$
and $\sigma_{-} \left\vert \psi_{n}^{s}\right\rangle \left\langle \psi_{m}^{t}\right\vert \sigma_{+}$
take states in the $(n,m)$ subspace to the $(n-1,m-1)$ subspace or
annihilate them if $n$ or $m=0$.
By taking the overlap between the
subspaces $\mathrm{span} \left\{
\left\vert \psi_{n}^{s}\right\rangle \left\langle \psi_{m}^{t}\right\vert \right\}$
and $\mathrm{span} \left\{
\left\vert \psi_{n-1}^{s}\right\rangle \left\langle \psi_{m-1}^{t}\right\vert \right\}$,
one can calculate transition rates
$T^{\Gamma_{\mu}}_{n^{s} \rightarrow (n-1)^{t}} = \left\vert \left\langle \psi_{n-1}^{t} \vert
\Gamma_{\mu} \vert  \psi_{n}^{s} \right\rangle \right\vert^{2}$.
For single photon decay,
\begin{widetext}
\begin{align}
\label{eq:transition-rate-photon}
n > 1 : \; &
T^{a}_{n^{-} \rightarrow (n-1)^{-}} =
(n-1) \cos^{2} \theta_{n} \cos^{2} \theta_{n-1}
+ n \sin^{2} \theta_{n} \sin^{2} \theta_{n-1}
 + 2 \sqrt{n(n-1)} \cos \theta_{n} \sin \theta_{n} \cos \theta_{n-1} \sin \theta_{n-1} ,
\nonumber \\
& T^{a}_{n^{-} \rightarrow (n-1)^{+}} =
(n-1) \cos^{2} \theta_{n} \sin^{2} \theta_{n-1}
+ n \sin^{2} \theta_{n} \cos^{2} \theta_{n-1}
 - 2 \sqrt{n(n-1)} \cos \theta_{n} \sin \theta_{n} \sin \theta_{n-1} \cos \theta_{n-1} , \nonumber \\
& T^{a}_{n^{+} \rightarrow (n-1)^{-}} =
(n-1) \sin^{2} \theta_{n} \cos^{2} \theta_{n-1}
+ n \cos^{2} \theta_{n} \sin^{2} \theta_{n-1}
 - 2 \sqrt{n(n-1)} \sin \theta_{n} \cos \theta_{n} \cos \theta_{n-1} \sin \theta_{n-1} ,
\nonumber \\
&T^{a}_{n^{+} \rightarrow (n-1)^{+}} =
(n-1) \sin^{2} \theta_{n} \sin^{2} \theta_{n-1}
+ n \cos^{2} \theta_{n} \cos^{2} \theta_{n-1}
 + 2 \sqrt{n(n-1)} \sin \theta_{n} \cos \theta_{n} \sin \theta_{n-1} \cos \theta_{n-1} , \nonumber \\
n = 1 : \; & T^{a}_{1^{-} \rightarrow 0} = \sin^{2} \theta_{1} , \quad T^{a}_{1^{+} \rightarrow 0} = \cos^{2} \theta_{1} .
\end{align}
\end{widetext}
Similarly, for spin flip,
\begin{align}
\label{eq:transition-rate-spin}
n > 1 : \; & T^{\sigma_{-}}_{n^{-} \rightarrow (n-1)^{-}} = \cos^{2} \theta_{n} \sin^{2} \theta_{n-1} , \nonumber \\
& T^{\sigma_{-}}_{n^{-} \rightarrow (n-1)^{+}} = \cos^{2} \theta_{n} \cos^{2} \theta_{n-1} ,
\nonumber \\
& T^{\sigma_{-}}_{n^{+} \rightarrow (n-1)^{-}} = \sin^{2} \theta_{n} \sin^{2} \theta_{n-1} , \nonumber \\
& T^{\sigma_{-}}_{n^{+} \rightarrow (n-1)^{+}} = \sin^{2} \theta_{n} \cos^{2} \theta_{n-1} ,
\nonumber \\
n = 1 : \; & T^{\sigma_{-}}_{1^{-} \rightarrow 0} = \cos^{2} \theta_{1} , \quad T^{\sigma_{-}}_{1^{+} \rightarrow 0} = \sin^{2} \theta_{1} .
\end{align}
At resonance $\omega_{c} = \omega_{q}$, where the first
two terms in $H_{JC}$ are the Casimir invariant of $u(1|1)$
and $\cos \theta_{n} = \sin \theta_{n} = \frac{1}{\sqrt{2}}$,
single photon decay rates simplify,
\begin{align}
T^{a}_{n^{-} \rightarrow (n-1)^{\mp}} &= \frac{1}{4} \left[ 2n-1 \pm 2\sqrt{n(n-1)} \right] ,
\nonumber \\
T^{a}_{n^{+} \rightarrow (n-1)^{\mp}} &= \frac{1}{4} \left[ 2n-1 \mp 2\sqrt{n(n-1)} \right] ,
\nonumber
\end{align}
\begin{align}
\label{eq:transition-rate-photon-resonance}
T^{a}_{1^{\mp} \rightarrow 0} &= \frac{1}{2} .
\end{align}
In particular, for $n=2$, $T^{a}_{2^{-} \rightarrow 1^{-}} = T^{a}_{2^{+} \rightarrow 1^{+}} = 1.46$
and $T^{a}_{2^{-} \rightarrow 1^{+}} = T^{a}_{2^{+} \rightarrow 1^{-}} = 0.04$.
The transition rates $1^{\mp} \rightarrow 0$, $2^{\mp} \rightarrow 1^{\mp}$ have
been measured in a cavity QED system \cite{fink} and agree with these calculations.
Particularly important is the observation
of the $2^{-} \rightarrow 1^{-}$ and $2^{+} \rightarrow 1^{+}$ transitions and the
non-observation of the $2^{-} \rightarrow 1^{+}$ and $2^{+} \rightarrow 1^{-}$
transitions, calculated to be $\sim 0$. For spin flip, at resonance all transitions
$n \rightarrow (n-1)$ are
\begin{equation}
\label{eq:transition-rate-spin-resonance}
T^{\sigma_{-}}_{n^{\mp} \rightarrow (n-1)^{\mp}} = \frac{1}{4} , \quad
T^{\sigma_{-}}_{1^{\mp} \rightarrow 0} = \frac{1}{2} .
\end{equation}

\section{Generalizations of the JC Model} \label{sec:jc-generalization}

\subsection{The Weyl reflected model} \label{sec:jc-weyl}

The same methods above can be applied to the system with Hamiltonian
\begin{equation}
\label{eq:weyl-jc-ham}
H_{\overline{JC}}=\omega_{c}a^{\dag}a-\omega_{q}\frac{\sigma_{z}}{2}
+g^{\prime}\left( \sigma_{+}a^{\dag}+\sigma_{-}a\right)
\end{equation}
(also called the anti-JC or $\mathrm{\overline{JC}}$ model) and its Liouvillian with
dissipators $\mathcal{D}[a]$ and $\mathcal{D}[\sigma_{+}]$. Here,
$H_{\overline{JC}}$ has a  $\overline{u(1|1)}$ dynamical superalgebra with elements
\begin{equation}
\label{eq:weyl-u11-alg}
\overline{u(1|1)} \doteqdot
\left( 
\begin{array}{cc}
G_{B}=a^{\dag } a & \bar{F}^{\dag }=\sigma_{+} a^{\dag } \\ 
\bar{F}=\sigma_{-} a & G_{F}=\sigma_{+}\sigma_{-}
\end{array}
\right) .
\end{equation}
$H_{\overline{JC}}$ commutes with the $\overline{u(1|1)}$ linear Casimir operator
$\bar{C}=a^{\dag}a-\sigma_{+}\sigma_{-} $
(or $\bar{C}^{\prime}=a^{\dag}a-\frac{\sigma_{z}}{2}$),
and $\overline{u(1|1)}$ is related to the $u(1|1)$ algebra
in Eq.~(\ref{eq:u11-alg}) by a change of basis for $su(2)$ operators,
$\sigma_{\pm} \rightarrow \sigma_{\mp}$ and $\sigma_{z} \rightarrow -\sigma_{z}$
(equivalently for fermion operators, $f^{\dag} \leftrightarrow f$).
In fact, this transformation takes $H_{JC} \rightarrow H_{\overline{JC}}$,
thus $H_{\overline{JC}}$ can be diagonalized analogously to $H_{JC}$
within subspaces characterized by eigenvalues $n$ of $\bar{C}$.
The $n=0$ subspace is one dimensional,
\begin{equation}
\label{eq:weyl-jc-eig-gnd}
\bar{E}_{0} = -\frac{\omega_{q}}{2} , \quad
\left\vert \bar{\psi}_{0} \right\rangle = \left\vert 0, \frac{1}{2} \right\rangle ,
\end{equation}
and diagonalizing $H_{\overline{JC}}$ within the two dimensional
$n>0$ subspaces yields
\begin{align}
\label{eq:weyl-jc-eig-exc}
\bar{E}_{n}^{\mp} &= \omega_{c}(n-\frac{1}{2})\mp \frac{1}{2}\sqrt{
\left( \omega_{c}-\omega_{q}\right)^{2}+4g^{\prime 2}n} \\
\left( 
\begin{array}{c}
\left\vert \bar{\psi}_{n}^{-}\right\rangle \\ 
\left\vert \bar{\psi}_{n}^{+}\right\rangle
\end{array}
\right)
&= \left( 
\begin{array}{cc}
\cos \bar{\theta}_{n} & \sin \bar{\theta}_{n} \\ 
-\sin \bar{\theta}_{n} & \cos \bar{\theta}_{n}
\end{array}
\right) \left( 
\begin{array}{c}
\left\vert n-1,-\frac{1}{2}\right\rangle \\ 
\left\vert n,+\frac{1}{2}\right\rangle
\end{array}
\right) . \nonumber
\end{align}
The eigenvalues of the Liouvillian for the $\mathrm{\overline{JC}}$ model with
dissipators $\mathcal{D}[a]$ and $\mathcal{D}[\sigma_{+}]$
can be obtained in a similar fashion to Eqs.~(\ref{eq:liou-block-00}-\ref{eq:liou-eigs-freq})
and in fact are the same as Eq.~(\ref{eq:liou-eigs-freq}).
The fact that the $\mathrm{\overline{JC}}$ model can be solved in the same manner as the JC model
is a general property of coupled systems, and the two are sometimes referred to
as ``Weyl reflected'' models since $\bar{C}$ is obtained from $C$ by a change of sign.

We also note that the eigenvalues of the Liouvillian of the JC and $\mathrm{\overline{JC}}$
models with inverted dissipators, $\mathcal{D}[a^{\dag}]$ and $\mathcal{D}[\sigma_{+}]$
for the JC model and $\mathcal{D}[a^{\dag}]$ and $\mathcal{D}[\sigma_{-}]$ for the
$\mathrm{\overline{JC}}$ model, can be obtained trivially from those with standard dissipators
$\mathcal{D}[a]$, $\mathcal{D}[\sigma_{-}]$ for JC and $\mathcal{D}[a]$, $\mathcal{D}[\sigma_{+}]$ for
$\mathrm{\overline{JC}}$, and are given in Appendix~\ref{appx:jc-liou-dissipators}. In this case, neither model
with inverted dissipators possesses a steady state solution, as there is no eigenvalue $\lambda$
of $\mathcal{L}$ with $\mathrm{Re}[\lambda] = 0$. A zero eigenvalue appears in numerical
simulations but is spurious and only due to the truncation of the Hilbert space.

\subsection{Multiphoton couplings} \label{sec:jc-multiphoton}

The methods above can also be generalized to obtain eigenvalues of the Liouvillian
for the Jaynes-Cummings model with multiphoton couplings,
\begin{equation}
\label{eq:multiphoton-jc-ham}
H_{JC}^{(m)}=\omega_{c}a^{\dag}a+\omega_{q}\frac{\sigma_{z}}{2}+g\left(
\sigma_{+}a^{m}+\sigma_{-}a^{\dag m}\right) .
\end{equation}
We outline here the case $m=2$, which has been experimentally studied \cite{felicetti1}.
In this case, one can introduce a generalized Heisenberg algebra
$h^{(2)}(2) \doteqdot a^{2},a^{\dag 2}, a^{\dag}a, I$, with commutation relations
$\left[ a^{2},a^{\dag}a\right] = 2a^{2}$, $\left[ a^{\dag 2},a^{\dag}a\right] = -2a^{\dag 2}$,
$\left[ a^{2},a^{\dag 2}\right] = 2I + 4 a^{\dag}a$, and an associated superalgebra
$u^{(2)}(1|1)$ with elements
\begin{equation}
u^{(2)}(1|1) \doteqdot
\label{eq:twophoton-u11-alg}
\left( 
\begin{array}{cc}
G_{B}=a^{\dag}a & F^{\dag}=\sigma_{+}a^{2} \\ 
F=\sigma_{-}a^{\dag 2} & G_{F}=\sigma_{+}\sigma_{-}
\end{array}
\right)
\end{equation}
and Casimir operator $C^{(2)} = a^{\dag}a + 2 \sigma_{+}\sigma_{-}$. $H_{JC}^{(2)}$
has a $u^{(2)}(1|1)$ supersymmetry as it commutes with $C^{(2)}$ and is block diagonal
in subspaces characterized by its eigenvalues, $n$.
Subspaces $n=0$, $\{ \left\vert 0,-1/2\right\rangle \}$, and $n=1$, $\{ \left\vert 1,-1/2\right\rangle \}$,
are one-dimensional, while two-dimensional subspaces $n \geq 2$ have basis
$\{ \left\vert n,-1/2\right\rangle , \left\vert n-2,+1/2\right\rangle \}$. Eigenvalues for
$n \geq 2$ are given by
\begin{equation}
\label{eq:twophoton-jc-ham-eig}
E_{n}^{\mp}=\omega_{c}(n-1)\mp \frac{1}{2}\sqrt{
\left(2\omega_{c}-\omega_{q}\right)^{2}+4g^{2}n(n-1)} .
\end{equation}
Eigenvalues of the Liouvillian with excitation-lowering and excitation-conserving
dissipators can then be calculated by diagonalizing excitation-conserving submatrices
of the Liouvillian with the same approach as
Eqs.~(\ref{eq:liou-block-00}-\ref{eq:liou-block-nm}).

\section{Generalizations of the Liouvillian} \label{sec:jc-liou-generalization}

In practical situations one may need dissipators other than single photon
loss, $\mathcal{D}[a]$, and qubit damping, $\mathcal{D}[\sigma_{-}]$.
We have developed a general method of treating excitation-lowering
and excitation-conserving dissipators for the JC
model, given in Appendix~\ref{appx:jc-framework}, and
similarly for the dissipative $\mathrm{\overline{JC}}$ model.
Because of its importance in analyzing experimental data,
we give here terms added to Liouvillian blocks for pure dephasing,
$\mathcal{D}[\sigma_{z}]$,
\begin{align}
&\mathcal{D}[\sigma_{z}] ^{(n,m)} = \nonumber \\
&\left( 
\begin{array}{cccc}
P_{n} P_{m} - 1 & P_{n} Q_{m} & Q_{n} P_{m} & Q_{n} Q_{m} \\
P_{n} Q_{m} & -P_{n} P_{m} - 1 & Q_{n} Q_{m} & -Q_{n} P_{m} \\
Q_{n} P_{m} & Q_{n} Q_{m} & - P_{n} P_{m} - 1 & -P_{n} Q_{m} \\
Q_{n} Q_{m} & -Q_{n} P_{m} &  -P_{n} Q_{m} & P_{n} P_{m} - 1
\end{array}
\right) , \nonumber \\
&\mathcal{D}[\sigma_{z}] ^{(n,0)}=
\left( 
\begin{array}{cc}
-P_{n} - 1 & Q_{n} \\
Q_{n} & P_{n} - 1
\end{array}
\right), \\
&\mathcal{D}[\sigma_{z}] ^{(0,m)}=
\left( 
\begin{array}{cc}
-P_{m} - 1 & Q_{m} \\
Q_{m} & P_{m} - 1
\end{array}
\right) , \quad \mathcal{D}[\sigma_{z}] ^{(0,0)} = 0 , \nonumber
\end{align}
with 
\begin{equation*}
P_{n} = \left( \sin^{2} \theta_{n} - \cos^{2} \theta_{n} \right) , \quad
Q_{n} = - 2 \cos \theta_{n} \sin \theta_{n} ,
\end{equation*}
and two photon loss $\mathcal{D}[a^2]$,
\begin{align}
&\mathcal{D}[a^2] ^{(n,m)}= 
-\frac{1}{2} \times \nonumber \\
&\left( 
\begin{array}{cccc}
\left(\mathcal{B}_{n}^{+} + \mathcal{B}_{m}^{+} \right) & \mathcal{A}_{m}^{-} & \mathcal{A}_{n}^{-} & 0 \\
\mathcal{A}_{m}^{+} & \left(\mathcal{B}_{n}^{+} + \mathcal{B}_{m}^{-} \right) & 0 & \mathcal{A}_{n}^{-} \\
\mathcal{A}_{n}^{+} & 0 & \left(\mathcal{B}_{n}^{-} + \mathcal{B}_{m}^{+} \right) & \mathcal{A}_{m}^{-} \\
0 & \mathcal{A}_{n}^{+} & \mathcal{A}_{m}^{+} &  \left(\mathcal{B}_{n}^{-} + \mathcal{B}_{m}^{-} \right)
\end{array}
\right) , \nonumber \\
&\mathcal{D}[a^2] ^{(n,0)} =
 -\frac{1}{2} \left( 
\begin{array}{cc}
\mathcal{B}_{n}^{+} & \mathcal{A}_{n}^{-} \\
\mathcal{A}_{n}^{+} & \mathcal{B}_{n}^{-}
\end{array}
\right) , \\
&\mathcal{D}[a^2] ^{(0,m)} = 
-\frac{1}{2} \left( 
\begin{array}{cc}
\mathcal{B}_{m}^{+}  & \mathcal{A}_{m}^{-} \\
\mathcal{A}_{m}^{+} & \mathcal{B}_{m}^{-} 
\end{array}
\right) , \quad
\mathcal{D}[a^2] ^{(0,0)}= 0 , \nonumber
\end{align}
with
\begin{align*}
\mathcal{B}_{n}^{+} &= \left( n(n-1) - 2(n-1) \sin^{2} \theta_{n} \right) , \\
\mathcal{B}_{n}^{-} &= \left( n(n-1) - 2(n-1) \cos^{2} \theta_{n} \right),  \\
\mathcal{A}_{n}^{+} &= \mathcal{A}_{n}^{-} = 2(n-1)\cos \theta_{n} \sin \theta_{n} .
\end{align*}

\section{Multiple qubits} \label{sec:mult-qubits}

We consider now $N$-identical qubits
interacting with a single-mode radiation field described by the Dicke
Hamiltonian \cite{dicke} under the rotating wave approximation,
\begin{equation}
\label{eq:dicke-ham-v2}
H = \omega_{c}a^{\dag}a + \omega_{q} \sum_{i=1}^{N} \frac{\sigma_{z,i}}{2}
+ g  \sum_{i=1}^{N} \left( \sigma_{+,i}a+\sigma_{-,i}a^{\dag}\right) .
\end{equation}
This Hamiltonian has a dynamical superalgebra $\sum_{i} u_{i}(1|1)$ where
each $u_{i}(1|1)$ contains the same bosonic mode, $a,a^{\dag}$. The linear
Casimir of $\sum_{i} u_{i}(1|1)$ is $C = a^{\dag}a +  \sum_{i} \frac{\sigma_{z,i}}{2}$,
which commutes with $H$. Thus, $H$ is block diagonal in subspaces with
fixed $\left\langle C \right\rangle$, which have dimension up to $2^{N}$.
In order to reduce the dimension of the basis further, it is
convenient to introduce collective spin operators
\begin{equation}
\label{eq:quasispin}
J_{z} = \sum_{i=1}^{N} \frac{\sigma_{z,i}}{2} , \quad
J_{+} = \sum_{i=1}^{N} \sigma_{+,i} , \quad
J_{-} = \sum_{i=1}^{N} \sigma_{-,i} ,
\end{equation}
in terms of which the Hamiltonian is
\begin{equation}
\label{eq:dicke-ham-collective}
H = \omega_{c}a^{\dag}a + \omega_{q} J_{z} + g  \left( J_{+} a + J_{-} a^{\dag} \right) .
\end{equation}
$H$ then also has a dynamical algebra $h(2) \oplus su(2)$.
States can be labelled
by $\left\vert n_{B} \right\rangle \otimes \left\vert J,M \right\rangle$ with
$n_{B} = 0,1,\ldots, \infty$ and $-J \leq M \leq J$,
where $J$ is an integer or half-integer for $N$ even or odd. The operators
$J_{z}, J_{\pm}$ satisfy the usual commutation relations of the $su(2)$
algebra $[J_{z},J_{\pm}] = \pm J_{\pm}$, $[J_{+},J_{-}] = 2J_{z}$. 
In this basis, operators act as
$J_{z} \left\vert J,M \right\rangle = M \left\vert J,M \right\rangle$ and
$J_{\pm} \left\vert J,M \right\rangle = \sqrt{J(J+1) - M(M \pm 1)} \left\vert J,M \pm 1 \right\rangle$.
In this collective basis, $C = a^{\dag}a + J_{z}$ and has eigenvalues
$\left\langle C \right\rangle = n = n_{B} + M$. Also, it is now easy to see that
$H$ commutes with the quadratic Casimir of $su(2)$ as well,
$\bm{J}^{2} = J_{z}^{2} + \left(J_{+}J_{-} + J_{-}J_{+}\right)/2$,
which acts as $\bm{J}^{2} \left\vert J,M \right\rangle = J(J+1) \left\vert J,M \right\rangle$.
$H$ is then block diagonal in subspaces with
fixed $J$ and $(n_{B} + M)$,
\begin{align}
\label{eq:quasispin-matels}
&\left\langle n_{B} ; J , M \vert H \vert n_{B}^{\prime} ; J , M^{\prime} \right \rangle = \\
&\left(\omega_{c} n + \left(\omega_{q}-\omega_{c}\right) M\right) \delta_{n_{B},n_{B}^{\prime}} \delta_{M,M^{\prime}} \nonumber \\
&+ g\sqrt{J(J+1) - M^{\prime}(M^{\prime} + 1)} \sqrt{n_{B}^{\prime}} \delta_{n_{B},n_{B}^{\prime}-1} \delta_{M,M^{\prime}+1} \nonumber \\
&+ g \sqrt{J(J+1) - M^{\prime}(M^{\prime} - 1)} \sqrt{n_{B}} \delta_{n_{B},n_{B}^{\prime}+1} \delta_{M,M^{\prime}-1} . \nonumber
\end{align}
To complete the calculation, one needs to know the allowed values of $J$ for
a given number of qubits, $N$. This is a standard group theoretical problem \cite{iac4,hamermesh},
which can be solved using several methods. We quote here the solution. The values of $J$
for a given $N$ are
\begin{equation}
\label{eq:quasispin-values}
J = \frac{N}{2} , \frac{N}{2} - 1 , \frac{N}{2} - 2 , \ldots , 0 \; \mathrm{or} \; \frac{1}{2} \;
(N \; \mathrm{even} \; \mathrm{or} \; \mathrm{odd}) .
\end{equation}
Here, each value of $J$ has multiplicity $N!(2J + 1) /$ $\left[ \left( \frac{N}{2} + J + 1 \right)! \left( \frac{N}{2} - J \right)!\right]$.
Each block of the Hamiltonian with fixed $\left\langle C \right\rangle$ decomposes
as a direct sum of block matrices with fixed $J$, of dimension $2J+1$.
There are $\frac{N}{2} + 1$ such matrices. Once the eigenvalues
and eigenfunctions of $H$ have been found, one can proceed to evaluate
the eigenvalues of the Liouvillian $\mathcal{L}$, where the reduction of dimension from
$2^{2N}$ to $(2J+1)^{2}$ is even more drastic. To illustrate the method, we consider here
explicitly the case of two qubits coupled to the radiation field, which can be
realized experimentally with cavity QED \cite{fink}.

\section{Two-qubit system} \label{sec:two-qubits}

\subsection{Hamiltonian eigenvalues} \label{sec:2q-ham}

\begin{figure*}[ht!]
    \centering
    \includegraphics[width=0.8\textwidth]{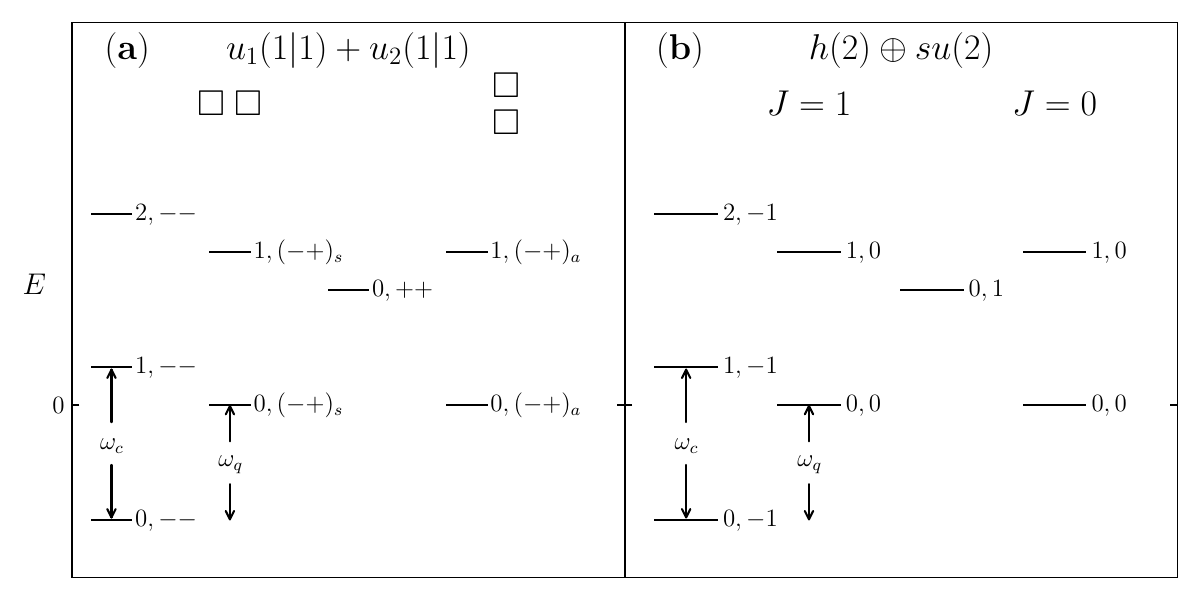}
    \caption{The uncoupled Dicke spectrum for two qubits.
    (a) States in a $u_{1}(1|1) + u_{2}(1|1)$ basis. (b) States in a $h(2) \oplus su(2)$ basis.
    The symmetric, $s$, and antisymmetric, $a$, states are labelled in (a)
    by the Young tableaux, $s = \Box \; \Box$, $a = \begin{array}{c} \Box \\ \Box \end{array}$.}
    \label{fig:uncoupled-2q-spectrum}
\end{figure*}

\begin{figure}[ht!]
    \centering
    \includegraphics[width=0.4\textwidth]{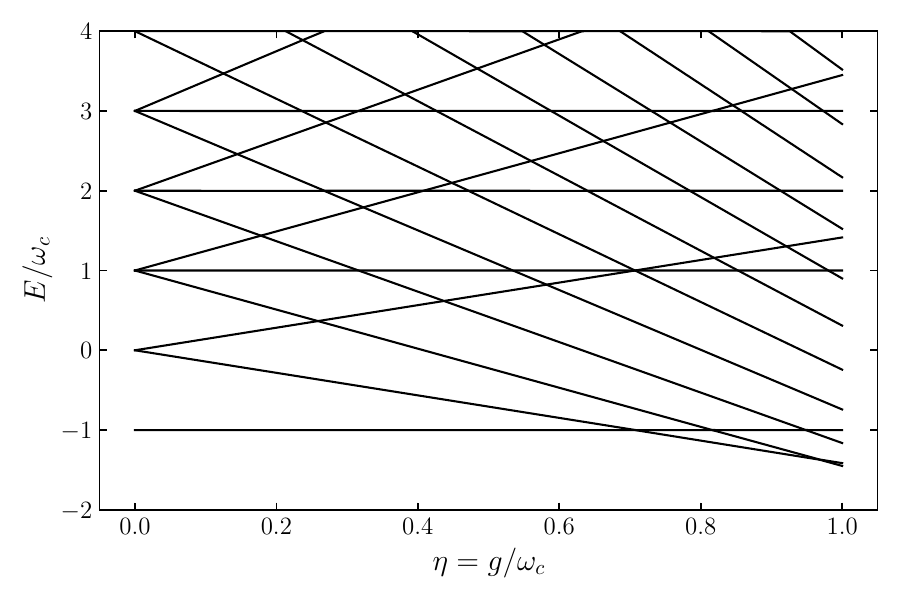}
    \caption{Behavior of energies for $J=1$ as a function
    of $g/\omega_{c}$ in the resonant case $\omega_{c}=\omega_{q}$.}
    \label{fig:2q-energy-g}
\end{figure}

For two qubits we have $\frac{1}{2} \otimes \frac{1}{2} = 0 \oplus 1$, so $J=0,1$. The uncoupled
spectrum is shown in Fig.~\ref{fig:uncoupled-2q-spectrum}. Since the
Hamiltonian, Eq.~(\ref{eq:dicke-ham-v2}) is invariant under permutation, states
must be properly symmetrized when expressed in a $\sum_{i} u_{i}(1|1)$  basis.
The states in Fig.~\ref{fig:uncoupled-2q-spectrum}
are the basis in which the coupled Hamiltonian, $g \neq 0$, is diagonalized. As
discussed above, we use the basis $\left\vert n_{B} ; J , M \right \rangle$ for the
diagonalization. For $J=0$, the spectrum is simply the harmonic oscillator spectrum,
$E_{n_{B}} = \omega_{c} n_{B}$. For $J=1$, we diagonalize $H$ in subspaces
labelled by $\left\langle C \right\rangle = n = n_{B} + M$. The ground state, $n = -1$, has
\begin{equation}
\label{eq:j1-ham-gnd}
\left\vert \psi_{g}\right\rangle = \left\vert 0,-1 \right\rangle , \quad
E_{g} = -\omega_{q} ,
\end{equation}
where we drop the $J$ label for simplicity and denote states
by $\left\vert n_{B} , M \right\rangle$. For $n=0$, there are two states, whose
energies and wavefunctions are obtained by diagonalizing a $2 \times 2$
matrix 
\begin{equation}
\label{eq:j1-ham-submat-1exc}
H_{n=0} = \left( 
\begin{array}{cc}
0 & g\sqrt{2} \\
g\sqrt{2} & \omega_{c} - \omega_{q}
\end{array}
\right)
\end{equation}
with result
\begin{align}
\label{eq:j1-ham-1exc}
&\left\vert \psi_{0}^{\pm}\right\rangle = \nonumber \\
&\frac{
\left[(\omega_{q} - \omega_{c}) \pm \sqrt{(\omega_{c} - \omega_{q})^{2} + 8g^{2}}\right]
\left\vert 0,0 \right\rangle
+2g\sqrt{2} \left\vert 1,-1 \right\rangle
}{\sqrt{2}\sqrt{ (\omega_{c} - \omega_{q})^{2}
+ 8g^{2} \pm (\omega_{q} - \omega_{c})\sqrt{(\omega_{c} - \omega_{q})^{2} + 8g^{2}}}} ,
\nonumber \\
&E_{0}^{\pm} =  \frac{\omega_{c} - \omega_{q}}{2} \pm
\frac{1}{2}\sqrt{(\omega_{c} - \omega_{q})^{2} + 8g^{2}} .
\end{align}
In the resonant case $\omega_{q} = \omega_{c}$, wavefunctions and energies reduce to
\begin{equation}
\label{eq:j1-ham-res-1exc}
\left\vert \psi_{0}^{\pm}\right\rangle = \frac{1}{\sqrt{2}}
\left( \pm \left\vert 0,0 \right\rangle + \left\vert 1,-1 \right\rangle \right) , \quad
E_{0}^{\pm} = \pm g \sqrt{2}.
\end{equation}
For $n \geq 1$, there are three states. The energies and wavefunctions
in the basis $\left( \left\vert n-1,1 \right\rangle , \left\vert n,0 \right\rangle ,
\left\vert n+1,-1 \right\rangle \right)$ can be obtained by diagonalizing a
$3 \times 3$ matrix
\begin{align}
\label{eq:j1-ham-submat-nexc}
&H_{n \geq 1} = \\
&\left( 
\begin{array}{ccc}
\omega_{c}n - (\omega_{c} - \omega_{q}) & g\sqrt{2n} & 0 \\
g\sqrt{2n} & \omega_{c}n & g\sqrt{2(n+1)} \\
0 & g\sqrt{2(n+1)} & \omega_{c}n + (\omega_{c} - \omega_{q})
\end{array}
\right) . \nonumber
\end{align}
General formulas for eigenvalues and eigenvectors are complicated,
however, in the resonant case they can be expressed in closed
form as
\begin{align}
\label{eq:j1-ham-res-nexc}
\left\vert \psi_{n}^{0}\right\rangle =&
-\sqrt{\frac{n+1}{2n+1}} \left\vert n-1,1 \right\rangle 
+ \sqrt{\frac{n}{2n+1}} \left\vert n+1,-1 \right\rangle ,
\nonumber \\
\left\vert \psi_{n}^{\pm}\right\rangle =&
\frac{1}{\sqrt{2}}\left(
\sqrt{\frac{n}{2n+1}} \left\vert n-1,1 \right\rangle \pm
\left\vert n,0 \right\rangle \right. \nonumber \\
&\left. + \sqrt{\frac{n+1}{2n+1}} \left\vert n+1,-1 \right\rangle
\right) ,
\nonumber \\
E_{n}^{0} =& \omega_{c}n , \quad E_{n}^{\pm} = \omega_{c}n \pm g \sqrt{2(2n+1)} .
\end{align}
The behavior of the energy eigenvalues as a function of $g$ is
shown in Fig.~\ref{fig:2q-energy-g}.

\subsection{Liouvillian eigenvalues} \label{sec:2q-liou}

We consider here single photon dissipation and collective qubit damping
\begin{align}
\label{eq:tc-collective-liouvillian}
\mathcal{L\rho} &= -\mathrm{i}\left[ H,\rho \right] +\gamma \mathcal{D}[a]\rho+
\gamma^{\prime}\mathcal{D}[J_{-}]\rho , \nonumber \\
\mathcal{D}[a]\rho &= a\rho a^{\dag}-\frac{1}{2}\left( a^{\dag}a\rho +\rho
a^{\dag}a\right) , \nonumber \\
\mathcal{D}[J_{-}]\rho &= J_{-}\rho J_{+}-\frac{1}{2}\left(
J_{+}J_{-}\rho +\rho J_{+}J_{-}\right) .
\end{align}
For $J=0$, the system is equivalent to a harmonic oscillator with photon dissipation.
The eigenvalues of the Liouvillian in this sector are
\begin{equation}
\label{eq:j0-liouvillian-eigs}
\lambda_{n,m} = -\mathrm{i}\omega_{c}(n-m) - \frac{\gamma}{2}(n+m) .
\end{equation}
The eigenvalues of the Liouvillian for $J=1$ can be obtained in a
similar way to those of $J=1/2$. When expressed in the eigenbasis
of the Hamiltonian $H$, eigenvalues of $\mathcal{L}$ can
be found by simply diagonalizing blocks along its diagonal,
$\mathcal{L}^{(s,t)}$, with $s,t = g,0,n$, $n>0$. A derivation
of blocks $\mathcal{L}^{(s,t)}$ for the resonant case,
$\omega_{q} = \omega_{c}$, is given in Appendix~\ref{appx:2q-liou}
and we quote the results here. 
The explicit forms of all blocks $\mathcal{L}^{(s,t)}$ are
\begin{widetext}
\begin{equation}
\label{eq:j1-liou-block-gg}
\mathcal{L}^{(g,g)} = (0) , \quad
\mathcal{L}^{(g,0)} = \left( 
\begin{array}{cc}
\Gamma_{g0}^{-} & G_{0} \\ 
G_{0} & \Gamma_{g0}^{+}
\end{array}
\right) , \quad
\mathcal{L}^{(0,g)}=\left( 
\begin{array}{cc}
\Gamma_{0g}^{-} & G_{0} \\ 
G_{0} & \Gamma_{0g}^{+}
\end{array}
\right) ,
\end{equation}
\begin{equation}
\label{eq:j1-liou-block-nggm}
\mathcal{L}^{(n,g)} = \left( 
\begin{array}{ccc}
\Gamma_{ng}^{0}  & F_{n} & F_{n} \\ 
F_{n} & \Gamma_{ng}^{-}  & G_{n} \\
F_{n} & G_{n} & \Gamma_{ng}^{+} 
\end{array}
\right) , \quad
\mathcal{L}^{(g,m)} = \left( 
\begin{array}{ccc}
\Gamma_{gm}^{0}  & F_{m} & F_{m} \\ 
F_{m} & \Gamma_{gm}^{-}  & G_{m} \\
F_{m} & G_{m} & \Gamma_{gm}^{+} 
\end{array}
\right) , \quad
\mathcal{L}^{(0,0)} = \left( 
\begin{array}{cccc}
\Gamma_{00}^{--} & G_{0} & G_{0} & 0 \\
G_{0} & \Gamma_{00}^{-+} & 0 & G_{0} \\
G_{0} & 0 & \Gamma_{00}^{+-} & G_{0} \\
0 & G_{0} & G_{0} & \Gamma_{00}^{++}
\end{array}
\right) ,
\end{equation}
\begin{equation}
\label{eq:j1-liou-block-n00m}
\mathcal{L}^{(n,0)} = \left( 
\begin{array}{cccccc}
\Gamma_{n0}^{0-}   & G_{0} & F_{n} & 0 & F_{n} & 0 \\
G_{0} & \Gamma_{n0}^{0+}   & 0 & F_{n} & 0 & F_{n} \\
F_{n} & 0 & \Gamma_{n0}^{--}   & G_{0} & G_{n} & 0 \\
0 & F_{n} & G_{0} & \Gamma_{n0}^{-+}   & 0 & G_{n} \\
F_{n} & 0 & G_{n} & 0 & \Gamma_{n0}^{+-}   & G_{0} \\
0 & F_{n} & 0 & G_{n} & G_{0} & \Gamma_{n0}^{++}  
\end{array}
\right) , \quad
\mathcal{L}^{(0,m)} = \left( 
\begin{array}{cccccc}
\Gamma_{0m}^{-0}   & G_{0} & F_{m} & 0 & F_{m} & 0 \\
G_{0} & \Gamma_{0m}^{+0}   & 0 & F_{m} & 0 & F_{m} \\
F_{m} & 0 & \Gamma_{0m}^{--}   & G_{0} & G_{m} & 0 \\
0 & F_{m} & G_{0} & \Gamma_{0m}^{-+}   & 0 & G_{m} \\
F_{m} & 0 & G_{m} & 0 & \Gamma_{0m}^{+-}   & G_{0} \\
0 & F_{m} & 0 & G_{m} & G_{0} & \Gamma_{0m}^{++}  
\end{array}
\right) , 
\end{equation}
\begin{equation}
\label{eq:j1-liou-block-nm}
\mathcal{L}^{(n,m)} = \left( 
\begin{array}{ccccccccc}
\Gamma_{nm}^{00}   & F_{m} & F_{m} & F_{n} & F_{n} & 0 & 0 & 0 & 0 \\
F_{m} & \Gamma_{nm}^{0-}   & G_{m} & 0 & 0 & F_{n} & 0 & F_{n} & 0 \\
F_{m} & G_{m} & \Gamma_{nm}^{0+}   & 0 & 0 & 0 & F_{n} & 0 & F_{n} \\
F_{n} & 0 & 0 & \Gamma_{nm}^{-0}   & G_{n} & F_{m} & 0 & F_{m} & 0 \\
F_{n} & 0 & 0 & G_{n} & \Gamma_{nm}^{+0}   & 0 & F_{m} & 0 & F_{m} \\
0 & F_{n} & 0 & F_{m} & 0 & \Gamma_{nm}^{--}   & G_{m} & G_{n} & 0 \\
0 & 0 & F_{n} & 0 & F_{m} & G_{m} & \Gamma_{nm}^{-+}   & 0 & G_{n} \\
0 & F_{n} & 0 & F_{m} & 0 & G_{n} & 0 & \Gamma_{nm}^{+-}   & G_{m} \\
0 & 0 & F_{n} & 0 & F_{m} & 0 & G_{n} & G_{m} & \Gamma_{nm}^{++}  
\end{array}
\right) ,
\end{equation}
\end{widetext}
where for clarity we have defined
\begin{align*}
&\Gamma_{g0}^{\pm} = \Delta_{g0}^{\pm} + B_{0} , \quad
\Gamma_{0g}^{\pm} = \Delta_{0g}^{\pm} + B_{0} \nonumber \\
&\left\{\begin{array}{l}
\Gamma_{ng}^{0} = \Delta_{ng}^{0} + A_{n} , \quad
\Gamma_{gm}^{0} = \Delta_{gm}^{0} + A_{m} 
\\
\Gamma_{ng}^{\pm} = \Delta_{ng}^{\pm} + B_{n} , \quad
\Gamma_{gm}^{\pm} = \Delta_{gm}^{\pm} + B_{m} 
\end{array}\right. \nonumber \\
&\Gamma_{00}^{st} = \Delta_{00}^{st} + 2B_{0} \\
&\left\{\begin{array}{l}
\Gamma_{n0}^{0\pm} = \Delta_{n0}^{0\pm} + A_{n} + B_{0} , \quad
\Gamma_{0m}^{\pm 0} = \Delta_{0m}^{\pm 0} + A_{m} + B_{0}
\\
\Gamma_{n0}^{st} = \Delta_{n0}^{st} + B_{n} + B_{0} , \quad
\Gamma_{0m}^{st} = \Delta_{0m}^{st} + B_{m} + B_{0}
\end{array}\right. \nonumber \\
&\left\{\begin{array}{l}
\Gamma_{nm}^{00} = \Delta_{nm}^{00} + A_{n} + A_{m}
\\
\Gamma_{nm}^{0\pm} = \Delta_{nm}^{0\pm} + A_{n} + B_{m} , \quad
\Gamma_{nm}^{\pm 0} = \Delta_{nm}^{\pm 0} + B_{n} + A_{m}
\\
\Gamma_{nm}^{st} = \Delta_{nm}^{st} + B_{n} + B_{m}
\end{array}\right. . \nonumber
\end{align*}
In all expressions, $\Delta_{st}^{pq}$ is the imaginary part of the
Liouvillian, $\Delta_{st}^{pq} = -\mathrm{i}\left(E_{s}^{p} - E_{t}^{q}\right)$, and
\begin{align*}
A_{n} &= -\frac{\gamma}{2}\left(n - \frac{1}{2n+1} \right) -
\frac{\gamma^{\prime}}{2}\left(\frac{2(n+1)}{2n+1} \right) , \nonumber \\
B_{n} &= -\frac{\gamma}{2}\left(n + \frac{1}{2}\frac{1}{2n+1} \right) -
\frac{\gamma^{\prime}}{2}\left(\frac{n}{2n+1} + 1\right) , \nonumber \\
F_{n} &= -\frac{(\gamma - \gamma^{\prime})}{2}\frac{\sqrt{2n(n+1)}}{2n+1} , \nonumber \\
G_{n} &= -\frac{\gamma}{4} \frac{1}{2n+1} - \frac{\gamma^{\prime}}{2}\left(\frac{n}{2n+1} - 1\right) .
\end{align*}
If $\gamma = \gamma^{\prime}$, $F_{m} = F_{n} = 0$ and $G_{m} = G_{n} = \gamma / 4$,
and $\mathcal{L}^{(n,0)}$, $\mathcal{L}^{(0,m)}$, $\mathcal{L}^{(n,m)}$ split
into submatrices with dimension at most $4 \times 4$, meaning they can be diagonalized
analytically in closed form,
\begin{widetext}
\begin{equation}
\label{eq:j1-liou-block-n00m-block}
\mathcal{L}^{(n,0)} =  \left( 
\begin{array}{cc}
\Gamma_{n0}^{0-} & G_{0} \\
G_{0} & \Gamma_{n0}^{0+}
\end{array}
\right) \oplus \left( 
\begin{array}{cccc}
\Gamma_{n0}^{--} & G_{0} & G_{n} & 0 \\
G_{0} & \Gamma_{n0}^{-+} & 0 & G_{n} \\
G_{n} & 0 & \Gamma_{n0}^{+-} & G_{0} \\
0 & G_{n} & G_{0} & \Gamma_{n0}^{++}
\end{array}
\right) , \;
\mathcal{L}^{(0,m)} = \left( 
\begin{array}{cc}
\Gamma_{0m}^{-0} & G_{0} \\
G_{0} & \Gamma_{0m}^{+0}
\end{array}
\right) \oplus \left( 
\begin{array}{cccc}
\Gamma_{0m}^{--} & G_{0} & G_{m} & 0 \\
G_{0} & \Gamma_{0m}^{-+} & 0 & G_{m} \\
G_{m} & 0 & \Gamma_{0m}^{+-} & G_{0} \\
0 & G_{m} & G_{0} & \Gamma_{0m}^{++}
\end{array}
\right) ,
\end{equation}
\begin{equation}
\label{eq:j1-liou-block-nm-block}
\mathcal{L}^{(n,m)} =
\Gamma_{nm}^{00}
\oplus \left( 
\begin{array}{cc}
\Gamma_{nm}^{0-} & G_{m} \\
G_{m} & \Gamma_{nm}^{0+}
\end{array}
\right) \oplus \left( 
\begin{array}{cc}
\Gamma_{nm}^{-0} & G_{n} \\
G_{n} & \Gamma_{nm}^{+0}
\end{array}
\right) \oplus \left( 
\begin{array}{cccc}
\Gamma_{nm}^{--} & G_{m} & G_{n} & 0 \\
G_{m} & \Gamma_{nm}^{-+} & 0 & G_{n} \\
G_{n} & 0 & \Gamma_{nm}^{+-} & G_{m} \\
0 & G_{n} & G_{m} & \Gamma_{nm}^{++}
\end{array}
\right)  .
\end{equation}
\end{widetext}
This simplification occurs because when $\gamma = \gamma^{\prime}$, the
excitation-conserving part of the dissipator gains an additional
symmetry and becomes
\begin{align}
\label{eq:j1-liou-susy}
&\gamma \left( a^{\dag}a\rho +\rho a^{\dag} a\right) +\gamma^{\prime}\left(
J_{+}J_{-}\rho +\rho J_{+}J_{-}\right) 
= \nonumber \\
&\gamma \left[ \left( a^{\dag}a+J_{+}J_{-}\right) \rho +\rho
\left( a^{\dag}a+J_{+}J_{-}\right) \right] ,
\end{align}
where $J_{+}J_{-} = \bm{J}^{2} - J_{z}(J_{z} - 1)$ and
$a^{\dag}a+J_{+}J_{-} = C + \bm{J}^{2} - J_{z}^{2}$. $J_{z}^{2}$
mixes $\left\vert \psi_{n}^{+}\right\rangle$ and
$\left\vert \psi_{n}^{-}\right\rangle$ for $n \geq 0$
but leaves $\left\vert \psi_{n}^{0}\right\rangle$ invariant,
further splitting the blocks as observed in
Eqs.~(\ref{eq:j1-liou-block-n00m-block}-\ref{eq:j1-liou-block-nm-block}).

In summary, the fact that the Hamiltonian $H$ of the Dicke model commutes
with the Casimir $C = a^{\dag}a + J_{z}$ reduces the eigenvalue problem for
$H$ of dimension $2^{N}$ to diagonalization of matrices of dimension at most $N+1$,
here $3 \times 3$. Consequently, the eigenvalue problem for $\mathcal{L}$ of
dimension $2^{2N}$ is reduced to diagonalization of matrices of dimension at most
$(N+1)^{2}$, here $9 \times 9$. Furthermore, when $\gamma = \gamma^{\prime}$,
the eigenvalue problem of $\mathcal{L}$ reduces to diagonalization of even smaller
matrices, at most $4 \times 4$ here.

\section{Beyond two qubits} \label{sec:beyond-2q}

\begin{figure}[ht!]
    \centering
    \includegraphics[width=0.45\textwidth]{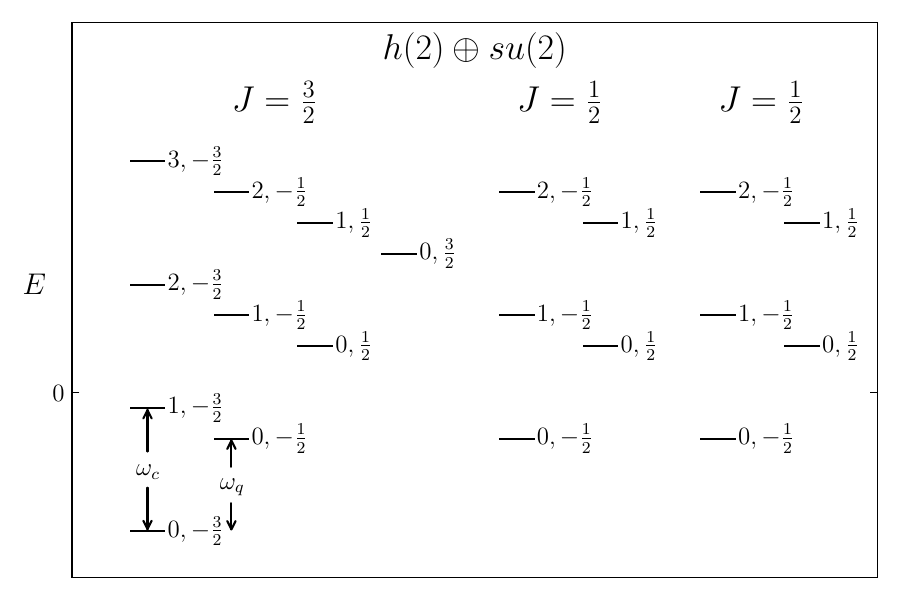}
    \caption{The uncoupled Dicke spectrum for three qubits in the
    $h(2) \oplus su(2)$ basis.}
    \label{fig:uncoupled-3q-spectrum}
\end{figure}

The explicit diagonalization of the Hamiltonian becomes more complex
as the number of qubits increases. For three qubits, the values of $J$ are
given by
\begin{equation}
\label{eq:3q-quasispin}
N = 3 : \frac{1}{2} \otimes \frac{1}{2} \otimes \frac{1}{2} = \frac{1}{2} \oplus \frac{1}{2} \oplus \frac{3}{2} .
\end{equation}
The uncoupled spectrum for three qubits is shown in Fig.~\ref{fig:uncoupled-3q-spectrum}.
The eigenvalues of $H$ and $\mathcal{L}$ for $J=1/2$ have been given in sections
\ref{sec:jc-ham}-\ref{sec:jc-liou}. The eigenvalues of $H$ for $J = 3/2$ are solutions
of at most $4 \times 4$ matrices which can still be explicitly diagonalized. The eigenvalues
of $\mathcal{L}$, Eq.~(\ref{eq:tc-collective-liouvillian}), are in general solutions
of at most $16 \times 16$ matrices which reduce to smaller dimensions
when the excitation-conserving part of $\mathcal{L}$ commutes with
$C = a^{\dag}a + J_{z}$. We note that $J = 3/2$ also has a non-trivial supersymmetry
$u(1|2)$, built from one boson creation and annihilation operator, $a^{\dag},a$, and two
fermion creation and annihilation operators $f_{1/2}^{\dag},f_{3/2}^{\dag}$ and $f_{1/2},f_{3/2}$.
(The case of $J=3/2$ is particularly interesting since its Hamiltonian problem has been investigated
using other methods \cite{braak}.) In the approach in terms of $h(2) \oplus su(2)$, the values
of $J$ are either integer (bosonic) or half-integer (fermionic). While for half-integer $J$ (fermion)
one can construct a superalgebra $u(1|J+1/2)$, for integer $J$ (boson) this construction
is not possible. A unified description can only be provided by the  $h(2) \oplus su(2)$ approach.

\begin{figure}[t!]
    \centering
    \includegraphics[width=0.4\textwidth]{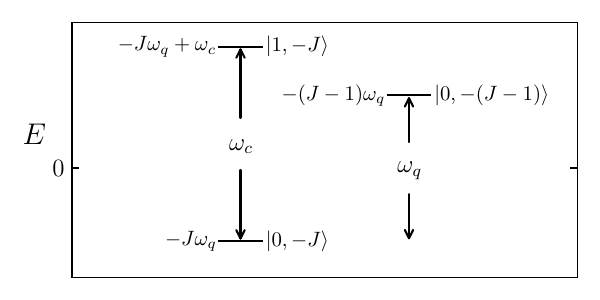}
    \caption{The lowest part of the uncoupled Dicke spectrum for arbitrary $J$.}
    \label{fig:uncoupled-J-spectrum}
\end{figure}

\begin{figure}[t!]
    \centering
    \includegraphics[width=0.4\textwidth]{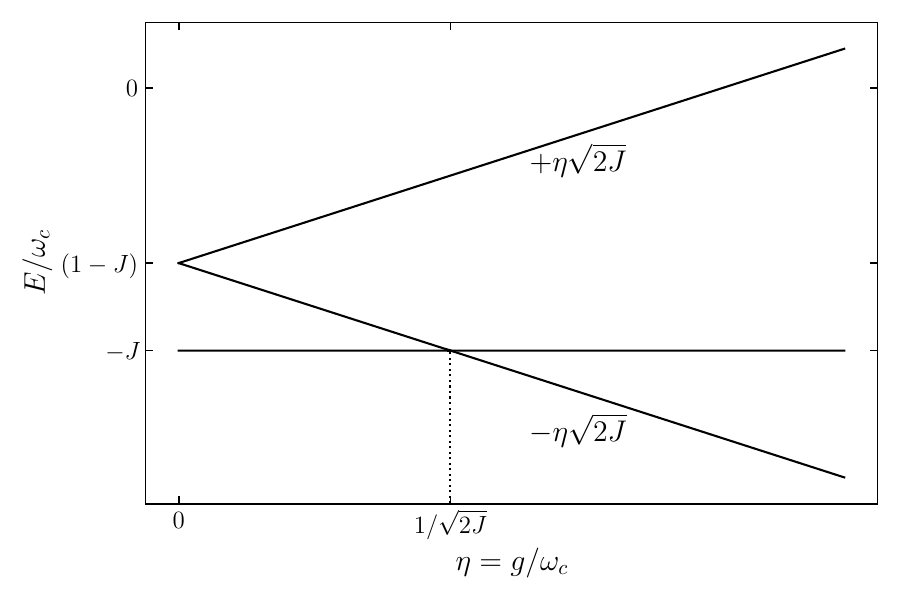}
    \caption{Behavior of the energies of the ground state and
    first excited state for arbitrary $J$. The value of $\left(\frac{g}{\omega_{c}}\right)_{cr}$
    at which the levels cross is $\left(\frac{g}{\omega_{c}}\right)_{cr} = \frac{1}{\sqrt{2J}} = \frac{1}{\sqrt{N}}$.}
    \label{fig:gnd-state-J}
\end{figure}

In spite of the complication to construct the entire spectrum of eigenvalues of $H$
and $\mathcal{L}$, it is possible to obtain explicit solutions for the eigenvalues of the
ground state and first excited states for arbitrary $J$, which are particularly important
for experimental applications in quantum optics and quantum computing.
The lowest part of uncoupled spectrum for arbitrary $J$ is shown in
Fig.~\ref{fig:uncoupled-J-spectrum}.
The eigenvalues of $H$ for coupling $g$ are given by
\begin{widetext}
\begin{eqnarray}
\label{eq:jj-energy-gnd}
E_{g} = -\omega_{q}J , \quad \left\vert \psi_{g}\right\rangle = \left\vert 0,-J \right\rangle , \quad \text{and} \quad
E_{-J+1}^{\pm} = \frac{\omega_{c} - (2J-1)\omega_{q}}{2} \pm
\frac{1}{2}\sqrt{(\omega_{c} - \omega_{q})^{2} + 8Jg^{2}}, \\
\label{eq:jj-eigenstates}
\left\vert \psi_{-J+1}^{\pm}\right\rangle =
\frac{\left(
\left[-(\omega_{c} - \omega_{q}) \pm \sqrt{(\omega_{c} - \omega_{q})^{2} + 8Jg^{2}}\right]
\left\vert 0,-J+1 \right\rangle +2g\sqrt{2J} \left\vert 1,-J \right\rangle \right)}{\sqrt{2 (\omega_{c} - \omega_{q}) }
\sqrt{ (\omega_{c} - \omega_{q})
+ \frac{8Jg^{2}}{(\omega_{c} - \omega_{q})} \mp \sqrt{(\omega_{c} - \omega_{q})^{2} + 8Jg^{2}}}} .
\nonumber
\end{eqnarray}
\end{widetext}
In the resonant case, $\omega_{c} = \omega_{q}$, this reduces to
\begin{align}
\label{eq:jj-ham-res-gnd}
E_{-J+1}^{\pm} = \omega_{c}(1-J) \pm g\sqrt{2J}, \nonumber \\
\left\vert \psi_{-J+1}^{\pm}\right\rangle = \frac{1}{\sqrt{2}}
\left(
\pm \left\vert 0,-J+1 \right\rangle + \left\vert 1,-J \right\rangle
\right) ,
\end{align}
whose behavior is shown in Fig.~\ref{fig:gnd-state-J}.
Also in the resonant case, the lowest-lying blocks along the diagonal of
$\mathcal{L}$, Eq.~(\ref{eq:tc-collective-liouvillian}), can be found
analogous to previous cases and are given by
\begin{widetext}
\begin{align}
\label{eq:jj-liou-gnd}
\mathcal{L}^{(1-J,g)} &= \left( 
\begin{array}{cc}
\mathrm{i}(g\sqrt{2J} - \omega_{c}) - \frac{\gamma}{4} - \frac{\gamma^\prime}{2}J &
- \frac{\gamma}{4} + \frac{\gamma^\prime}{2}J \\
- \frac{\gamma}{4} + \frac{\gamma^\prime}{2}J &
-\mathrm{i}(g\sqrt{2J} + \omega_{c}) - \frac{\gamma}{4} - \frac{\gamma^\prime}{2}J
\end{array}
\right) ,
\nonumber \\
\mathcal{L}^{(g,1-J)} &= \left( 
\begin{array}{cc}
\mathrm{i}(\omega_{c}-g\sqrt{2J}) - \frac{\gamma}{4} - \frac{\gamma^\prime}{2}J &
- \frac{\gamma}{4} + \frac{\gamma^\prime}{2}J \\
- \frac{\gamma}{4} + \frac{\gamma^\prime}{2}J &
\mathrm{i}(\omega_{c}+g\sqrt{2J}) - \frac{\gamma}{4} - \frac{\gamma^\prime}{2}J
\end{array}
\right) ,
\nonumber \\
 \mathcal{L}^{(1-J,1-J)} &= \left( 
\begin{array}{cccc}
-\frac{\gamma}{2} - \gamma^\prime J & - \frac{\gamma}{4} + \frac{\gamma^\prime}{2}J &
- \frac{\gamma}{4} + \frac{\gamma^\prime}{2}J & 0 \\
- \frac{\gamma}{4} + \frac{\gamma^\prime}{2}J & \mathrm{i}\sqrt{8Jg^{2}} -\frac{\gamma}{2} - \gamma^\prime J & 0 & - \frac{\gamma}{4} + \frac{\gamma^\prime}{2}J \\
- \frac{\gamma}{4} + \frac{\gamma^\prime}{2}J & 0 & -\mathrm{i}\sqrt{8Jg^{2}} -\frac{\gamma}{2} - \gamma^\prime J & - \frac{\gamma}{4} + \frac{\gamma^\prime}{2}J \\
0 & - \frac{\gamma}{4} + \frac{\gamma^\prime}{2}J &
- \frac{\gamma}{4} + \frac{\gamma^\prime}{2}J & -\frac{\gamma}{2} - \gamma^\prime J
\end{array}
\right) , \quad \mathcal{L}^{(g,g)} = 0 .
\end{align}
\end{widetext}
The corresponding eigenvalues of $\mathcal{L}$ are then
\begin{align}
\label{eq:jj-liou-eigs-gnd}
\lambda_{(1-J,g)}^{\pm} =& -\mathrm{i}\omega_{c} - \frac{\gamma}{4} - \frac{\gamma^\prime}{2}J \pm \frac{\mathrm{i}}{2} \sqrt{8Jg^{2} - \left(\gamma^\prime J - \frac{\gamma}{2}\right)^{2}} , \nonumber \\
\lambda_{(g,1-J)}^{\pm} =& \mathrm{i}\omega_{c} - \frac{\gamma}{4} - \frac{\gamma^\prime}{2}J \pm \frac{\mathrm{i}}{2} \sqrt{8Jg^{2} - \left(\gamma^\prime J - \frac{\gamma}{2}\right)^{2}} , \nonumber \\
\lambda_{(1-J,1-J)}^{s,t} &= \lambda_{(1-J,g)}^{s} + \lambda_{(g,1-J)}^{t} , \quad \lambda_{(g,g)} = 0 .
\end{align}

\section{Qubit-qubit coupling} \label{sec:qq-coupling}

In cavity QED experiments and applications to quantum
computing one also cares about Hamiltonians which
couple qubits to each other \cite{blais1}.
We consider here the Hamiltonian with isotropic two-qubit coupling
\begin{align}
\label{eq:dicke-ham-qq}
H_{D} =& \omega_{c}a^{\dag}a + \omega_{q} \sum_{i=1}^{N} \frac{\sigma_{z,i}}{2}
+ g  \sum_{i=1}^{N} \left( \sigma_{+,i}a+\sigma_{-,i}a^{\dag}\right) \nonumber \\
&+ g^{\prime}  \sum_{i \neq j}^{N} \left( \sigma_{+,i}\sigma_{-,j}\right) .
\end{align}
Noting that $\sigma_{+}\sigma_{-} - \frac{1}{2} = \frac{\sigma_{z}}{2}$, the coupling
term becomes
\begin{align}
\label{eq:qq-coupling}
&\sum_{i \neq j} \left( \sigma_{+,i}\sigma_{-,j}\right) =
\sum_{i, j} \left( \sigma_{+,i}\sigma_{-,j}\right) - \sum_{i} \left( \sigma_{+,i}\sigma_{-,i}\right) =
\nonumber \\
&\left( \sum_{i} \sigma_{+,i} \right) \left( \sum_{j} \sigma_{-,j} \right) -
\sum_{i} \left( \frac{\sigma_{z,i}}{2} + \frac{1}{2} \right) .
\end{align}
Introducing collective operators as in Eq.~(\ref{eq:quasispin}),
with $J_{+}J_{-} = \bm{J}^{2} - J_{z}(J_{z} - 1)$,
one may rewrite $H_{D}$ as
\begin{align}
\label{eq:dicke-ham-qq-quasispin}
H_{D} &= \omega_{c}a^{\dag}a + \omega_{q} J_{z}
+ g \left( J_{+}a+J_{-}a^{\dag}\right) \nonumber \\
&+ g^{\prime} \left(J(J+1) - J_{z}^{2} - \frac{N}{2}\right) .
\end{align}
This Hamiltonian can be diagonalized in the same way as in
Section \ref{sec:mult-qubits}, the only difference between Eq.~(\ref{eq:quasispin-matels})
and matrix elements of Eq.~(\ref{eq:dicke-ham-qq-quasispin}) is
the renormalization of the term $\omega_{q} M$ to
$\omega_{q} M + g^{\prime} \left(J(J+1) - M^{2} - \frac{N}{2}\right)$.
For the lowest band, $J = \frac{N}{2}$, the normalization is
$g^{\prime} \left(J^{2} - M^{2} \right)$, with $-J \leq M \leq +J$.
For the ground state band $M = -J$ the renormalization is zero,
while for the first excited band $M = -J + 1$ it is
$g^{\prime} \left(2J - 1\right)$. One can then follow the same
procedures as applied previously to obtain eigenvalues of
the Liouvillian, Eq.~(\ref{eq:tc-collective-liouvillian}).

\section{Summary and conclusions} \label{sec:conclusion}

In the first part of this paper we highlighted
the $u(1|1)$ supersymmetry of the Jaynes-Cummings model Hamiltonian
and showed that when dissipators cannot increase the number of
excitations $(n,m)$, the Liouvillian, $\mathcal{L}$, of the dissipative
Jaynes-Cummings model can be expressed in block triangular form,
each block being a finite-dimensional matrix due to the
Hamiltonian's supersymmetry. The eigenvalues of
the Liouvillian can then be obtained in explicit analytic form by
diagonalizing $1\times 1$, $2\times 2$, and $4\times 4$ submatrices
along the diagonal of $\mathcal{L}$. Additionally, if the excitation-conserving terms
of the dissipators of $\mathcal{L}$ contain only the Casimir operator $C$ of $u(1|1)$,
the Liouvillian has a dynamical supersymmetry and analytic expressions
for eigenvalues can be further simplified. These simple analytic solutions and the
more general procedure for obtaining eigenvalues of $\mathcal{L}$ from
small submatrices are the main results of this paper. We also demonstrated
that $u(1|1)$ supersymmetry can be used to analytically solve a variety of
Jaynes-Cummings-like model Hamiltonians and block triangularize their
Liouvillians in the same manner as for the JC model. These models include
the Weyl reflected JC model and models with multiphoton couplings.
In the second part of this paper, we studied the case of a
bosonic degree of freedom coupled to several qubits described by the Dicke
Hamiltonian \cite{dicke} with counter-rotating terms removed by the rotating
wave approximation \cite{tavis}, providing a solution of the eigenvalues of the
Hamiltonian and the Liouvillian for two qubits, $N=2$, and partial solutions
for any number of qubits $N$. In solving the Dicke Hamiltonian,
it is convenient to exploit other symmetries
in addition to the supersymmetry $\sum_{i} u_{i}(1|1)$, namely
permutational symmetry and ``rotational'' symmetry \cite{braak}. 
In doing so, the dynamical algebra $h(2) \oplus su(2)$ further reduces
the dimension of matrices to diagonalize.

The methods in the first part of this article can be extended to obtain eigenvalues
of the Hamiltonian $H$ and Liouvillian $\mathcal{L}$ for all coupled Bose-Fermi
systems of the form of Eq.~(\ref{eq:jc-susy}), namely those with $n$ bosonic and $m$ fermionic
degrees of freedom and spectrum generating algebra $u(n|m)$. The methods
described here can be used to solve dissipative systems with $u(n|m)$ dynamical
superalgebras in situations when the combined Hamiltonian
$H = H_{B} + H_{F} + V_{BF}$ commutes with the linear invariant $C$ of $u(n|m)$,
\begin{equation}
\label{eq:unm-casimir}
C = \sum_{\alpha} a^{\dag}_{\alpha} a_{\alpha} + \sum_{i} f^{\dag}_{i} f_{i} , \quad [H, C] = 0 ,
\end{equation}
and the jump operators are functionals $\Gamma_{\mu}$ of only the annihilation operators
$a_{\alpha}$ and $f_{i}$,
\begin{equation}
\label{eq:dissipator-functional}
\Gamma_{\mu} = \Gamma_{\mu} \left( a_{\alpha}, f_{i} \right) .
\end{equation}
The methods discussed in the second part of this article can be used
to obtain eigenvalues of the Hamiltonian $H$ and Liouvillian $\mathcal{L}$
of all dissipative systems with dynamical algebra $h(n+1) \oplus su(m+1)$. The conditions
here are that the combined Hamiltonian $H = H_{B} + H_{F} + V_{BF}$ commutes
with the Casimir operators of $h(n+1) \oplus su(m+1)$, and that the dissipators
are functionals $\Gamma_{\mu}$ of only the lowering operators of $h(n+1)$ and $su(m+1)$.
For the Dicke model, these are $a$ and $J_{-}$. An interesting case
here is that of a spin-$1/2$ fermion (qubit)
coupled to a $n$-dimensional isotropic oscillator with Hamiltonian
\begin{equation}
\label{eq:many-boson-spin-ham}
H_{D} = \omega_{c} \sum_{i=1}^{n} a_{i}^{\dag}a_{i} + \omega_{q} \frac{\sigma_{z}}{2}
+ g  \sum_{i=1}^{n} \left( \sigma_{+}a_{i}+\sigma_{-}a_{i}^{\dag}\right)
\end{equation}
and dissipators $\mathcal{D}[a_{i}]$, $\mathcal{D}[\sum_{i=1}^{n}a_{i}]$, and $\mathcal{D}[\sigma_{-}]$.

In conclusion, this article offers specific examples of the general quantum
mechanical principle that whenever a Hamiltonian commutes with the
invariant (Casimir) operators of some algebra, $[H,C] = 0$, its eigenvalue
problem can be cast in terms of diagonalization of finite matrices of dimension
$\dim \left[\Lambda\right]$, where $\Lambda$ denotes an irreducible representation
of the algebra. For rotationally invariant problems, $\Lambda \equiv J$ and
$\dim \left[J\right] = 2J + 1$. The main contribution of this paper is to
show explicitly that this general principle applies also to the eigenvalues of the Liouvillian
when dissipators do not increase the value of a conserved quantum number, which,
for the Jaynes-Cummings model is $n + \left\langle \frac{\sigma_{z}}{2} \right\rangle$,
and for the Dicke model is $n + \left\langle J_{z} \right\rangle$. The explicit analytic
solutions are also useful in the analysis of experimental data,
\cite{pedernales,leibfried,walther,blais1,fink,wallraff,gambetta,mezzacapo,felicetti1,blais2},
especially in the spectroscopy of Jaynes-Cummings and
Dicke models \cite{blais1,fink,wallraff}.

\bibliography{bibliography}

\appendix

\onecolumngrid

\section{Derivation of Jaynes-Cummings model Liouvillian eigenvalues} \label{appx:jc-liou}

Consider the Liouvillian of the Jaynes-Cummings model with bosonic and fermionic dissipators,
\begin{align}
\mathcal{L}\rho = -\mathrm{i}\left[ H,\rho \right] &+\gamma \mathcal{D}[a]\rho+
\gamma^{\prime }\mathcal{D}[\sigma_{-}]\rho , \\
\mathcal{D}[a]\rho = a\rho a^{\dag }-\frac{1}{2}\left( a^{\dag }a\rho +\rho
a^{\dag }a\right) &, \quad
\mathcal{D}[\sigma_{-}]\rho = \sigma_{-}\rho \sigma_{+}-\frac{1}{2}\left(
\sigma_{+}\sigma_{-}\rho +\rho \sigma_{+}\sigma_{-}\right) . \nonumber
\end{align}
To find the eigenvalues of $\mathcal{L}$, we first construct a basis for its eigenmatrices,
then find its matrix elements in this basis. Any linear operator acting on the Jaynes-Cummings
model Hilbert space can be expanded onto a basis of eigenstates of $H$, 
$\left\{ \left\vert \psi_{n}^{s}\right\rangle \right\} =
\left\{ \left\vert \psi_{0} \right\rangle \right\} \cup \left\{ \left\vert \psi_{n>0}^{\pm}\right\rangle  \right\}$,
where $\left\vert \psi_{0}^{-} \right\rangle \equiv \left\vert \psi_{0} \right\rangle$ and
$n \geq 0$ counts total bosonic plus fermionic excitations
(eigenvalues of the Casimir operator $C$) and $s=\pm$.
An operator $\rho$ can be written in this basis as
\begin{equation}
\rho =\sum \rho_{nm}^{st}\left\vert \psi_{n}^{s}\right\rangle \left\langle
\psi_{m}^{t}\right\vert \; , \; n,m=0,1,2,...; s,t= \pm .
\end{equation}
$n>0$ states are expressed in the $\left\{ \left\vert n_{B},s\right\rangle \right\}$ basis in Eq.~(\ref{eq:jc-ham-eig-exc}).
Jump operators act on $\left\vert \psi_{n}^{\mp}\right\rangle$ as
\begin{align}
a\left\vert \psi_{n}^{-}\right\rangle &=
\left( \sqrt{n-1} \cos \theta_{n} \cos \theta_{n-1}
+ \sqrt{n} \sin \theta_{n} \sin \theta_{n-1} \right)
\left\vert \psi_{n-1}^{-}\right\rangle
\nonumber \\
&+ \left( -\sqrt{n-1} \cos \theta_{n} \sin \theta_{n-1}
+ \sqrt{n} \sin \theta_{n} \cos \theta_{n-1} \right)
\left\vert \psi_{n-1}^{+}\right\rangle ,
\\
a\left\vert \psi_{n}^{+}\right\rangle &=
\left( -\sqrt{n-1} \sin \theta_{n} \cos \theta_{n-1}
+ \sqrt{n} \cos \theta_{n} \sin \theta_{n-1} \right)
\left\vert \psi_{n-1}^{-}\right\rangle
\nonumber \\
&+ \left( \sqrt{n-1} \sin \theta_{n} \sin \theta_{n-1}
+ \sqrt{n} \cos \theta_{n} \cos \theta_{n-1} \right)
\left\vert \psi_{n-1}^{+}\right\rangle ,
\nonumber \\
a\left\vert \psi_{0}\right\rangle &= 0 , \nonumber \\
\sigma_{-}\left\vert \psi_{n}^{-}\right\rangle &=
\cos \theta_{n} \sin \theta_{n-1}
\left\vert \psi_{n-1}^{-}\right\rangle
+\cos \theta_{n} \cos \theta_{n-1}
\left\vert \psi_{n-1}^{+}\right\rangle ,
\\
\sigma_{-}\left\vert \psi_{n}^{+}\right\rangle &=
-\sin \theta_{n} \sin \theta_{n-1}
\left\vert \psi_{n-1}^{-}\right\rangle
-\sin \theta_{n} \cos \theta_{n-1}
\left\vert \psi_{n-1}^{+}\right\rangle ,
\nonumber \\
\sigma_{-}\left\vert \psi_{0}\right\rangle &= 0 . \nonumber
\end{align}
One can se that the first terms in $\mathcal{D}[a]$ and $\mathcal{D}[\sigma_{-}]$,
$a\left\vert \psi_{n}^{s}\right\rangle \left\langle \psi_{m}^{t}\right\vert a^{\dag}$ and
$\sigma_{-}\left\vert\psi_{n}^{s}\right\rangle \left\langle \psi_{m}^{t}\right\vert \sigma_{+}$,
take states in the $(n,m)$ subspace into the $(n-1,m-1)$ subspace or annihilate them if $n$ or $m=0$.
However, also notice that
\begin{align}
a^{\dag }a\left\vert \psi_{n}^{-}\right\rangle &=
n \left\vert \psi_{n}^{-}\right\rangle -\cos \theta_{n}\left\vert n-1,+\frac{1}{2}
\right\rangle = \left( n - \cos^{2} \theta_{n} \right) \left\vert \psi_{n}^{-}\right\rangle +
\cos \theta_{n} \sin \theta_{n} \left\vert \psi_{n}^{+}\right\rangle , \nonumber \\
a^{\dag }a\left\vert \psi_{n}^{+}\right\rangle &=
n\left\vert \psi_{n}^{+}\right\rangle +\sin \theta_{n}\left\vert n-1,+\frac{1}{2}
\right\rangle = \left( n - \sin^{2} \theta_{n} \right) \left\vert \psi_{n}^{+}\right\rangle +
\cos \theta_{n} \sin \theta_{n} \left\vert \psi_{n}^{-}\right\rangle ,
\\
\sigma_{+}\sigma_{-}\left\vert \psi_{n}^{-}\right\rangle &=\cos \theta
_{n}\left\vert n-1,+\frac{1}{2}\right\rangle 
= \cos^{2} \theta_{n} \left\vert \psi_{n}^{-}\right\rangle -
\cos \theta_{n} \sin \theta_{n} \left\vert \psi_{n}^{+}\right\rangle ,
\nonumber \\
\sigma_{+}\sigma_{-}\left\vert \psi_{n}^{+}\right\rangle &=-\sin \theta
_{n}\left\vert n-1,+\frac{1}{2}\right\rangle 
= \sin^{2} \theta_{n} \left\vert \psi_{n}^{+}\right\rangle -
\cos \theta_{n} \sin \theta_{n} \left\vert \psi_{n}^{-}\right\rangle ,
\end{align}
and $a^{\dag}a\left\vert \psi_{0}\right\rangle = \sigma_{+}\sigma_{-}\left\vert \psi_{0}\right\rangle = 0$.
Therefore, the terms
$a^{\dag }a\left\vert \psi_{n}^{s}\right\rangle\left\langle \psi_{m}^{t}\right\vert
+\left\vert \psi_{n}^{s}\right\rangle\left\langle \psi_{m}^{t}\right\vert a^{\dag }a$
and
$\sigma_{+}\sigma_{-}\left\vert \psi_{n}^{s}\right\rangle \left\langle \psi_{m}^{t}\right\vert
+\left\vert \psi_{n}^{s}\right\rangle \left\langle \psi_{m}^{t}\right\vert \sigma_{+}\sigma_{-}$
act only within the $(n,m)$ subspace. Also,
\begin{equation}
\left[ H,\left\vert \psi_{n}^{s}\right\rangle \left\langle \psi_{m}^{t}\right\vert \right] =
(E_{n}^{s}-E_{m}^{t})\left\vert \psi_{n}^{s}\right\rangle \left\langle\psi_{m}^{t}\right\vert
\end{equation}
is diagonal and acts within the $(n,m)$ subspace. The Liouvillian acts overall as
\begin{align}
 \mathcal{L} \left\vert \psi_{n}^{s}\right\rangle \left\langle \psi_{m}^{t}\right\vert &=
-\mathrm{i} (E_{n}^{s}-E_{m}^{t})\left\vert \psi_{n}^{s}\right\rangle \left\langle\psi_{m}^{t}\right\vert + \gamma \left( a\left\vert \psi_{n}^{s}\right\rangle \left\langle \psi_{m}^{t}\right\vert a^{\dag }
- \frac{1}{2} \left( a^{\dag }a\left\vert \psi_{n}^{s}\right\rangle\left\langle \psi_{m}^{t}\right\vert 
+ \left\vert \psi_{n}^{s}\right\rangle\left\langle \psi_{m}^{t}\right\vert a^{\dag }a \right) \right) \nonumber \\
&+ \gamma^\prime \left( \sigma_{-}\left\vert\psi_{n}^{s}\right\rangle \left\langle \psi_{m}^{t}\right\vert \sigma_{+}
- \frac{1}{2} \left( \sigma_{+}\sigma_{-}\left\vert \psi_{n}^{s}\right\rangle \left\langle \psi_{m}^{t}\right\vert 
+ \left\vert \psi_{n}^{s}\right\rangle \left\langle \psi_{m}^{t}\right\vert \sigma_{+}\sigma_{-} \right) \right)
\end{align}
and one can see that all superoperator terms in $\mathcal{L}$ either conserve the total
number of bosonic plus fermionic excitations in the system, preserving
eigenvalues of the $u(1|1)$ Casimir operator $C$ and transforming states in the
subspace $(n,m) \rightarrow (n,m)$, or decrease the number of bra- and ket-excitations each by one,
$(n,m) \rightarrow (n-1,m-1)$. Thus the Liouvillian takes on a block triangular form in the 
$\left\vert \psi_{n}^{s}\right\rangle \left\langle \psi_{m}^{t}\right\vert$ basis, and its
eigenvalues can be found by simply diagonalizing the blocks of matrix elements
which conserve the total number of excitations $(n,m) \rightarrow (n,m)$, denoted
$\mathcal{L}^{(n,m)}$. In the vectorized representation of $\mathcal{L}$ and $\rho$,
\begin{equation}
\mathcal{L} \rho = \left( 
\begin{array}{ccccc}
\mathcal{L}^{(0,0)} & \times & & & \\
0 & \mathcal{L}^{(0,1)} & \times & & \\
 & 0 & \ddots & \times & \\
  & & 0 & \mathcal{L}^{(n,m)} & \times \\
  & & & 0 & \ddots
\end{array}
\right)
 \left( 
\begin{array}{c}
\rho_{(0,0)} \\
\rho_{(0,1)} \\
\vdots \\
\rho_{(n,m)} \\
\vdots
\end{array}
\right) .
\end{equation}
Each $(n,m)$ subspace is degenerate based on the possible spin values $s$ and $t$ that states
$\left\vert \psi_{n}^{s}\right\rangle \left\langle \psi_{m}^{t}\right\vert$ can take.
For $(n,m) = (0,0)$, $s=t=-$, and there is a single state $\left\vert \psi_{0}\right\rangle \left\langle \psi_{0}\right\vert$.
For $(n,m)$ with $n,m > 0$, $s = \pm$ and $t = \pm$ and there are four states. For $(n,m) = (n,0)$,
$s = \pm$ and $t = -$, so there are two states $\left\{ \left\vert \psi_{n}^{s}\right\rangle \left\langle \psi_{0}\right\vert \right\}$,
and lastly for $(n,m) = (0,m)$ there are also two states,
$\left\{ \left\vert \psi_{0}\right\rangle \left\langle \psi_{m}^{t}\right\vert \right\}$, since $s = -$ and $t = \pm$.

Starting with $(n,m) = (0,0)$, it is easy to see
$\mathcal{L} \left\vert \psi_{0}\right\rangle \left\langle \psi_{0}\right\vert = 0 = \mathcal{L}^{(0,0)} $.
For the remaining subspaces, introducing the following constants simplifies notation,
\begin{align}
 B^{st}_{nm} &=-\mathrm{i}(E_{n}^{s}-E_{m}^{t})-\frac{\gamma }{2}(n+m)-
\frac{\gamma^{\prime }-\gamma }{2}\left( \left\{ 
\begin{array}{c}
\delta_{s,+} \sin^{2}\theta_{n} + \\ 
 \delta_{s,-} \cos^{2}\theta_{n}
\end{array}
\right\} +\left\{ 
\begin{array}{c}
\delta_{t,+} \sin^{2}\theta_{m} + \\ 
\delta_{t,-} \cos^{2}\theta_{m}
\end{array}
\right\} \right) , \nonumber \\
K_{n}^{s} &=-\mathrm{i}(E_{n}^{s}-E_{0})-\frac{\gamma }{2}n-\frac{\gamma
^{\prime }-\gamma }{2}\left\{ 
\begin{array}{c}
\delta_{s,+} \sin^{2}\theta_{n} + \\ 
\delta_{s,-} \cos^{2}\theta_{n}
\end{array}
\right\} , \quad
K_{m}^{t} =-\mathrm{i}(E_{0}-E_{m}^{t})-\frac{\gamma }{2}m-\frac{\gamma
^{\prime }-\gamma }{2}\left\{ 
\begin{array}{c}
\delta_{t,+} \sin^{2}\theta_{m} + \\ 
\delta_{t,-} \cos^{2}\theta_{m}
\end{array}
\right\} , \nonumber \\
A_{n} &=\frac{\gamma^{\prime }-\gamma }{2}\cos \theta_{n}\sin \theta_{n}
, \quad A_{m}=\frac{\gamma^{\prime }-\gamma }{2}\cos \theta_{m}\sin \theta_{m} ,
\end{align}
where $\delta_{s/t,\pm}$ are Kronecker deltas.
For the four-dimensional subspaces $(n,m)$, $n,m > 0$, we consider the action of $\mathcal{L}$
on each state individually.
Using the above expressions, this yields
\begin{align}
 \mathcal{L} \left\vert \psi_{n}^{+}\right\rangle \left\langle \psi_{m}^{+}\right\vert &=
B^{++}_{nm}
\left\vert \psi_{n}^{+}\right\rangle \left\langle \psi_{m}^{+}\right\vert
+ A_{n}\left\vert  \psi_{n}^{-}\right\rangle \left\langle \psi_{m}^{+}\right\vert
+ A_{m}\left\vert \psi_{n}^{+}\right\rangle \left\langle \psi_{m}^{-}\right\vert
+ \rho_{\perp}^{++} , \nonumber \\
\mathcal{L} \left\vert \psi_{n}^{-}\right\rangle \left\langle \psi_{m}^{-}\right\vert &=
B^{--}_{nm}
\left\vert \psi_{n}^{-}\right\rangle \left\langle \psi_{m}^{-}\right\vert
+ A_{n}\left\vert  \psi_{n}^{+}\right\rangle \left\langle \psi_{m}^{-}\right\vert
+ A_{m}\left\vert \psi_{n}^{-}\right\rangle \left\langle \psi_{m}^{+}\right\vert
+ \rho_{\perp}^{--} , \nonumber \\
 \mathcal{L} \left\vert \psi_{n}^{+}\right\rangle \left\langle \psi_{m}^{-}\right\vert &=
B^{+-}_{nm}
\left\vert \psi_{n}^{+}\right\rangle \left\langle \psi_{m}^{-}\right\vert
+ A_{n}\left\vert  \psi_{n}^{-}\right\rangle \left\langle \psi_{m}^{-}\right\vert
+ A_{m}\left\vert \psi_{n}^{+}\right\rangle \left\langle \psi_{m}^{+}\right\vert
+ \rho_{\perp}^{+-} , \nonumber \\
 \mathcal{L} \left\vert \psi_{n}^{-}\right\rangle \left\langle \psi_{m}^{+}\right\vert &=
B^{-+}_{nm}
\left\vert \psi_{n}^{-}\right\rangle \left\langle \psi_{m}^{+}\right\vert
+ A_{n}\left\vert  \psi_{n}^{+}\right\rangle \left\langle \psi_{m}^{+}\right\vert
+ A_{m}\left\vert \psi_{n}^{-}\right\rangle \left\langle \psi_{m}^{-}\right\vert
+ \rho_{\perp}^{-+} .
\end{align}
where $\rho_{\perp}^{st} = \left(  \gamma a\left\vert \psi_{n}^{s}\right\rangle \left\langle \psi_{m}^{t}\right\vert a^{\dag}
+ \gamma^\prime  \sigma_{-} \left\vert\psi_{n}^{s}\right\rangle \left\langle \psi_{m}^{t}\right\vert \sigma_{+} \right)$
takes $(n,m) \rightarrow (n-1,m-1)$. We can now describe how $\mathcal{L}$ acts on a $(n,m)$ subspace,
\begin{equation}
\mathcal{L}
 \left( 
\begin{array}{c}
\left\vert \psi_{n}^{+}\right\rangle \left\langle \psi_{m}^{+}\right\vert \\
\left\vert \psi_{n}^{+}\right\rangle \left\langle \psi_{m}^{-}\right\vert \\
\left\vert \psi_{n}^{-}\right\rangle \left\langle \psi_{m}^{+}\right\vert \\
\left\vert \psi_{n}^{-}\right\rangle \left\langle \psi_{m}^{-}\right\vert 
\end{array}
\right) = \mathcal{L}^{(n,m)}  \left( 
\begin{array}{c}
\left\vert \psi_{n}^{+}\right\rangle \left\langle \psi_{m}^{+}\right\vert \\
\left\vert \psi_{n}^{+}\right\rangle \left\langle \psi_{m}^{-}\right\vert \\
\left\vert \psi_{n}^{-}\right\rangle \left\langle \psi_{m}^{+}\right\vert \\
\left\vert \psi_{n}^{-}\right\rangle \left\langle \psi_{m}^{-}\right\vert 
\end{array}
\right) + \rho_{\perp}^{(n,m)} ,
\end{equation}
and construct submatrices $\mathcal{L}^{(n,m)}$. As mentioned previously,
terms $\rho_{\perp}^{st}$ in $\rho_{\perp}^{(n,m)}$ are not needed to find
eigenvalues of $\mathcal{L}$. The $4 \times 4$ blocks $\mathcal{L}^{(n,m)}$ take the form
\begin{equation}
\mathcal{L}^{(n,m)}=\left( 
\begin{array}{cccc}
B^{++}_{nm} & A_{m} & A_{n} & 0 \\ 
A_{m} & B^{+-}_{nm} & 0 & A_{n} \\ 
A_{n} & 0 & B^{-+}_{nm} & A_{m} \\ 
0 & A_{n} & A_{m} & B^{--}_{nm}
\end{array}
\right) .
\end{equation}
We repeat the same analysis as above for the $2 \times 2$ $(n,0)$ and $(0,m)$ blocks.
Here, results are considerably simpler as $a \left\vert \psi_{0}\right\rangle = \sigma_{-} \left\vert \psi_{0}\right\rangle = 0$.
For $(n,0)$ subspaces, we have
\begin{align}
\mathcal{L} \left\vert \psi_{n}^{+}\right\rangle \left\langle \psi_{0}\right\vert &=
K_{n}^{+}
\left\vert \psi_{n}^{+}\right\rangle \left\langle \psi_{0}\right\vert
+ A_{n} \left\vert  \psi_{n}^{-}\right\rangle \left\langle \psi_{0}\right\vert
+ \rho_{\perp}^{+0} ,\nonumber \\
\mathcal{L} \left\vert \psi_{n}^{-}\right\rangle \left\langle \psi_{0}\right\vert &=
K_{n}^{-}
\left\vert \psi_{n}^{-}\right\rangle \left\langle \psi_{0}\right\vert
+ A_{n}\left\vert  \psi_{n}^{+}\right\rangle \left\langle \psi_{0}\right\vert 
+ \rho_{\perp}^{-0} .
\end{align}
For $(0,m)$ subspaces, we have
\begin{align}
\mathcal{L} \left\vert \psi_{0}\right\rangle \left\langle \psi_{m}^{+}\right\vert &=
K_{m}^{+}
\left\vert \psi_{0}\right\rangle \left\langle \psi_{m}^{+}\right\vert
+ A_{m}
\left\vert  \psi_{0}\right\rangle \left\langle \psi_{m}^{-}\right\vert
+ \rho_{\perp}^{0+} , \nonumber \\
\mathcal{L} \left\vert \psi_{0}\right\rangle \left\langle \psi_{m}^{-}\right\vert &=
K_{m}^{-}]
\left\vert \psi_{0}\right\rangle \left\langle \psi_{m}^{-}\right\vert
+ A_{m}\left\vert  \psi_{0}\right\rangle \left\langle \psi_{m}^{+}\right\vert
+ \rho_{\perp}^{0-} .
\end{align}
From these matrix elements, we can construct the $\mathcal{L}^{(n,0)}$
and $\mathcal{L}^{(0,m)}$ blocks,
\begin{equation}
\mathcal{L}^{(n,0)}=\left( 
\begin{array}{cc}
K_{n}^{+} & A_{n} \\ 
A_{n} & K_{n}^{-}
\end{array}
\right) , \quad \mathcal{L}^{(0,m)}=\left( 
\begin{array}{cc}
K_{m}^{+} & A_{m} \\ 
A_{m} & K_{m}^{-}
\end{array}
\right) .
\end{equation}
We can diagonalize each block individually, $\mathcal{L}^{(0,0)}$, $\mathcal{L}^{(n,0)}$, $\mathcal{L}^{(0,m)}$,
and $\mathcal{L}^{(n,m)}$, to obtain eigenvalues of $\mathcal{L}$. Blocks are small enough that
diagonalization may be performed analytically. Denote eigenvalues
corresponding to the $(n,m)$ subspace as $\lambda^{s,t}_{(n,m)}$ where $s,t = \pm, 0$.
Diagonalization yields, after some substitutions,
\begin{align}
 \lambda^{0,0}_{(0,0)} &= 0 \\
 \lambda^{\pm,0}_{(n,0)} &= 
-\mathrm{i}\left[ \omega_{c}n - \frac{1}{2}\left(\omega_{c}-\omega_{q}\right)
\pm \frac{1}{2} \sqrt{\left(\omega_{c}-\omega_{q} + \mathrm{i} \frac{\gamma^{\prime}-\gamma}{2} \right)^{2} + 4g^{2}n} \right]
- \frac{\gamma^{\prime}-\gamma}{4} - \frac{\gamma}{2}n \nonumber \\
 \lambda^{0,\pm}_{(0,m)} &= 
\mathrm{i}\left[ \omega_{c}m - \frac{1}{2}\left(\omega_{c}-\omega_{q}\right)
\pm \frac{1}{2} \sqrt{\left(\omega_{c}-\omega_{q} - \mathrm{i} \frac{\gamma^{\prime}-\gamma}{2} \right)^{2} + 4g^{2}m} \right]
- \frac{\gamma^{\prime}-\gamma}{4} - \frac{\gamma}{2}m \nonumber \\
 \lambda^{\pm,\pm}_{(n,m)} &= 
-\mathrm{i}\left[ \omega_{c}\left(n-m\right)
\pm \frac{1}{2} \sqrt{\left(\omega_{c}-\omega_{q} + \mathrm{i} \frac{\gamma^{\prime}-\gamma}{2} \right)^{2} + 4g^{2}n}
\right. \nonumber \\
 &\mp \left. \frac{1}{2} \sqrt{\left(\omega_{c}-\omega_{q} - \mathrm{i} \frac{\gamma^{\prime}-\gamma}{2} \right)^{2} + 4g^{2}m} \right]
- \frac{\gamma^{\prime}-\gamma}{2} - \frac{\gamma}{2} \left(n + m\right) . \nonumber
\end{align}
After some manipulations, we can also see that
\begin{align}
\lambda^{\pm,0}_{(n,0)} &= 
-\mathrm{i}\left[ \left(\omega_{c} -\mathrm{i}\frac{\gamma}{2} \right)\left(n - \frac{1}{2}\right) + \left(\omega_{q} -\mathrm{i}\frac{\gamma^{\prime}}{2} \right)\frac{1}{2}
\pm \frac{1}{2} \sqrt{\left[\left(\omega_{c} - \mathrm{i} \frac{\gamma}{2}\right) - \left(\omega_{q} - \mathrm{i} \frac{\gamma^{\prime}}{2} \right)\right]^{2} + 4g^{2}n} \right] , \nonumber \\
\lambda^{0,\pm}_{(0,m)} &= 
\mathrm{i}\left[ \left(\omega_{c} +\mathrm{i}\frac{\gamma}{2} \right)\left(m - \frac{1}{2}\right) + \left(\omega_{q} +\mathrm{i}\frac{\gamma^{\prime}}{2} \right)\frac{1}{2}
\pm \frac{1}{2} \sqrt{\left[\left(\omega_{c} + \mathrm{i} \frac{\gamma}{2}\right) - \left(\omega_{q} + \mathrm{i} \frac{\gamma^{\prime}}{2} \right)\right]^{2} + 4g^{2}m} \right] , \nonumber \\
\lambda^{\pm,\pm}_{(n,m)} &= \lambda^{\pm,0}_{(n,0)} + \lambda^{0,\pm}_{(0,m)} ,
\quad \lambda^{0,0}_{(0,0)} = 0 .
\end{align}
These results can be further simplified by introducing complex frequencies
$\widetilde{\omega}_{c} \equiv \omega_{c} -\mathrm{i}\frac{\gamma}{2}$ and
$\widetilde{\omega}_{q} \equiv \omega_{q} -\mathrm{i}\frac{\gamma^{\prime}}{2}$,
\begin{align}
\lambda^{\pm,0}_{(n,0)} &= 
-\mathrm{i}\left[ \widetilde{\omega}_{c}\left(n - \frac{1}{2}\right) + \frac{\widetilde{\omega}_{q}}{2}
\pm \frac{1}{2} \sqrt{\left(\widetilde{\omega}_{c} - \widetilde{\omega}_{q}\right)^{2} + 4g^{2}n} \right] , \nonumber \\
\lambda^{0,\pm}_{(0,m)} &= 
\mathrm{i}\left[ \widetilde{\omega}_{c}^{\ast}\left(m - \frac{1}{2}\right) + \frac{\widetilde{\omega}_{q}^{\ast}}{2}
\pm \frac{1}{2} \sqrt{\left(\widetilde{\omega}_{c}^{\ast} - \widetilde{\omega}_{q}^{\ast}\right)^{2} + 4g^{2}m} \right] , \nonumber \\
\lambda^{\pm,\pm}_{(n,m)} &= \lambda^{\pm,0}_{(n,0)} + \lambda^{0,\pm}_{(0,m)} ,
\quad \lambda^{0,0}_{(0,0)} = 0 .
\end{align}
Here, we observe that $\lambda^{\pm,0}_{(n,0)} = -\mathrm{i} \left( \widetilde{E}_{n}^{\pm} - \widetilde{E}_{0} \right)$
and $\lambda^{0,\pm}_{(0,m)} = \mathrm{i} \left( \widetilde{E}_{m}^{\pm \ast} - \widetilde{E}_{0}^{\ast} \right)$ where
$\widetilde{E}_{n}^{\pm}, \widetilde{E}_{0}$ are eigenvalues of the JC model Hamiltonian with 
complex frequencies $\widetilde{\omega}_{c}$ and $\widetilde{\omega}_{q}$.

Because the Weyl-reflected JC model, 
\begin{equation}
H_{\overline{JC}}=\omega_{c}a^{\dag}a-\omega_{q}\frac{\sigma_{z}}{2}
+g \left( \sigma_{+}a^{\dag}+\sigma_{-}a\right) ,
\end{equation}
is isomorphic to the JC model with a change of basis for spin operators,
$\sigma_{\pm} \rightarrow \sigma_{\mp}$ and $\sigma_{z} \rightarrow -\sigma_{z}$,
its Liouvillian with dissipators $\mathcal{D}[a]$ and$\mathcal{D}[\sigma_{+}]$,
\begin{align}
\mathcal{L}\rho = -\mathrm{i}\left[ H_{\overline{JC}},\rho \right] &+\gamma \mathcal{D}[a]\rho+
\gamma^{\prime }\mathcal{D}[\sigma_{+}]\rho , \\
\mathcal{D}[a]\rho = a\rho a^{\dag }-\frac{1}{2}\left( a^{\dag }a\rho +\rho
a^{\dag }a\right) &, \quad
\mathcal{D}[\sigma_{+}] \rho = \sigma_{+}\rho \sigma_{-}-\frac{1}{2}\left(
\sigma_{-}\sigma_{+}\rho +\rho \sigma_{-}\sigma_{+}\right) , \nonumber
\end{align}
can be solved identically to that of the JC model above and has the same eigenvalues.

\subsection{Jaynes-Cummings model with inverted dissipators} \label{appx:jc-liou-dissipators}

Now, we consider the standard JC model again,
\begin{equation}
H_{JC}=\omega_{c}a^{\dag }a+\omega_{q}\frac{\sigma_{z}}{2}+g\left(
\sigma_{+}a+\sigma_{-}a^{\dag }\right) ,
\end{equation}
and its Liouvillian with inverted dissipators, $\mathcal{D}[a^{\dag}]$ and $\mathcal{D}[\sigma_{+}]$,
\begin{align}
\mathcal{L}\rho = -\mathrm{i}\left[ H,\rho \right] &+\gamma \mathcal{D}[a^{\dag}]\rho+\gamma
^{\prime }\mathcal{D}[\sigma_{+}]\rho , \\
\mathcal{D}[a^{\dag}]\rho = a^{\dag}\rho a-\frac{1}{2}\left( aa^{\dag}\rho +\rho
aa^{\dag}\right) &, \quad
\mathcal{D}[\sigma_{+}]\rho = \sigma_{+}\rho \sigma_{-}-\frac{1}{2}\left(
\sigma_{-}\sigma_{+}\rho +\rho \sigma_{-}\sigma_{+}\right) . \nonumber
\end{align}
Here, note that $aa^{\dag} = 1 + a^{\dag}a$ and $\sigma_{-}\sigma_{+} = 1 - \sigma_{+}\sigma_{-}$,
so we can use similar relations to those above to explicitly compute matrix elements of the Liouvillian.
Now, $a^{\dag} \left\vert \psi_{n}^{s}\right\rangle \left\langle \psi_{m}^{t}\right\vert a$ and
$\sigma_{+}\left\vert \psi_{n}^{s}\right\rangle \left\langle \psi_{m}^{+}\right\vert \sigma_{-}$
take states from subspace $(n,m) \rightarrow (n+1,m+1)$. Because of this, the matrix representation
of the Liouvillian is lower block triangular (recall the original Liouvillian was upper block triangular).
We can then find eigenvalues in the same way as before, by diagonalizing blocks of matrix elements along
the diagonal which conserve excitations $(n,m) \rightarrow (n,m)$. Denote these blocks $\mathcal{L}^{(n,m)}$
similar to before. Explicitly calculating blocks,
\begin{equation}
\mathcal{L}^{(0,0)} = -(\gamma + \gamma^\prime) , \;
\mathcal{L}^{(n,0)}=\left( 
\begin{array}{cc}
K_{n}^{+} & A_{n} \\ 
A_{n} & K_{n}^{-}
\end{array}
\right) , \;
\mathcal{L}^{(0,m)}=\left( 
\begin{array}{cc}
K_{m}^{+} & A_{m} \\ 
A_{m} & K_{m}^{-}
\end{array}
\right) , \;
\mathcal{L}^{(n,m)}=\left( 
\begin{array}{cccc}
B^{++}_{nm} & A_{m} & A_{n} & 0 \\ 
A_{m} & B^{+-}_{nm} & 0 & A_{n} \\ 
A_{n} & 0 & B^{-+}_{nm} & A_{m} \\ 
0 & A_{n} & A_{m} & B^{--}_{nm}
\end{array}
\right) ,
\end{equation}
\begin{align*}
B^{st}_{nm} &=-\mathrm{i}(E_{n}^{s}-E_{m}^{t})-\frac{\gamma }{2}(n+m)-(\gamma + \gamma^\prime)
+\frac{\gamma+\gamma^{\prime }}{2}\left( \left\{ 
\begin{array}{c}
\delta_{s,+} \sin^{2}\theta_{n} + \\ 
 \delta_{s,-} \cos^{2}\theta_{n}
\end{array}
\right\} +\left\{ 
\begin{array}{c}
\delta_{t,+} \sin^{2}\theta_{m} + \\ 
\delta_{t,-} \cos^{2}\theta_{m}
\end{array}
\right\} \right) , \\
K_{n}^{s} &=-\mathrm{i}(E_{n}^{s}-E_{0})-\frac{\gamma }{2}n-(\gamma + \gamma^\prime)
+\frac{\gamma+\gamma^{\prime }}{2}\left\{ 
\begin{array}{c}
\delta_{s,+} \sin^{2}\theta_{n} + \\ 
\delta_{s,-} \cos^{2}\theta_{n}
\end{array}
\right\}, \quad A_{n} = -\frac{\gamma+\gamma^{\prime} }{2}\cos \theta_{n}\sin \theta_{n} , \\
K_{m}^{t} &=-\mathrm{i}(E_{0}-E_{m}^{t})-\frac{\gamma }{2}m-(\gamma + \gamma^\prime)
+\frac{\gamma+\gamma^{\prime }}{2}\left\{ 
\begin{array}{c}
\delta_{t,+} \sin^{2}\theta_{m} + \\ 
\delta_{t,-} \cos^{2}\theta_{m}
\end{array}
\right\} , \quad A_{m}=-\frac{\gamma+\gamma^{\prime }}{2}\cos \theta_{m}\sin
\theta_{m} .
\end{align*}
These terms are similar to those for the JC model with standard dissipators,
except with $\gamma^\prime \leftrightarrow - \gamma^\prime$ flipped and a constant shift
$-(\gamma + \gamma^\prime)$ added along the diagonal. This can be derived explicitly
using $aa^{\dag} = 1 + a^{\dag}a$ and $\sigma_{-}\sigma_{+} = 1 - \sigma_{+}\sigma_{-}$,
as
\begin{align}
\left( \mathcal{D}[a^{\dag}] \left\vert \psi_{n}^{s}\right\rangle \left\langle \psi_{m}^{t}\right\vert \right) \Bigg\vert_{(n,m)} &= \left( (-1 + \mathcal{D}[a] ) \left\vert \psi_{n}^{s}\right\rangle \left\langle \psi_{m}^{t}\right\vert \right) \Bigg\vert_{(n,m)} \nonumber \\
\left( \mathcal{D}[\sigma_{+}] \left\vert \psi_{n}^{s}\right\rangle \left\langle \psi_{m}^{t}\right\vert \right) \Bigg\vert_{(n,m)} &= \left( (-1 - \mathcal{D}[\sigma_{-}] ) \left\vert \psi_{n}^{s}\right\rangle \left\langle \psi_{m}^{t}\right\vert \right) \Bigg\vert_{(n,m)}
\end{align}
when restricted to a $(n,m)$ subspace. Thus, eigenvalues can be found by simply swapping
$\gamma^\prime \leftrightarrow - \gamma^\prime$ in the results derived for the standard dissipators
and shifting by $-(\gamma + \gamma^\prime)$. Introducing complex
frequencies $\widetilde{\omega}_{c} \equiv \omega_{c} -\mathrm{i}\frac{\gamma}{2}$ and
$\widetilde{\omega}_{q} \equiv \omega_{q} +\mathrm{i}\frac{\gamma^{\prime}}{2}$,
eigenvalues are
\begin{align}
\lambda^{\pm,0}_{(n,0)} &= 
-\mathrm{i}\left[ \widetilde{\omega}_{c}\left(n - \frac{1}{2}\right) + \frac{\widetilde{\omega}_{q}}{2}
\pm \frac{1}{2} \sqrt{\left(\widetilde{\omega}_{c} - \widetilde{\omega}_{q}\right)^{2} + 4g^{2}n} \right]
-(\gamma + \gamma^\prime) \nonumber \\
\lambda^{0,\pm}_{(0,m)} &= 
\mathrm{i}\left[ \widetilde{\omega}_{c}^{\ast}\left(m - \frac{1}{2}\right) + \frac{\widetilde{\omega}_{q}^{\ast}}{2}
\pm \frac{1}{2} \sqrt{\left(\widetilde{\omega}_{c}^{\ast} - \widetilde{\omega}_{q}^{\ast}\right)^{2} + 4g^{2}m} \right]
-(\gamma + \gamma^\prime) \nonumber \\
\lambda^{\pm,\pm}_{(n,m)} &= \lambda^{\pm,0}_{(n,0)} + \lambda^{0,\pm}_{(0,m)} + (\gamma + \gamma^\prime) ,
\quad \lambda^{0,0}_{(0,0)} = -(\gamma + \gamma^\prime) ,
\end{align}
where $(\gamma + \gamma^\prime)$ is added in $\lambda^{\pm,\pm}_{(n,m)}$
to ensure the shift in each eigenvalue is constant.

Similar to what was derived previously, the Weyl-reflected JC model with dissipators
$\mathcal{D}[a^{\dag}]$ and $\mathcal{D}[\sigma_{-}]$,
\begin{align}
\mathcal{L}\rho = -\mathrm{i}\left[ H_{\overline{JC}},\rho \right] &+\gamma \mathcal{D}[a]\rho+
\gamma^{\prime }\mathcal{D}[\sigma_{+}]\rho , \\
\mathcal{D}[a^{\dag}]\rho = a^{\dag}\rho a-\frac{1}{2}\left( aa^{\dag}\rho +\rho
aa^{\dag}\right) &, \quad
\mathcal{D}[\sigma_{-}]\rho = \sigma_{-}\rho \sigma_{+}-\frac{1}{2}\left(
\sigma_{+}\sigma_{-}\rho +\rho \sigma_{+}\sigma_{-}\right) , \nonumber
\end{align}
can be solved identically to the JC model with dissipators
$\mathcal{D}[a^{\dag}]$ and $\mathcal{D}[\sigma_{+}]$ above and has the same eigenvalues.

\section{General Framework For Evaluating Additional Dissipators} \label{appx:jc-framework}

In this section, we provide a general formulation of Liouvillian submatrices which
conserve excitations $(n,m) \rightarrow (n,m)$ for dissipators of operators $\Gamma$
which either conserve or lower excitations, $\Gamma : n \rightarrow \bigoplus_{n^{\prime} \leq n} n^{\prime}$.
These include powers of $a$, $\sigma_{-}$, and $\sigma_{z}$, as well as products and linear combinations
of these operators. We start by writing how $\Gamma$ acts on basis states within the $n$ subspace,
\begin{align}
\Gamma \left\vert \psi_{n}^{-}\right\rangle &=
\mathcal{P}_{n}^{-} \left\vert \psi_{n}^{-}\right\rangle +
\mathcal{Q}_{n}^{-} \left\vert \psi_{n}^{+}\right\rangle + \left\vert \psi_{\perp}\right\rangle , \nonumber \\
\Gamma \left\vert \psi_{n}^{+}\right\rangle &=
\mathcal{P}_{n}^{+} \left\vert \psi_{n}^{+}\right\rangle +
\mathcal{Q}_{n}^{+} \left\vert \psi_{n}^{-}\right\rangle + \left\vert \psi_{\perp}\right\rangle ,
\end{align}
where $\left\vert \psi_{\perp}\right\rangle$ denotes any component of 
$\Gamma \left\vert \psi_{n}^{\mp}\right\rangle$ not in the $n$ subspace
and $\mathcal{P}_{n}^{\pm}$, $\mathcal{Q}_{n}^{\pm}$ are coefficients characterizing
how $\Gamma$ acts within this subspace. The dissipator $\mathcal{D}[\Gamma]\rho =
\Gamma \rho \Gamma^{\dag} - \frac{1}{2}\left(\Gamma^{\dag} \Gamma \rho +
\rho \Gamma^{\dag}\Gamma \right)$ also contains $\Gamma^{\dag}\Gamma$
which strictly conserves excitations,
\begin{align}
\Gamma^{\dag}\Gamma \left\vert \psi_{n}^{-}\right\rangle &=
\mathcal{B}_{n}^{-} \left\vert \psi_{n}^{-}\right\rangle +
\mathcal{A}_{n}^{-} \left\vert \psi_{n}^{+}\right\rangle , \nonumber \\
\Gamma^{\dag}\Gamma \left\vert \psi_{n}^{+}\right\rangle &=
\mathcal{B}_{n}^{+} \left\vert \psi_{n}^{+}\right\rangle +
\mathcal{A}_{n}^{+} \left\vert \psi_{n}^{-}\right\rangle .
\end{align}
Here, $\mathcal{B}_{n}^{-}$ and $\mathcal{A}_{n}^{-}$ are coefficients. We can now write
down how $\mathcal{D}[\Gamma]$ acts on basis states within each $(n,m)$ subspace. Starting
with $(n>0,m>0)$, $\mathcal{D}[\Gamma]$ acts as
\begin{align}
\mathcal{D}[\Gamma] \left\vert \psi_{n}^{+}\right\rangle \left\langle \psi_{m}^{+}\right\vert &=
\Gamma \left\vert \psi_{n}^{+}\right\rangle \left\langle \psi_{m}^{+}\right\vert \Gamma^{\dag}
- \frac{1}{2}\left[
\left(\mathcal{B}_{n}^{+} + \mathcal{B}_{m}^{+} \right) \left\vert \psi_{n}^{+}\right\rangle \left\langle \psi_{m}^{+}\right\vert
+\mathcal{A}_{n}^{+} \left\vert \psi_{n}^{-}\right\rangle \left\langle \psi_{m}^{+}\right\vert
+ \mathcal{A}_{m}^{+} \left\vert \psi_{n}^{+}\right\rangle \left\langle \psi_{m}^{-}\right\vert
\right] \nonumber \\
\mathcal{D}[\Gamma] \left\vert \psi_{n}^{+}\right\rangle \left\langle \psi_{m}^{-}\right\vert &=
\Gamma \left\vert \psi_{n}^{+}\right\rangle \left\langle \psi_{m}^{-}\right\vert \Gamma^{\dag}
- \frac{1}{2}\left[
\left(\mathcal{B}_{n}^{+} + \mathcal{B}_{m}^{-} \right) \left\vert \psi_{n}^{+}\right\rangle \left\langle \psi_{m}^{-}\right\vert
+\mathcal{A}_{n}^{+} \left\vert \psi_{n}^{-}\right\rangle \left\langle \psi_{m}^{-}\right\vert
+ \mathcal{A}_{m}^{-} \left\vert \psi_{n}^{+}\right\rangle \left\langle \psi_{m}^{+}\right\vert
\right] \nonumber \\
\mathcal{D}[\Gamma] \left\vert \psi_{n}^{-}\right\rangle \left\langle \psi_{m}^{+}\right\vert &=
\Gamma \left\vert \psi_{n}^{-}\right\rangle \left\langle \psi_{m}^{+}\right\vert \Gamma^{\dag}
- \frac{1}{2}\left[
\left(\mathcal{B}_{n}^{-} + \mathcal{B}_{m}^{+} \right) \left\vert \psi_{n}^{-}\right\rangle \left\langle \psi_{m}^{+}\right\vert
+\mathcal{A}_{n}^{-} \left\vert \psi_{n}^{+}\right\rangle \left\langle \psi_{m}^{+}\right\vert
+ \mathcal{A}_{m}^{+} \left\vert \psi_{n}^{-}\right\rangle \left\langle \psi_{m}^{-}\right\vert
\right] \nonumber \\
\mathcal{D}[\Gamma] \left\vert \psi_{n}^{-}\right\rangle \left\langle \psi_{m}^{-}\right\vert &=
\Gamma \left\vert \psi_{n}^{-}\right\rangle \left\langle \psi_{m}^{-}\right\vert \Gamma^{\dag}
- \frac{1}{2}\left[
\left(\mathcal{B}_{n}^{-} + \mathcal{B}_{m}^{-} \right) \left\vert \psi_{n}^{-}\right\rangle \left\langle \psi_{m}^{-}\right\vert
+\mathcal{A}_{n}^{-} \left\vert \psi_{n}^{+}\right\rangle \left\langle \psi_{m}^{-}\right\vert
+ \mathcal{A}_{m}^{-} \left\vert \psi_{n}^{-}\right\rangle \left\langle \psi_{m}^{+}\right\vert
\right] .
\end{align}
Letting $\rho_{\perp}$ denote any terms not within the $(n,m)$ subspace, we have
\begin{align}
\Gamma \left\vert \psi_{n}^{+}\right\rangle \left\langle \psi_{m}^{+}\right\vert \Gamma^{\dag} &=
\mathcal{P}_{n}^{+}  \mathcal{P}_{m}^{+} \left\vert \psi_{n}^{+}\right\rangle \left\langle \psi_{m}^{+}\right\vert
+ \mathcal{Q}_{n}^{+}  \mathcal{P}_{m}^{+} \left\vert \psi_{n}^{-}\right\rangle \left\langle \psi_{m}^{+}\right\vert
+ \mathcal{P}_{n}^{+}  \mathcal{Q}_{m}^{+} \left\vert \psi_{n}^{+}\right\rangle \left\langle \psi_{m}^{-}\right\vert
+ \mathcal{Q}_{n}^{+}  \mathcal{Q}_{m}^{+} \left\vert \psi_{n}^{-}\right\rangle \left\langle \psi_{m}^{-}\right\vert
+ \rho_{\perp}
\nonumber \\
\Gamma \left\vert \psi_{n}^{+}\right\rangle \left\langle \psi_{m}^{-}\right\vert \Gamma^{\dag} &=
\mathcal{P}_{n}^{+}  \mathcal{P}_{m}^{-} \left\vert \psi_{n}^{+}\right\rangle \left\langle \psi_{m}^{-}\right\vert
+ \mathcal{Q}_{n}^{+}  \mathcal{P}_{m}^{-} \left\vert \psi_{n}^{-}\right\rangle \left\langle \psi_{m}^{-}\right\vert
+ \mathcal{P}_{n}^{+}  \mathcal{Q}_{m}^{-} \left\vert \psi_{n}^{+}\right\rangle \left\langle \psi_{m}^{+}\right\vert
+ \mathcal{Q}_{n}^{+}  \mathcal{Q}_{m}^{-} \left\vert \psi_{n}^{-}\right\rangle \left\langle \psi_{m}^{+}\right\vert
+ \rho_{\perp}
\nonumber \\
\Gamma \left\vert \psi_{n}^{-}\right\rangle \left\langle \psi_{m}^{+}\right\vert \Gamma^{\dag} &=
\mathcal{P}_{n}^{-}  \mathcal{P}_{m}^{+} \left\vert \psi_{n}^{-}\right\rangle \left\langle \psi_{m}^{+}\right\vert
+ \mathcal{Q}_{n}^{-}  \mathcal{P}_{m}^{+} \left\vert \psi_{n}^{+}\right\rangle \left\langle \psi_{m}^{+}\right\vert
+ \mathcal{P}_{n}^{-}  \mathcal{Q}_{m}^{+} \left\vert \psi_{n}^{-}\right\rangle \left\langle \psi_{m}^{-}\right\vert
+ \mathcal{Q}_{n}^{-}  \mathcal{Q}_{m}^{+} \left\vert \psi_{n}^{+}\right\rangle \left\langle \psi_{m}^{-}\right\vert
+ \rho_{\perp}
\nonumber \\
\Gamma \left\vert \psi_{n}^{-}\right\rangle \left\langle \psi_{m}^{-}\right\vert \Gamma^{\dag} &=
\mathcal{P}_{n}^{-}  \mathcal{P}_{m}^{-} \left\vert \psi_{n}^{-}\right\rangle \left\langle \psi_{m}^{-}\right\vert
+ \mathcal{Q}_{n}^{-}  \mathcal{P}_{m}^{-} \left\vert \psi_{n}^{+}\right\rangle \left\langle \psi_{m}^{-}\right\vert
+ \mathcal{P}_{n}^{-}  \mathcal{Q}_{m}^{-} \left\vert \psi_{n}^{-}\right\rangle \left\langle \psi_{m}^{+}\right\vert
+ \mathcal{Q}_{n}^{-}  \mathcal{Q}_{m}^{-} \left\vert \psi_{n}^{+}\right\rangle \left\langle \psi_{m}^{+}\right\vert
+ \rho_{\perp} .
\end{align}
We can now express the dissipator acting on $(n,m)$ basis states in matrix form,
\begin{equation}
\mathcal{D}[\Gamma]
 \left( 
\begin{array}{c}
\left\vert \psi_{n}^{+}\right\rangle \left\langle \psi_{m}^{+}\right\vert \\
\left\vert \psi_{n}^{+}\right\rangle \left\langle \psi_{m}^{-}\right\vert \\
\left\vert \psi_{n}^{-}\right\rangle \left\langle \psi_{m}^{+}\right\vert \\
\left\vert \psi_{n}^{-}\right\rangle \left\langle \psi_{m}^{-}\right\vert 
\end{array}
\right) = 
\mathcal{D}[\Gamma]^{(n,m)} 
\left( 
\begin{array}{c}
\left\vert \psi_{n}^{+}\right\rangle \left\langle \psi_{m}^{+}\right\vert \\
\left\vert \psi_{n}^{+}\right\rangle \left\langle \psi_{m}^{-}\right\vert \\
\left\vert \psi_{n}^{-}\right\rangle \left\langle \psi_{m}^{+}\right\vert \\
\left\vert \psi_{n}^{-}\right\rangle \left\langle \psi_{m}^{-}\right\vert 
\end{array}
\right) + \rho_{\perp}^{(n,m)} .
\end{equation}
Since $\mathcal{D}[\Gamma] : (n,m) \rightarrow \bigoplus_{n^\prime , m^\prime} (n^\prime \leq n,m^\prime \leq m)$
is block triangular, we need only consider the excitation-conserving submatrices along the
diagonal of $\mathcal{D}[\Gamma]$, which take $(n,m) \rightarrow (n,m)$,
\begin{equation}
 \mathcal{D}[\Gamma] ^{(n,m)} =
\left( 
\begin{array}{cccc}
\mathcal{P}_{n}^{+}  \mathcal{P}_{m}^{+} & \mathcal{P}_{n}^{+}  \mathcal{Q}_{m}^{-} &
\mathcal{Q}_{n}^{-}  \mathcal{P}_{m}^{+} & \mathcal{Q}_{n}^{-}  \mathcal{Q}_{m}^{-} \\
\mathcal{P}_{n}^{+}  \mathcal{Q}_{m}^{+} & \mathcal{P}_{n}^{+}  \mathcal{P}_{m}^{-} &
\mathcal{Q}_{n}^{-}  \mathcal{Q}_{m}^{+} & \mathcal{Q}_{n}^{-}  \mathcal{P}_{m}^{-} \\
\mathcal{Q}_{n}^{+}  \mathcal{P}_{m}^{+} & \mathcal{Q}_{n}^{+}  \mathcal{Q}_{m}^{-} &
\mathcal{P}_{n}^{-}  \mathcal{P}_{m}^{+} & \mathcal{P}_{n}^{-}  \mathcal{Q}_{m}^{-} \\
\mathcal{Q}_{n}^{+}  \mathcal{Q}_{m}^{+} & \mathcal{Q}_{n}^{+}  \mathcal{P}_{m}^{-} &
\mathcal{P}_{n}^{-}  \mathcal{Q}_{m}^{+} & \mathcal{P}_{n}^{-}  \mathcal{P}_{m}^{-}
\end{array}
\right)
-\frac{1}{2} \left( 
\begin{array}{cccc}
\left(\mathcal{B}_{n}^{+} + \mathcal{B}_{m}^{+} \right) & \mathcal{A}_{m}^{-} & \mathcal{A}_{n}^{-} & 0 \\
\mathcal{A}_{m}^{+} & \left(\mathcal{B}_{n}^{+} + \mathcal{B}_{m}^{-} \right) & 0 & \mathcal{A}_{n}^{-} \\
\mathcal{A}_{n}^{+} & 0 & \left(\mathcal{B}_{n}^{-} + \mathcal{B}_{m}^{+} \right) & \mathcal{A}_{m}^{-}\\
0 & \mathcal{A}_{n}^{+} & \mathcal{A}_{m}^{+} &  \left(\mathcal{B}_{n}^{-} + \mathcal{B}_{m}^{-} \right)
\end{array}
\right) .
\end{equation}
To obtain the $(n,0)$, $(0,m)$, and $(0,0)$ excitation-conserving submatrices,
we need to know how $\Gamma$ acts on $\left\vert \psi_{0} \right\rangle$,
\begin{equation}
\Gamma^{\dag}\Gamma \left\vert \psi_{0} \right\rangle = f_0 \left\vert \psi_{0} \right\rangle, \quad
\Gamma \left\vert \psi_{0} \right\rangle = g_0 \left\vert \psi_{0} \right\rangle + \left\vert \psi_{\perp}\right\rangle ,
\end{equation}
where $f_0, g_0$ are constants. Thus, $\mathcal{D}[\Gamma]$ acts on states in the
$(n,0)$ and $(0,m)$ subspaces as
\begin{align}
 \mathcal{D}[\Gamma] \left\vert \psi_{n}^{+}\right\rangle \left\langle \psi_{0}\right\vert &=
\Gamma \left\vert \psi_{n}^{+}\right\rangle \left\langle \psi_{0}\right\vert \Gamma^{\dag}
- \frac{1}{2}\left[
\mathcal{B}_{n}^{+} \left\vert \psi_{n}^{+}\right\rangle \left\langle \psi_{0}\right\vert
+ \mathcal{A}_{n}^{+} \left\vert \psi_{n}^{-}\right\rangle \left\langle \psi_{0}\right\vert
+ f_{0} \left\vert \psi_{n}^{+}\right\rangle \left\langle \psi_{0}\right\vert
\right] \nonumber \\
 \mathcal{D}[\Gamma] \left\vert \psi_{n}^{-}\right\rangle \left\langle \psi_{0}\right\vert &=
\Gamma \left\vert \psi_{n}^{-}\right\rangle \left\langle \psi_{0}\right\vert \Gamma^{\dag}
- \frac{1}{2}\left[
\mathcal{B}_{n}^{-}  \left\vert \psi_{n}^{-}\right\rangle \left\langle \psi_{0}\right\vert
+ \mathcal{A}_{n}^{-} \left\vert \psi_{n}^{+}\right\rangle \left\langle \psi_{0}\right\vert
+ f_{0} \left\vert \psi_{n}^{-}\right\rangle \left\langle \psi_{0}\right\vert
\right] \nonumber \\
 \mathcal{D}[\Gamma] \left\vert \psi_{0}\right\rangle \left\langle \psi_{m}^{+}\right\vert &=
\Gamma \left\vert \psi_{0}\right\rangle \left\langle \psi_{m}^{+}\right\vert \Gamma^{\dag}
- \frac{1}{2}\left[
\mathcal{B}_{m}^{+} \left\vert \psi_{0}\right\rangle \left\langle \psi_{m}^{+}\right\vert
+ \mathcal{A}_{m}^{+} \left\vert \psi_{0}\right\rangle \left\langle \psi_{m}^{-}\right\vert
+ f_{0} \left\vert \psi_{0}\right\rangle \left\langle \psi_{m}^{+}\right\vert
\right] \nonumber \\
 \mathcal{D}[\Gamma] \left\vert \psi_{0}\right\rangle \left\langle \psi_{m}^{-}\right\vert &=
\Gamma \left\vert \psi_{0}\right\rangle \left\langle \psi_{m}^{-}\right\vert \Gamma^{\dag}
- \frac{1}{2}\left[
\mathcal{B}_{m}^{-} \left\vert \psi_{0}\right\rangle \left\langle \psi_{m}^{-}\right\vert
+\mathcal{A}_{m}^{-} \left\vert \psi_{0}\right\rangle \left\langle \psi_{m}^{+}\right\vert
+ f_{0} \left\vert \psi_{0}\right\rangle \left\langle \psi_{m}^{-}\right\vert
\right] ,
\end{align}
\begin{align}
\Gamma \left\vert \psi_{n}^{+}\right\rangle \left\langle \psi_{0}\right\vert \Gamma^{\dag}  &=
\mathcal{P}_{n}^{+} g_{0} \left\vert \psi_{n}^{+}\right\rangle \left\langle \psi_{0}\right\vert
+\mathcal{Q}_{n}^{+} g_{0} \left\vert \psi_{n}^{-}\right\rangle \left\langle \psi_{0}\right\vert
+ \rho_{\perp}
 \nonumber \\
\Gamma \left\vert \psi_{n}^{-}\right\rangle \left\langle \psi_{0}\right\vert \Gamma^{\dag} &=
\mathcal{P}_{n}^{-} g_{0} \left\vert \psi_{n}^{-}\right\rangle \left\langle \psi_{0}\right\vert
+\mathcal{Q}_{n}^{-} g_{0} \left\vert \psi_{n}^{+}\right\rangle \left\langle \psi_{0}\right\vert
+ \rho_{\perp}
 \nonumber \\
\Gamma \left\vert \psi_{0}\right\rangle \left\langle \psi_{m}^{+}\right\vert \Gamma^{\dag} &=
\mathcal{P}_{m}^{+} g_{0} \left\vert \psi_{0}\right\rangle \left\langle \psi_{m}^{+}\right\vert
+\mathcal{Q}_{m}^{+} g_{0} \left\vert \psi_{0}\right\rangle \left\langle \psi_{m}^{-}\right\vert
+ \rho_{\perp}
 \nonumber \\
\Gamma \left\vert \psi_{0}\right\rangle \left\langle \psi_{m}^{-}\right\vert \Gamma^{\dag} &=
\mathcal{P}_{m}^{-} g_{0} \left\vert \psi_{0}\right\rangle \left\langle \psi_{m}^{0}\right\vert
+\mathcal{Q}_{m}^{-} g_{0} \left\vert \psi_{0}\right\rangle \left\langle \psi_{m}^{+}\right\vert
+ \rho_{\perp} ,
\end{align}
and on the $(0,0)$ subspace as
\begin{equation}
\mathcal{D}[\Gamma] \left\vert \psi_{0}\right\rangle \left\langle \psi_{0}\right\vert =
g_{0} g_{0} \left\vert \psi_{0}\right\rangle \left\langle \psi_{0}\right\vert
-  f_{0} \left\vert \psi_{0}\right\rangle \left\langle \psi_{0}\right\vert + \rho_{\perp} .
\end{equation}
The above expressions yield the matrix form of the excitation-conserving
$(n,0) \rightarrow (n,0)$, $(0,m) \rightarrow (0,m)$, and $(0,0) \rightarrow (0,0)$
dissipator submatrices,
\begin{align}
\mathcal{D}[\Gamma] ^{(n,0)} &=
g_{0} \left(
\begin{array}{cc}
\mathcal{P}_{n}^{+} & \mathcal{Q}_{n}^{-} \\
\mathcal{Q}_{n}^{+} & \mathcal{P}_{n}^{-}
\end{array}
\right)
 -\frac{1}{2} \left( 
\begin{array}{cc}
\left(\mathcal{B}_{n}^{+} + f_{0} \right) & \mathcal{A}_{n}^{-} \\
\mathcal{A}_{n}^{+} & \left(\mathcal{B}_{n}^{-} + f_{0} \right)
\end{array}
\right) ,
\nonumber \\
\mathcal{D}[\Gamma] ^{(0,m)} &= 
g_{0} \left(
\begin{array}{cc}
\mathcal{P}_{m}^{+} & \mathcal{Q}_{m}^{-} \\
\mathcal{Q}_{m}^{+} & \mathcal{P}_{m}^{-}
\end{array}
\right)
-\frac{1}{2} \left( 
\begin{array}{cc}
\left(\mathcal{B}_{m}^{+} + f_{0} \right) & \mathcal{A}_{m}^{-} \\
\mathcal{A}_{m}^{+} & \left(\mathcal{B}_{m}^{-} + f_{0} \right)
\end{array}
\right) , \quad
\mathcal{D}[\Gamma] ^{(0,0)}= \left( 
g_{0}^{2} - f_{0}
\right) .
\end{align}
Therefore, to find the block diagonal terms of $\mathcal{D}[\Gamma]^{(n,m)}$ for a given
Lindblad operator $\Gamma$ which takes
$(n,m) \rightarrow \bigoplus_{n^\prime , m^\prime} (n^\prime \leq n,m^\prime \leq m)$, 
one simply needs to identify the set of coefficients
\begin{equation}
\left \{ \mathcal{B}_{n}^{\pm}, \mathcal{A}_{n}^{\pm},
\mathcal{P}_{n}^{\pm},  \mathcal{Q}_{n}^{\pm}, f_{0}, g_{0} \right \}
\end{equation}
and use the above expressions for submatrices $\mathcal{D}[\Gamma]^{(n,m)}$.

\subsection{Coefficients for $\mathcal{D}[a]$ block matrices} \label{appx:Da-mats}

Submatrices for $\mathcal{D}[a]$, are described in previous sections.
Coefficients are then
\begin{equation}
\mathcal{B}_{n}^{+} = \left( n - \sin^{2} \theta_{n} \right), \quad \mathcal{B}_{n}^{-} = \left( n - \cos^{2} \theta_{n} \right), \quad
\mathcal{A}_{n}^{+} = \mathcal{A}_{n}^{-} = \cos \theta_{n} \sin \theta_{n}, \quad
\mathcal{P}_{n}^{\pm} = \mathcal{Q}_{n}^{\pm} = f_{0} = g_{0} = 0 .
\end{equation}

\subsection{Coefficients for $\mathcal{D}[\sigma_{-}]$ block matrices} \label{appx:Dsm-mats}

Similar to $a$, submatrices for $\mathcal{D}[\sigma_{-}]$, are described
in previous sections. Coefficients are
\begin{equation}
\mathcal{B}_{n}^{+} = \sin^{2} \theta_{n}, \quad \mathcal{B}_{n}^{-} = \cos^{2} \theta_{n}, \quad
\mathcal{A}_{n}^{+} = \mathcal{A}_{n}^{-} = -\cos \theta_{n} \sin \theta_{n}, \quad
\mathcal{P}_{n}^{\pm} = \mathcal{Q}_{n}^{\pm} = f_{0} = g_{0} = 0 .
\end{equation}

\subsection{$\mathcal{D}[\sigma_{z}]$ block matrices} \label{appx:Dsz-mats}

Since $\sigma_{z}^{2} = I$,
$\mathcal{D}[\sigma_{z}] \rho = \sigma_{z} \rho \sigma_{z} - \rho$.
Since $\sigma_{z} = 2 \sigma_{+}\sigma_{-} - 1$, it acts on $n>0$ basis states as
\begin{align}
\sigma_{z}\left\vert \psi_{n}^{-}\right\rangle &= \left( 2\cos^{2} \theta_{n} - 1 \right) \left\vert \psi_{n}^{-}\right\rangle -
2 \cos \theta_{n} \sin \theta_{n} \left\vert \psi_{n}^{+}\right\rangle
= \left( \cos^{2} \theta_{n} - \sin^{2} \theta_{n} \right) \left\vert \psi_{n}^{-}\right\rangle -
2 \cos \theta_{n} \sin \theta_{n} \left\vert \psi_{n}^{+}\right\rangle
\nonumber \\
\sigma_{z}\left\vert \psi_{n}^{+}\right\rangle &=\left( 2\sin^{2} \theta_{n} - 1 \right) \left\vert \psi_{n}^{+}\right\rangle -
2 \cos \theta_{n} \sin \theta_{n} \left\vert \psi_{n}^{-}\right\rangle
= -\left( \cos^{2} \theta_{n} - \sin^{2} \theta_{n} \right) \left\vert \psi_{n}^{+}\right\rangle -
2 \cos \theta_{n} \sin \theta_{n} \left\vert \psi_{n}^{-}\right\rangle .
\end{align}
Furthermore, $\sigma_{z} \left\vert \psi_{0}\right\rangle = -\left\vert \psi_{0}\right\rangle$.
Thus, we have coefficients
\begin{equation}
\mathcal{B}_{n}^{+} = \mathcal{B}_{n}^{-} = 1, \quad
\mathcal{A}_{n}^{+} = \mathcal{A}_{n}^{-} = 0, \quad
\mathcal{P}_{n}^{\pm} = \pm \left( \sin^{2} \theta_{n} - \cos^{2} \theta_{n} \right) , \quad
\mathcal{Q}_{n}^{\pm} = - 2 \cos \theta_{n} \sin \theta_{n}, \quad f_{0} = 1, \quad g_{0} = -1.
\end{equation}
Since $\mathcal{P}_{n}^{\pm}, \mathcal{Q}_{n}^{\pm} \neq 0$, denoting
$\pm \mathcal{P}_{n}^{\pm} = P_{n}, \mathcal{Q}_{n}^{\pm} = Q_n$, we
have submatrices of the form
\begin{align}
\mathcal{D}[\sigma_{z}] ^{(0,0)} &= 0 , \quad
\mathcal{D}[\sigma_{z}] ^{(n,m)}=
\left( 
\begin{array}{cccc}
P_{n} P_{m} - 1 & P_{n} Q_{m} & Q_{n} P_{m} & Q_{n} Q_{m} \\
P_{n} Q_{m} & -P_{n} P_{m} - 1 & Q_{n} Q_{m} & -Q_{n} P_{m} \\
Q_{n} P_{m} & Q_{n} Q_{m} & - P_{n} P_{m} - 1 & -P_{n} Q_{m} \\
Q_{n} Q_{m} & -Q_{n} P_{m} &  -P_{n} Q_{m} & P_{n} P_{m} - 1
\end{array}
\right) , \\
 \mathcal{D}[\sigma_{z}] ^{(n,0)} &=
\left( 
\begin{array}{cc}
-P_{n} - 1 & Q_{n} \\
Q_{n} & P_{n} - 1
\end{array}
\right),
\quad
\mathcal{D}[\sigma_{z}] ^{(0,m)}=
\left( 
\begin{array}{cc}
-P_{m} - 1 & Q_{m} \\
Q_{m} & P_{m} - 1
\end{array}
\right) .
\end{align}

\subsection{$\mathcal{D}[a^2]$ block matrices} \label{appx:Da2-mats}

The operator $a^2$ strictly decreases excitations $n \rightarrow n-2$. Also,
since $a^{\dag 2}a^2 = a^{\dag }a \left( a^{\dag }a - 1 \right)$, excitation-conserving
terms relevant to $\mathcal{D}[a^2]$ are
\begin{align}
a^{\dag 2}a^2 \left\vert \psi_{n}^{-}\right\rangle &=
\left( n(n-1) - (2n-2) \cos^{2} \theta_{n} \right) \left\vert \psi_{n}^{-}\right\rangle 
+ (2n-2)\cos \theta_{n} \sin \theta_{n} \left\vert \psi_{n}^{+}\right\rangle \nonumber \\
a^{\dag 2}a^2 \left\vert \psi_{n}^{+}\right\rangle &=
\left( n(n-1) - (2n-2) \sin^{2} \theta_{n} \right) \left\vert \psi_{n}^{+}\right\rangle
+ (2n-2)\cos \theta_{n} \sin \theta_{n} \left\vert \psi_{n}^{-}\right\rangle
\end{align}
and $a^{\dag 2}a^2 \left\vert \psi_{0}\right\rangle = 0 \left\vert \psi_{0}\right\rangle$.
This yields the coefficients
\begin{align}
\mathcal{B}_{n}^{+} &= \left( n(n-1) - 2(n-1) \sin^{2} \theta_{n} \right), \quad
\mathcal{B}_{n}^{-} = \left( n(n-1) - 2(n-1) \cos^{2} \theta_{n} \right), \nonumber \\
\mathcal{A}_{n}^{+} &= \mathcal{A}_{n}^{-} = 2(n-1)\cos \theta_{n} \sin \theta_{n}, \quad
\mathcal{P}_{n}^{\pm} = \mathcal{Q}_{n}^{\pm} = f_{0} = g_{0} = 0 .
\end{align}
Explicitly, submatrices are
\begin{align}
\mathcal{D}[a^2] ^{(0,0)} &= 0 , \quad
\mathcal{D}[a^2] ^{(n,m)} =
 -\frac{1}{2} \left( 
\begin{array}{cccc}
\left(\mathcal{B}_{n}^{+} + \mathcal{B}_{m}^{+} \right) & \mathcal{A}_{m}^{-} & \mathcal{A}_{n}^{-} & 0 \\
\mathcal{A}_{m}^{+} & \left(\mathcal{B}_{n}^{+} + \mathcal{B}_{m}^{-} \right) & 0 & \mathcal{A}_{n}^{-} \\
\mathcal{A}_{n}^{+} & 0 & \left(\mathcal{B}_{n}^{-} + \mathcal{B}_{m}^{+} \right) & \mathcal{A}_{m}^{-} \\
0 & \mathcal{A}_{n}^{+} & \mathcal{A}_{m}^{+} &  \left(\mathcal{B}_{n}^{-} + \mathcal{B}_{m}^{-} \right)
\end{array}
\right) , \\
\mathcal{D}[a^2] ^{(n,0)} &=
 -\frac{1}{2} \left( 
\begin{array}{cc}
\mathcal{B}_{n}^{+} & \mathcal{A}_{n}^{-} \\
\mathcal{A}_{n}^{+} & \mathcal{B}_{n}^{-}
\end{array}
\right),
\quad
\mathcal{D}[a^2] ^{(0,m)} = 
-\frac{1}{2} \left( 
\begin{array}{cc}
\mathcal{B}_{m}^{+}  & \mathcal{A}_{m}^{-} \\
\mathcal{A}_{m}^{+} & \mathcal{B}_{m}^{-} 
\end{array}
\right) .
\end{align}

\section{Derivation of Liouvillian eigenvalues for two qubits} \label{appx:2q-liou}

Consider the Liouvillian of the Tavis-Cummings model with
single-boson dissipation and collective spin damping,
\begin{align}
\mathcal{L\rho} &= -\mathrm{i}\left[ H,\rho \right] +\gamma \mathcal{D}[a]\rho+
\gamma^{\prime}\mathcal{D}[J_{-}]\rho , \quad
H = \omega_{c}\left(a^{\dag}a + J_{z}\right) + g  \left( J_{+} a + J_{-} a^{\dag} \right) , \\
\mathcal{D}[a]\rho &= a\rho a^{\dag}-\frac{1}{2}\left( a^{\dag}a\rho +\rho
a^{\dag}a\right) , \quad
\mathcal{D}[J_{-}]\rho = J_{-}\rho J_{+}-\frac{1}{2}\left(
J_{+}J_{-}\rho +\rho J_{+}J_{-}\right) . \nonumber
\end{align}
For two qubits, $J=0$, $1$. Since all terms in $\mathcal{L}$ preserve $J$,
diagonalization can be performed separately in the $J=0$ and $J=1$ sectors.
For the $J=0$ sector, the system is just a harmonic oscillator with single-boson
dissipation. The eigenvalues of the Liouvillian are
\begin{equation}
\lambda_{n,m}^{J=0} = -\mathrm{i}\omega_{c}(n-m) - \frac{\gamma}{2}(n+m) .
\end{equation}
For the $J=1$ sector, to find the eigenvalues of $\mathcal{L}$, we first
construct a basis for its eigenmatrices, then find its matrix elements in this basis.
Any linear operator can be expanded onto a basis of eigenstates of $H$,
$\left\{ \left\vert \psi_{n}^{s}\right\rangle \right\}$, where $n \in \{g,0,\mathbb{Z}_{+}\}$ is
the eigenvalue of $C = a^{\dag}a + J_{z}$ and $s=0,\pm$ labels states with a given $n$.
Note that $n=g$ is synonymous with $n=-1$ here, for $n = g$, $s = 0$ only,
for $n = 0$, $s = \pm$ only, and for $n>0$, $s = 0,\pm$.
An operator $\rho$ can be written in this basis as
\begin{equation}
\rho =\sum \rho_{nm}^{st}\left\vert \psi_{n}^{s}\right\rangle \left\langle
\psi_{m}^{t}\right\vert \; , \; n,m=g,0,1,2,...; s,t= 0,\pm ,
\end{equation}
where explicitly,
\begin{align}
\left\vert \psi_{g}\right\rangle &= \left\vert 0,-1 \right\rangle , \quad
\left\vert \psi_{0}^{\pm}\right\rangle = \frac{1}{\sqrt{2}}
\left( \pm \left\vert 0,0 \right\rangle + \left\vert 1,-1 \right\rangle \right) , \quad
\left\vert \psi_{n}^{0}\right\rangle =
-\sqrt{\frac{n+1}{2n+1}} \left\vert n-1,1 \right\rangle 
+ \sqrt{\frac{n}{2n+1}} \left\vert n+1,-1 \right\rangle ,
\nonumber \\
\left\vert \psi_{n}^{\pm}\right\rangle &=
\frac{1}{\sqrt{2}}\left(
\sqrt{\frac{n}{2n+1}} \left\vert n-1,1 \right\rangle \pm
\left\vert n,0 \right\rangle + \sqrt{\frac{n+1}{2n+1}} \left\vert n+1,-1 \right\rangle
\right) .
\end{align}
Consider the action of operators in $\mathcal{L}$ on $\left\vert \psi_{n}^{s}\right\rangle$.
First, note that since $\left\vert \psi_{n}^{s}\right\rangle \left\langle \psi_{m}^{t}\right\vert$
are eigenstates of $H$, 
\begin{equation}
\left[ H,\left\vert \psi_{n}^{s}\right\rangle \left\langle \psi_{m}^{t}\right\vert \right] =
(E_{n}^{s}-E_{m}^{t})\left\vert \psi_{n}^{s}\right\rangle \left\langle\psi_{m}^{t}\right\vert
\end{equation}
is diagonal and acts within the subspace of states
$\left\{\left\vert \psi_{n}^{s}\right\rangle \left\langle \psi_{m}^{t}\right\vert\right\}_{s,t}$ with
a definite number of excitations $(n,m)$. Next, observe that since each
$\left\vert \psi_{n}^{s}\right\rangle$ is an eigenstate of $C$,
\begin{equation}
\left\vert \psi_{n}^{s}\right\rangle = \sum_{n_{B},M : n_{B} + M = n} c^{s}_{n_{B},M} \left\vert n_{B},M \right\rangle ,
\end{equation}
 $a$ and $J_{-}$ both decrease $n \rightarrow n-1$ when acting on $\left\vert \psi_{n}^{s}\right\rangle$.
Thus, $a\left\vert \psi_{n}^{s}\right\rangle \left\langle \psi_{m}^{t}\right\vert a^{\dag}$ and
$J_{-}\left\vert\psi_{n}^{s}\right\rangle \left\langle \psi_{m}^{t}\right\vert J_{+}$ both
take states in the $(n,m)$ subspace into the $(n-1,m-1)$ subspace, or annihilate them
if $n=g$ or $m=g$. Now, we can look at how the non-diagonal, excitation-conserving
operators $a^{\dag}a$ and $J_{+}J_{-}$ act on states $\left\vert \psi_{n}^{s}\right\rangle$. First,
observe for $n=g,0$,
\begin{align}
a^{\dag}a \left\vert \psi_{g}\right\rangle &= J_{+}J_{-} \left\vert \psi_{g}\right\rangle = 0 , \quad
a^{\dag}a \left\vert \psi_{0}^{\pm}\right\rangle = \frac{1}{\sqrt{2}}
\left( \left\vert 1,-1 \right\rangle \right) = \frac{1}{2}
\left( \left\vert \psi_{0}^{+}\right\rangle + \left\vert \psi_{0}^{-}\right\rangle \right) , \nonumber \\
J_{+}J_{-} \left\vert \psi_{0}^{\pm}\right\rangle &= \frac{1}{\sqrt{2}}
\left( \pm 2 \left\vert 0,0 \right\rangle \right) =
\pm \left( \left\vert \psi_{0}^{+}\right\rangle - \left\vert \psi_{0}^{-}\right\rangle \right) =
\left( \left\vert \psi_{0}^{\pm}\right\rangle - \left\vert \psi_{0}^{\mp}\right\rangle \right) .
\end{align}
For $n>0$, $a^{\dag}a$ and $J_{+}J_{-}$ act as
\begin{align}
a^{\dag}a \left\vert \psi_{n}^{0}\right\rangle &= \left(n - \frac{1}{2n+1} \right)
\left\vert \psi_{n}^{0}\right\rangle + \frac{\sqrt{2n(n+1)}}{2n+1}
\left(\left\vert \psi_{n}^{+}\right\rangle + \left\vert \psi_{n}^{-}\right\rangle\right) , \nonumber \\
a^{\dag}a \left\vert \psi_{n}^{\pm}\right\rangle &= \left(n + \frac{1}{2}\frac{1}{2n+1} \right)
\left\vert \psi_{n}^{\pm}\right\rangle + \frac{\sqrt{2n(n+1)}}{2n+1} \left\vert \psi_{n}^{0}\right\rangle
+\frac{1}{2} \frac{1}{2n+1} \left\vert \psi_{n}^{\mp}\right\rangle , \nonumber \\
J_{+}J_{-} \left\vert \psi_{n}^{0}\right\rangle &= \frac{2(n+1)}{2n+1}
\left\vert \psi_{n}^{0}\right\rangle - \frac{\sqrt{2n(n+1)}}{2n+1}
\left(\left\vert \psi_{n}^{+}\right\rangle + \left\vert \psi_{n}^{-}\right\rangle\right) , \nonumber \\
J_{+}J_{-} \left\vert \psi_{n}^{\pm}\right\rangle &= - \frac{\sqrt{2n(n+1)}}{2n+1}
\left\vert \psi_{n}^{0}\right\rangle + \left(\frac{n}{2n+1} \pm 1\right) \left\vert \psi_{n}^{+}\right\rangle
+ \left(\frac{n}{2n+1} \mp 1\right) \left\vert \psi_{n}^{-}\right\rangle .
\end{align}
For all cases, $a^{\dag}a$ and $J_{+}J_{-}$ act only within the $(n,m)$ subspace.
The Liouvillian acts as
\begin{align}
 \mathcal{L} \left\vert \psi_{n}^{s}\right\rangle \left\langle \psi_{m}^{t}\right\vert &=
-\mathrm{i} (E_{n}^{s}-E_{m}^{t})\left\vert \psi_{n}^{s}\right\rangle \left\langle\psi_{m}^{t}\right\vert
+ \gamma \left( a\left\vert \psi_{n}^{s}\right\rangle \left\langle \psi_{m}^{t}\right\vert a^{\dag }
- \frac{1}{2} \left( a^{\dag }a\left\vert \psi_{n}^{s}\right\rangle\left\langle \psi_{m}^{t}\right\vert 
+ \left\vert \psi_{n}^{s}\right\rangle\left\langle \psi_{m}^{t}\right\vert a^{\dag }a \right) \right) \nonumber \\
&+ \gamma^\prime \left( J_{-}\left\vert\psi_{n}^{s}\right\rangle \left\langle \psi_{m}^{t}\right\vert J_{+}
- \frac{1}{2} \left( J_{+}J_{-}\left\vert \psi_{n}^{s}\right\rangle \left\langle \psi_{m}^{t}\right\vert 
+ \left\vert \psi_{n}^{s}\right\rangle \left\langle \psi_{m}^{t}\right\vert J_{+}J_{-} \right) \right)
\end{align}
and one can see that all superoperator terms in $\mathcal{L}$ either preserve
eigenvalues of the Casimir operator $C$ and transform states in the
subspace $(n,m) \rightarrow (n,m)$, or decrease the number of bra- and ket-excitations each by one,
$(n,m) \rightarrow (n-1,m-1)$. Thus, the Liouvillian takes on a block triangular form in the 
$\left\vert \psi_{n}^{s}\right\rangle \left\langle \psi_{m}^{t}\right\vert$ basis, and its
eigenvalues can be found by diagonalizing the blocks of matrix elements
which conserve the total number of excitations $(n,m) \rightarrow (n,m)$, denoted
$\mathcal{L}^{(n,m)}$. In the vectorized representation of $\mathcal{L}$ and $\rho$,
\begin{equation}
\mathcal{L} \rho = \left( 
\begin{array}{ccccc}
\mathcal{L}^{(g,g)} & \times & & & \\
0 & \mathcal{L}^{(g,0)} & \times & & \\
 & 0 & \ddots & \times & \\
  & & 0 & \mathcal{L}^{(n,m)} & \times \\
  & & & 0 & \ddots
\end{array}
\right)
 \left( 
\begin{array}{c}
\rho_{(g,g)} \\
\rho_{(g,0)} \\
\vdots \\
\rho_{(n,m)} \\
\vdots
\end{array}
\right) .
\end{equation}
Each $(n,m)$ subspace is degenerate based on the possible spin values $s$ and $t$ that states
$\left\vert \psi_{n}^{s}\right\rangle \left\langle \psi_{m}^{t}\right\vert$ can take. All subspaces and
a relevant basis for each subspace are given by
\begin{align}
 (g,g) &= \left\{ \left\vert \psi_{g} \right\rangle \left\langle \psi_{g} \right\vert \right\} , \quad
(g,0) = \mathrm{span} \left\{ \left\vert \psi_{g} \right\rangle \left\langle \psi_{0}^{t} \right\vert \right\}_{t=\pm} , \quad
(0,g) = \mathrm{span} \left\{ \left\vert \psi_{0}^{s} \right\rangle \left\langle \psi_{g} \right\vert \right\}_{s=\pm} , \nonumber \\
 (n,g) &= \mathrm{span} \left\{ \left\vert \psi_{n}^{s} \right\rangle \left\langle \psi_{g} \right\vert \right\}_{s=0,\pm} , \quad
(g,m) = \mathrm{span} \left\{ \left\vert \psi_{g} \right\rangle \left\langle \psi_{m}^{t} \right\vert \right\}_{t=0,\pm} , \nonumber \\
 (n,0) &= \mathrm{span} \left\{ \left\vert \psi_{n}^{s} \right\rangle \left\langle \psi_{0}^{t} \right\vert \right\}_{s=0,\pm,t=\pm} , \quad
(0,m) = \mathrm{span} \left\{ \left\vert \psi_{0}^{s} \right\rangle \left\langle \psi_{m}^{t} \right\vert \right\}_{s=\pm,t=0,\pm} , \nonumber \\
 (0,0) &= \mathrm{span} \left\{ \left\vert \psi_{0}^{s} \right\rangle \left\langle \psi_{0}^{t} \right\vert \right\}_{s,t=\pm} , \quad
(n,m) = \mathrm{span} \left\{ \left\vert \psi_{n}^{s} \right\rangle \left\langle \psi_{m}^{t} \right\vert \right\}_{s,t=0,\pm} .
\end{align}
We can then study the action of the excitation-conserving terms of $\mathcal{L}$ on each
basis state for each subspace. For the $(g,g)$ subspace, it is easy to see that
$\mathcal{L} \left\vert \psi_{g}\right\rangle \left\langle \psi_{g}\right\vert = 0 = \mathcal{L}^{(0,0)}$.
For the remaining subspaces, it is useful to introduce the following constants
to simplify notation, valid for $n \geq 0$,
\begin{align}
A_{n} &= -\frac{\gamma}{2}\left(n - \frac{1}{2n+1} \right) -
\frac{\gamma^{\prime}}{2}\left(\frac{2(n+1)}{2n+1} \right) , \quad
B_{n} = -\frac{\gamma}{2}\left(n + \frac{1}{2}\frac{1}{2n+1} \right) -
\frac{\gamma^{\prime}}{2}\left(\frac{n}{2n+1} + 1\right) , \nonumber \\
F_{n} &= -\frac{(\gamma - \gamma^{\prime})}{2}\frac{\sqrt{2n(n+1)}}{2n+1} , \quad
G_{n} = -\frac{\gamma}{4} \frac{1}{2n+1} - \frac{\gamma^{\prime}}{2}\left(\frac{n}{2n+1} - 1\right) .
\end{align}
Also, let $\Delta_{st}^{pq} = -\mathrm{i}\left(E_{s}^{p} - E_{t}^{q}\right)$ be the
imaginary part of the Liouvillian in all future expressions.
For $(g,0)$, $\mathcal{L}$ acts on basis states as
\begin{equation}
\mathcal{L} \left\vert \psi_{g}\right\rangle \left\langle \psi_{0}^{\mp}\right\vert =
\left[ -\mathrm{i}(E_{g}-E_{0}^{\mp}) + B_{0} \right]
\left\vert \psi_{g}\right\rangle \left\langle \psi_{0}^{\mp}\right\vert + G_{0}
\left\vert \psi_{g}\right\rangle \left\langle \psi_{0}^{\pm}\right\vert ,
\end{equation}
and similarly for $(0,g)$, $\mathcal{L}$ acts as
\begin{equation}
\mathcal{L} \left\vert \psi_{0}^{\mp}\right\rangle \left\langle \psi_{g}\right\vert =
\left[ -\mathrm{i}(E_{0}^{\mp}-E_{g}) + B_{0} \right]
\left\vert \psi_{0}^{\mp}\right\rangle \left\langle \psi_{g}\right\vert + G_{0}
\left\vert \psi_{0}^{\pm}\right\rangle \left\langle \psi_{g}\right\vert .
\end{equation}
Submatrices describing how $\mathcal{L}$ acts on each subspace,
\begin{equation}
\mathcal{L}
 \left( 
\begin{array}{c}
\left\vert \psi_{g}\right\rangle \left\langle \psi_{0}^{-}\right\vert \\
\left\vert \psi_{g}\right\rangle \left\langle \psi_{0}^{+}\right\vert
\end{array}
\right) = \mathcal{L}^{(g,0)}  \left( 
\begin{array}{c}
\left\vert \psi_{g}\right\rangle \left\langle \psi_{0}^{-}\right\vert \\
\left\vert \psi_{g}\right\rangle \left\langle \psi_{0}^{+}\right\vert
\end{array}
\right) , \quad
\mathcal{L}
 \left( 
\begin{array}{c}
\left\vert \psi_{0}^{-}\right\rangle \left\langle \psi_{g}\right\vert \\
\left\vert \psi_{0}^{+}\right\rangle \left\langle \psi_{g}\right\vert
\end{array}
\right) = \mathcal{L}^{(0,g)}  \left( 
\begin{array}{c}
\left\vert \psi_{0}^{-}\right\rangle \left\langle \psi_{g}\right\vert \\
\left\vert \psi_{0}^{+}\right\rangle \left\langle \psi_{g}\right\vert
\end{array}
\right) ,
\end{equation}
can then be written as
\begin{equation}
\mathcal{L}^{(g,0)} = \left( 
\begin{array}{cc}
\Gamma_{g0}^{-} & G_{0} \\ 
G_{0} & \Gamma_{g0}^{+}
\end{array}
\right) , \quad
\mathcal{L}^{(0,g)}=\left( 
\begin{array}{cc}
\Gamma_{0g}^{-} & G_{0} \\ 
G_{0} & \Gamma_{0g}^{+}
\end{array}
\right) ,
\end{equation}
where $\Gamma_{g0}^{\pm} = \Delta_{g0}^{\pm} + B_{0}$
and $\Gamma_{0g}^{\pm} = \Delta_{0g}^{\pm} + B_{0}$.
For $(n,g)$, $\mathcal{L}$ acts on basis states as
\begin{align}
 \mathcal{L} \left\vert \psi_{n}^{0}\right\rangle \left\langle \psi_{g}\right\vert &=
\left[ -\mathrm{i}(E_{n}^{0}-E_{g}) + A_{n} \right]
\left\vert \psi_{n}^{0}\right\rangle \left\langle \psi_{g}\right\vert
+ F_{n}
\left\vert \psi_{n}^{-}\right\rangle \left\langle \psi_{g}\right\vert + F_{n}
\left\vert \psi_{n}^{+}\right\rangle \left\langle \psi_{g}\right\vert  , \nonumber \\
 \mathcal{L} \left\vert \psi_{n}^{\mp}\right\rangle \left\langle \psi_{g}\right\vert &=
\left[ -\mathrm{i}(E_{n}^{\mp}-E_{g}) + B_{n} \right]
\left\vert \psi_{n}^{\mp}\right\rangle \left\langle \psi_{g}\right\vert
+ G_{n}
\left\vert \psi_{n}^{\pm}\right\rangle \left\langle \psi_{g}\right\vert + F_{n}
\left\vert \psi_{n}^{0}\right\rangle \left\langle \psi_{g}\right\vert ,
\end{align}
and for $(g,m)$, $\mathcal{L}$ acts on basis states as
\begin{align}
 \mathcal{L} \left\vert \psi_{g}\right\rangle \left\langle \psi_{m}^{0}\right\vert &=
\left[ -\mathrm{i}(E_{g}-E_{m}^{0}) + A_{m} \right]
\left\vert \psi_{g}\right\rangle \left\langle \psi_{m}^{0}\right\vert
+ F_{m}
\left\vert \psi_{g}\right\rangle \left\langle \psi_{m}^{-}\right\vert + F_{m}
\left\vert \psi_{g}\right\rangle \left\langle \psi_{m}^{+}\right\vert  , \nonumber \\
 \mathcal{L} \left\vert \psi_{g}\right\rangle \left\langle \psi_{m}^{\mp}\right\vert &=
\left[ -\mathrm{i}(E_{g}-E_{m}^{\mp}) + B_{m} \right]
\left\vert \psi_{g}\right\rangle \left\langle \psi_{m}^{\mp}\right\vert
+ G_{m}
\left\vert \psi_{g}\right\rangle \left\langle \psi_{m}^{\pm}\right\vert + F_{m}
\left\vert \psi_{g}\right\rangle \left\langle \psi_{m}^{0}\right\vert .
\end{align}
Submatrices can be written as
\begin{align}
 \mathcal{L}
\left( 
\begin{array}{c}
\left\vert \psi_{n}^{0}\right\rangle \left\langle \psi_{g}\right\vert \\
\left\vert \psi_{n}^{-}\right\rangle \left\langle \psi_{g}\right\vert \\
\left\vert \psi_{n}^{+}\right\rangle \left\langle \psi_{g}\right\vert
\end{array}
\right) = \mathcal{L}^{(n,g)}  \left( 
\begin{array}{c}
\left\vert \psi_{n}^{0}\right\rangle \left\langle \psi_{g}\right\vert \\
\left\vert \psi_{n}^{-}\right\rangle \left\langle \psi_{g}\right\vert \\
\left\vert \psi_{n}^{+}\right\rangle \left\langle \psi_{g}\right\vert
\end{array}
\right) , &\quad
\mathcal{L}
\left( 
\begin{array}{c}
\left\vert \psi_{g}\right\rangle \left\langle \psi_{m}^{0}\right\vert \\
\left\vert \psi_{g}\right\rangle \left\langle \psi_{m}^{-}\right\vert \\
\left\vert \psi_{g}\right\rangle \left\langle \psi_{m}^{+}\right\vert
\end{array}
\right) = \mathcal{L}^{(g,m)}  \left( 
\begin{array}{c}
\left\vert \psi_{g}\right\rangle \left\langle \psi_{m}^{0}\right\vert \\
\left\vert \psi_{g}\right\rangle \left\langle \psi_{m}^{-}\right\vert \\
\left\vert \psi_{g}\right\rangle \left\langle \psi_{m}^{+}\right\vert
\end{array}
\right) , \nonumber \\
\mathcal{L}^{(n,g)} = \left( 
\begin{array}{ccc}
\Gamma_{ng}^{0}  & F_{n} & F_{n} \\ 
F_{n} & \Gamma_{ng}^{-}  & G_{n} \\
F_{n} & G_{n} & \Gamma_{ng}^{+} 
\end{array}
\right) , &\quad
\mathcal{L}^{(g,m)} = \left( 
\begin{array}{ccc}
\Gamma_{gm}^{0}  & F_{m} & F_{m} \\ 
F_{m} & \Gamma_{gm}^{-}  & G_{m} \\
F_{m} & G_{m} & \Gamma_{gm}^{+} 
\end{array}
\right) ,
\end{align}
with $\Gamma_{ng}^{0} = \Delta_{ng}^{0} + A_{n}$, $\Gamma_{gm}^{0} = \Delta_{gm}^{0} + A_{m}$,
$\Gamma_{ng}^{\pm} = \Delta_{ng}^{\pm} + B_{n}$, and $\Gamma_{gm}^{\pm} = \Delta_{gm}^{\pm} + B_{m}$.

For the remaining subspaces, there will be nonzero off-block diagonal matrix elements
which take $(n,m) \rightarrow (n-1,m-1)$, namely
$a\left\vert\psi_{n}^{s}\right\rangle \left\langle \psi_{m}^{t}\right\vert a^{\dag}$ and
$J_{-}\left\vert\psi_{n}^{s}\right\rangle \left\langle \psi_{m}^{t}\right\vert J_{+}$ terms, however
we need not consider them in writing down the relevant submatrices to obtain eigenvalues,
and will encapsulate all such terms by writing the state $\rho_{\perp}$. 
These terms only are important when trying to obtain eigenvectors of $\mathcal{L}$.
For $(0,0)$, $\mathcal{L}$ acts on basis states as
\begin{equation}
 \mathcal{L} \left\vert \psi_{n}^{\mp}\right\rangle \left\langle \psi_{m}^{\mp}\right\vert =
\left[ -\mathrm{i}(E_{n}^{\mp}-E_{m}^{\mp}) + 2B_{0} \right]
\left\vert \psi_{n}^{\mp}\right\rangle \left\langle \psi_{m}^{\mp}\right\vert
+ G_{0}
\left\vert \psi_{n}^{\pm}\right\rangle \left\langle \psi_{m}^{\mp}\right\vert + G_{0}
\left\vert \psi_{n}^{\mp}\right\rangle \left\langle \psi_{m}^{\pm}\right\vert ,
\end{equation}
and the submatrix can be written as
\begin{equation}
\mathcal{L} \left( 
\begin{array}{c}
\left\vert \psi_{0}^{-}\right\rangle \left\langle \psi_{0}^{-}\right\vert \\
\left\vert \psi_{0}^{-}\right\rangle \left\langle \psi_{0}^{+}\right\vert \\
\left\vert \psi_{0}^{+}\right\rangle \left\langle \psi_{0}^{-}\right\vert \\
\left\vert \psi_{0}^{+}\right\rangle \left\langle \psi_{0}^{+}\right\vert
\end{array}
\right) = \mathcal{L}^{(0,0)}  \left( 
\begin{array}{c}
\left\vert \psi_{0}^{-}\right\rangle \left\langle \psi_{0}^{-}\right\vert \\
\left\vert \psi_{0}^{-}\right\rangle \left\langle \psi_{0}^{+}\right\vert \\
\left\vert \psi_{0}^{+}\right\rangle \left\langle \psi_{0}^{-}\right\vert \\
\left\vert \psi_{0}^{+}\right\rangle \left\langle \psi_{0}^{+}\right\vert
\end{array}
\right) + \rho_{\perp}^{(0,0)} , \quad 
\mathcal{L}^{(0,0)} = \left( 
\begin{array}{cccc}
\Gamma_{00}^{--} & G_{0} & G_{0} & 0 \\
G_{0} & \Gamma_{00}^{-+} & 0 & G_{0} \\
G_{0} & 0 & \Gamma_{00}^{+-} & G_{0} \\
0 & G_{0} & G_{0} & \Gamma_{00}^{++}
\end{array}
\right) ,
\end{equation}
with $\Gamma_{00}^{st} = \Delta_{00}^{st} + 2B_{0}$. For $(n,0)$,
$\mathcal{L}$ acts on basis states as
\begin{align}
\mathcal{L} \left\vert \psi_{n}^{0}\right\rangle \left\langle \psi_{0}^{\mp}\right\vert &=
\left[ -\mathrm{i}(E_{n}^{0}-E_{0}^{\mp}) + A_{n} + B_{0} \right]
\left\vert \psi_{n}^{0}\right\rangle \left\langle \psi_{0}^{\mp}\right\vert
+ F_{n}
\left\vert \psi_{n}^{-}\right\rangle \left\langle \psi_{0}^{\mp}\right\vert + F_{n}
\left\vert \psi_{n}^{+}\right\rangle \left\langle \psi_{0}^{\mp}\right\vert + G_{0}
\left\vert \psi_{n}^{0}\right\rangle \left\langle \psi_{0}^{\pm}\right\vert, \nonumber \\
\mathcal{L} \left\vert \psi_{n}^{\mp}\right\rangle \left\langle \psi_{0}^{\mp}\right\vert &=
\left[ -\mathrm{i}(E_{n}^{\mp}-E_{0}^{\mp}) + B_{n} + B_{0} \right]
\left\vert \psi_{n}^{\mp}\right\rangle \left\langle \psi_{0}^{\mp}\right\vert 
+ G_{n}
\left\vert \psi_{n}^{\pm}\right\rangle \left\langle \psi_{0}^{\mp}\right\vert + F_{n}
\left\vert \psi_{n}^{0}\right\rangle \left\langle \psi_{0}^{\mp}\right\vert + G_{0}
\left\vert \psi_{n}^{\mp}\right\rangle \left\langle \psi_{0}^{\pm}\right\vert ,
\end{align}
and for $(0,m)$, $\mathcal{L}$ acts on basis states as
\begin{align}
\mathcal{L} \left\vert \psi_{0}^{\mp}\right\rangle \left\langle \psi_{m}^{0}\right\vert &=
\left[ -\mathrm{i}(E_{0}^{\mp}-E_{m}^{0}) + A_{m} + B_{0} \right]
\left\vert \psi_{0}^{\mp}\right\rangle \left\langle \psi_{m}^{0}\right\vert
+ F_{m}
\left\vert \psi_{0}^{\mp}\right\rangle \left\langle \psi_{m}^{-}\right\vert + F_{m}
\left\vert \psi_{0}^{\mp}\right\rangle \left\langle \psi_{m}^{+}\right\vert + G_{0}
\left\vert \psi_{0}^{\pm}\right\rangle \left\langle \psi_{m}^{0}\right\vert , \nonumber \\
\mathcal{L} \left\vert \psi_{0}^{\mp}\right\rangle \left\langle \psi_{m}^{\mp}\right\vert &=
\left[ -\mathrm{i}(E_{0}^{\mp}-E_{m}^{\mp}) + B_{m} + B_{0} \right]
\left\vert \psi_{0}^{\mp}\right\rangle \left\langle \psi_{m}^{\mp}\right\vert
+ G_{m}
\left\vert \psi_{0}^{\mp}\right\rangle \left\langle \psi_{m}^{\pm}\right\vert + F_{m}
\left\vert \psi_{0}^{\mp}\right\rangle \left\langle \psi_{m}^{0}\right\vert + G_{0}
\left\vert \psi_{0}^{\pm}\right\rangle \left\langle \psi_{m}^{\mp}\right\vert .
\end{align}
Submatrices can be written as
\begin{align}
\mathcal{L}
\left( 
\begin{array}{c}
\left\vert \psi_{n}^{0}\right\rangle \left\langle \psi_{0}^{-}\right\vert \\
\left\vert \psi_{n}^{0}\right\rangle \left\langle \psi_{0}^{+}\right\vert \\
\left\vert \psi_{n}^{-}\right\rangle \left\langle \psi_{0}^{-}\right\vert \\
\left\vert \psi_{n}^{-}\right\rangle \left\langle \psi_{0}^{+}\right\vert \\
\left\vert \psi_{n}^{+}\right\rangle \left\langle \psi_{0}^{-}\right\vert \\
\left\vert \psi_{n}^{+}\right\rangle \left\langle \psi_{0}^{+}\right\vert
\end{array}
\right) = \mathcal{L}^{(n,0)} \left( 
\begin{array}{c}
\left\vert \psi_{n}^{0}\right\rangle \left\langle \psi_{0}^{-}\right\vert \\
\left\vert \psi_{n}^{0}\right\rangle \left\langle \psi_{0}^{+}\right\vert \\
\left\vert \psi_{n}^{-}\right\rangle \left\langle \psi_{0}^{-}\right\vert \\
\left\vert \psi_{n}^{-}\right\rangle \left\langle \psi_{0}^{+}\right\vert \\
\left\vert \psi_{n}^{+}\right\rangle \left\langle \psi_{0}^{-}\right\vert \\
\left\vert \psi_{n}^{+}\right\rangle \left\langle \psi_{0}^{+}\right\vert
\end{array}
\right) + \rho_{\perp}^{(n,0)} , &\quad
\mathcal{L}
\left( 
\begin{array}{c}
\left\vert \psi_{0}^{-}\right\rangle \left\langle \psi_{m}^{0}\right\vert \\
\left\vert \psi_{0}^{+}\right\rangle \left\langle \psi_{m}^{0}\right\vert \\
\left\vert \psi_{0}^{-}\right\rangle \left\langle \psi_{m}^{-}\right\vert \\
\left\vert \psi_{0}^{+}\right\rangle \left\langle \psi_{m}^{-}\right\vert \\
\left\vert \psi_{0}^{-}\right\rangle \left\langle \psi_{m}^{+}\right\vert \\
\left\vert \psi_{0}^{+}\right\rangle \left\langle \psi_{m}^{+}\right\vert
\end{array}
\right) = \mathcal{L}^{(0,m)} \left( 
\begin{array}{c}
\left\vert \psi_{0}^{-}\right\rangle \left\langle \psi_{m}^{0}\right\vert \\
\left\vert \psi_{0}^{+}\right\rangle \left\langle \psi_{m}^{0}\right\vert \\
\left\vert \psi_{0}^{-}\right\rangle \left\langle \psi_{m}^{-}\right\vert \\
\left\vert \psi_{0}^{+}\right\rangle \left\langle \psi_{m}^{-}\right\vert \\
\left\vert \psi_{0}^{-}\right\rangle \left\langle \psi_{m}^{+}\right\vert \\
\left\vert \psi_{0}^{+}\right\rangle \left\langle \psi_{m}^{+}\right\vert
\end{array}
\right) + \rho_{\perp}^{(0,m)} , \nonumber \\
\mathcal{L}^{(n,0)} = \left( 
\begin{array}{cccccc}
\Gamma_{n0}^{0-}   & G_{0} & F_{n} & 0 & F_{n} & 0 \\
G_{0} & \Gamma_{n0}^{0+}   & 0 & F_{n} & 0 & F_{n} \\
F_{n} & 0 & \Gamma_{n0}^{--}   & G_{0} & G_{n} & 0 \\
0 & F_{n} & G_{0} & \Gamma_{n0}^{-+}   & 0 & G_{n} \\
F_{n} & 0 & G_{n} & 0 & \Gamma_{n0}^{+-}   & G_{0} \\
0 & F_{n} & 0 & G_{n} & G_{0} & \Gamma_{n0}^{++}  
\end{array}
\right) , &\quad
\mathcal{L}^{(0,m)} = \left( 
\begin{array}{cccccc}
\Gamma_{0m}^{-0}   & G_{0} & F_{m} & 0 & F_{m} & 0 \\
G_{0} & \Gamma_{0m}^{+0}   & 0 & F_{m} & 0 & F_{m} \\
F_{m} & 0 & \Gamma_{0m}^{--}   & G_{0} & G_{m} & 0 \\
0 & F_{m} & G_{0} & \Gamma_{0m}^{-+}   & 0 & G_{m} \\
F_{m} & 0 & G_{m} & 0 & \Gamma_{0m}^{+-}   & G_{0} \\
0 & F_{m} & 0 & G_{m} & G_{0} & \Gamma_{0m}^{++}  
\end{array}
\right) ,
\end{align}
with $\Gamma_{n0}^{0\pm} = \Delta_{n0}^{0\pm} + A_{n} + B_{0}$,
$\Gamma_{0m}^{\pm 0} = \Delta_{0m}^{\pm 0} + A_{m} + B_{0}$, 
$\Gamma_{n0}^{st} = \Delta_{n0}^{st} + B_{n} + B_{0}$, and
$\Gamma_{0m}^{st} = \Delta_{0m}^{st} + B_{m} + B_{0}$.
Lastly, for $(n,m)$, $\mathcal{L}$ acts on basis states as
\begin{align}
 \mathcal{L} \left\vert \psi_{n}^{0}\right\rangle \left\langle \psi_{m}^{0}\right\vert &=
\left[ -\mathrm{i}(E_{n}^{0}-E_{m}^{0}) + A_{n} + A_{m} \right]
\left\vert \psi_{n}^{0}\right\rangle \left\langle \psi_{m}^{0}\right\vert
\nonumber \\
&+ F_{n}
\left(\left\vert \psi_{n}^{-}\right\rangle + \left\vert \psi_{n}^{+}\right\rangle \right)
\left\langle \psi_{m}^{0}\right\vert + F_{m}
\left\vert \psi_{n}^{0}\right\rangle
\left( \left\langle \psi_{m}^{-}\right\vert + \left\langle \psi_{m}^{+}\right\vert \right) , \nonumber \\
 \mathcal{L} \left\vert \psi_{n}^{0}\right\rangle \left\langle \psi_{m}^{\mp}\right\vert &=
\left[ -\mathrm{i}(E_{n}^{0}-E_{m}^{\mp}) + A_{n} + B_{m} \right]
\left\vert \psi_{n}^{0}\right\rangle \left\langle \psi_{m}^{\mp}\right\vert
\nonumber \\
&+ F_{m}
\left\vert \psi_{n}^{0}\right\rangle \left\langle \psi_{m}^{0}\right\vert + F_{n}
\left(\left\vert \psi_{n}^{-}\right\rangle + \left\vert \psi_{n}^{+}\right\rangle \right)
\left\langle \psi_{m}^{\mp}\right\vert + G_{m}
\left\vert \psi_{n}^{0}\right\rangle \left\langle \psi_{m}^{\pm}\right\vert , \nonumber \\
 \mathcal{L} \left\vert \psi_{n}^{\mp}\right\rangle \left\langle \psi_{m}^{0}\right\vert &=
\left[ -\mathrm{i}(E_{n}^{\mp}-E_{m}^{0}) + B_{n} + A_{m} \right]
\left\vert \psi_{n}^{\mp}\right\rangle \left\langle \psi_{m}^{0}\right\vert
\nonumber \\
&+ F_{n}
\left\vert \psi_{n}^{0}\right\rangle \left\langle \psi_{m}^{0}\right\vert + F_{m}
\left\vert \psi_{n}^{\mp}\right\rangle
\left( \left\langle \psi_{m}^{-}\right\vert + \left\langle \psi_{m}^{+}\right\vert \right) + G_{m}
\left\vert \psi_{n}^{\pm}\right\rangle \left\langle \psi_{m}^{0}\right\vert , \nonumber \\
 \mathcal{L} \left\vert \psi_{n}^{\mp}\right\rangle \left\langle \psi_{m}^{\mp}\right\vert &=
\left[ -\mathrm{i}(E_{n}^{\mp}-E_{m}^{\mp}) + B_{n} + B_{m} \right]
\left\vert \psi_{n}^{\mp}\right\rangle \left\langle \psi_{m}^{\mp}\right\vert
\nonumber \\
&+ F_{n}
\left\vert \psi_{n}^{0}\right\rangle \left\langle \psi_{m}^{\mp}\right\vert + F_{m}
\left\vert \psi_{n}^{\mp}\right\rangle \left\langle \psi_{m}^{0}\right\vert
+ G_{n} \left\vert \psi_{n}^{\pm}\right\rangle \left\langle \psi_{m}^{\mp}\right\vert
+ G_{m} \left\vert \psi_{n}^{\mp}\right\rangle \left\langle \psi_{m}^{\pm}\right\vert .
\end{align}
Submatrices can then be written as
\begin{eqnarray}
\mathcal{L}
\left( 
\begin{array}{c}
\left\vert \psi_{n}^{0}\right\rangle \left\langle \psi_{m}^{0}\right\vert \\
\left\vert \psi_{n}^{0}\right\rangle \left\langle \psi_{m}^{-}\right\vert \\
\left\vert \psi_{n}^{0}\right\rangle \left\langle \psi_{m}^{+}\right\vert \\
\left\vert \psi_{n}^{-}\right\rangle \left\langle \psi_{m}^{0}\right\vert \\
\left\vert \psi_{n}^{+}\right\rangle \left\langle \psi_{m}^{0}\right\vert \\
\left\vert \psi_{n}^{-}\right\rangle \left\langle \psi_{m}^{-}\right\vert \\
\left\vert \psi_{n}^{-}\right\rangle \left\langle \psi_{m}^{+}\right\vert \\
\left\vert \psi_{n}^{+}\right\rangle \left\langle \psi_{m}^{-}\right\vert \\
\left\vert \psi_{n}^{+}\right\rangle \left\langle \psi_{m}^{+}\right\vert
\end{array}
\right) = \mathcal{L}^{(n,m)} \left( 
\begin{array}{c}
\left\vert \psi_{n}^{0}\right\rangle \left\langle \psi_{m}^{0}\right\vert \\
\left\vert \psi_{n}^{0}\right\rangle \left\langle \psi_{m}^{-}\right\vert \\
\left\vert \psi_{n}^{0}\right\rangle \left\langle \psi_{m}^{+}\right\vert \\
\left\vert \psi_{n}^{-}\right\rangle \left\langle \psi_{m}^{0}\right\vert \\
\left\vert \psi_{n}^{+}\right\rangle \left\langle \psi_{m}^{0}\right\vert \\
\left\vert \psi_{n}^{-}\right\rangle \left\langle \psi_{m}^{-}\right\vert \\
\left\vert \psi_{n}^{-}\right\rangle \left\langle \psi_{m}^{+}\right\vert \\
\left\vert \psi_{n}^{+}\right\rangle \left\langle \psi_{m}^{-}\right\vert \\
\left\vert \psi_{n}^{+}\right\rangle \left\langle \psi_{m}^{+}\right\vert
\end{array}
\right) + \rho_{\perp}^{(n,m)} ,
\nonumber \\
\mathcal{L}^{(n,m)} = \left( 
\begin{array}{ccccccccc}
\Gamma_{nm}^{00}   & F_{m} & F_{m} & F_{n} & F_{n} & 0 & 0 & 0 & 0 \\
F_{m} & \Gamma_{nm}^{0-}   & G_{m} & 0 & 0 & F_{n} & 0 & F_{n} & 0 \\
F_{m} & G_{m} & \Gamma_{nm}^{0+}   & 0 & 0 & 0 & F_{n} & 0 & F_{n} \\
F_{n} & 0 & 0 & \Gamma_{nm}^{-0}   & G_{n} & F_{m} & 0 & F_{m} & 0 \\
F_{n} & 0 & 0 & G_{n} & \Gamma_{nm}^{+0}   & 0 & F_{m} & 0 & F_{m} \\
0 & F_{n} & 0 & F_{m} & 0 & \Gamma_{nm}^{--}   & G_{m} & G_{n} & 0 \\
0 & 0 & F_{n} & 0 & F_{m} & G_{m} & \Gamma_{nm}^{-+}   & 0 & G_{n} \\
0 & F_{n} & 0 & F_{m} & 0 & G_{n} & 0 & \Gamma_{nm}^{+-}   & G_{m} \\
0 & 0 & F_{n} & 0 & F_{m} & 0 & G_{n} & G_{m} & \Gamma_{nm}^{++}  
\end{array}
\right) ,
\end{eqnarray}
with $\Gamma_{nm}^{00} = \Delta_{nm}^{00} + A_{n} + A_{m}$,
$\Gamma_{nm}^{0\pm} = \Delta_{nm}^{0\pm} + A_{n} + B_{m}$,
$\Gamma_{nm}^{\pm 0} = \Delta_{nm}^{\pm 0} + B_{n} + A_{m}$, and
$\Gamma_{nm}^{st} = \Delta_{nm}^{st} + B_{n} + B_{m}$.
Each block $\mathcal{L}^{(n,m)}$ can be diagonalized individually to obtain
eigenvalues of $\mathcal{L}$ corresponding to the $J=1$ sector,
and all blocks are at most $9 \times 9$ matrices. 

\end{document}